\documentclass[aps,prb,amsmath,amssymb,floatfix,twocolumn]{revtex4-1}
\usepackage{graphicx}
\usepackage{subfigure}
\usepackage{amsmath}
\usepackage{bbold}
\usepackage{slashed}
\usepackage{amssymb}
\usepackage{braket}
\usepackage{array}
\usepackage[colorlinks]{hyperref}
\usepackage{float}
\usepackage[margin=1in]{geometry}
\usepackage{times}

\newcommand{\ncmd}{\newcommand}
\ncmd{\nn}{\nonumber}
\ncmd{\pg}[1]{\textcolor{red}{#1}}
\ncmd{\mbf}[1]{\bs{#1}}
\ncmd{\Lam}{\Lambda}
\ncmd{\lam}{\lambda}
\ncmd{\Gam}{\Gamma}
\ncmd{\gam}{\gamma}
\ncmd{\sig}{\sigma}
\ncmd{\Dl}{\Delta}
\ncmd{\dl}{\delta}
\ncmd{\kap}{\kappa}
\ncmd{\Om}{\Omega}
\ncmd{\om}{\omega}
\ncmd{\mc}{\mathcal}
\ncmd{\eps}{\epsilon}
\ncmd{\veps}{\varepsilon}
\ncmd{\vphi}{\varphi}
\ncmd{\vtheta}{\vartheta}
\ncmd{\note}[1]{{\color{red}{#1}}}
\ncmd{\new}[1]{{\texttt{#1}  } }
\ncmd{\eq}[1]{Eq. \eqref{#1}}
\ncmd{\bs}{\boldsymbol}
\ncmd{\pll}{\parallel}
\ncmd{\dsty}{\displaystyle}

\begin{document}

\title{Part I: Staggered index and 3D winding number of Kramers-degenerate bands}
\author{Alexander C. Tyner$^{1}$ and Pallab Goswami$^{1,2}$}
\affiliation{$^{1}$ Graduate Program in Applied Physics, Northwestern University, Evanston, Illinois, 60208, USA}
\affiliation{$^{2}$ Department of Physics and Astronomy, Northwestern University, Evanston, Illinois, 60208, USA}

\date{\today}

\begin{abstract} 
For three-dimensional (3D) crystalline insulators, preserving space-inversion ($\mathcal{P}$) and time-reversal ($\mathcal{T}$) symmetries, the third homotopy class of two-fold, Kramers-degenerate bands is described by a 3D winding number $n_{3,j} \in \mathbb{Z}$, where $j$ is the band index. It governs space group symmetry-protected, instanton or tunneling configurations of $SU(2)$ Berry connection, and the quantization of magneto-electric coefficient $\theta_j = n_{3,j} \pi$. We show that $|n_{3,j}|$ for realistic, \emph{ab initio} band structures can be identified from a staggered symmetry-indicator $\kappa_{AF,j} \in \mathbb{Z}$ and the gauge-invariant spectrum of $SU(2)$ Wilson loops. The procedure is elucidated for $4$-band and $8$-band tight-binding models and \emph{ab initio} band structure of Bi, which is a $\mathbb{Z}_2$-trivial, higher-order, topological crystalline insulator. When the tunneling is protected by $C_{nh}$ and $D_{nh}$ point groups, the proposed method can also identify the signed winding number $n_{3,j}$. Our analysis distinguishes between magneto-electrically trivial ($\theta=0$) and non-trivial ($\theta=2 s \pi$, with $s \neq 0$) topological crystalline insulators. In Part II, we demonstrate $\mathbb{Z}$-classification of $\theta$ by computing induced electric charge (Witten effect) on magnetic Dirac monopoles. \end{abstract}

\maketitle  

\section{Introduction}\label{intro}
Band structures of $\mathcal{PT}$ symmetric materials are described by $2N \times 2N$ Bloch Hamiltonian matrix
$\hat{H}(\mathbf{k})=\sum_{j=1}^{N} E_j(\bs{k}) \hat{P}_j(\bs{k})$, where $N$, $E_j(\bs{k})$, $\hat{P}_j(\bs{k})$ respectively correspond to the total number, the energy eigenvalues, and the projection operators of two-fold Kramers-degenerate bands, and $\bs{k}$ is the wave vector. Since $E_j(\bs{k})$ and $\hat{P}_j(\bs{k})$ remain unchanged by $U(2)$ gauge transformations of Bloch wave functions of Kramers pairs, $\hat{H}(\bs{k})$ describes maps from \emph{crystalline space groups} to the coset space 
$\frac{U(2N)}{U(2) \times ...\times U(2)}=\frac{U(2N)}{[U(2)]^N }$.
The objective of topological band theory is to classify such maps with appropriate bulk invariants.~\cite{Kane2005,bernevig2006quantum,FuKaneMele2007,FuKane,Moore2007,ZhangFT,RyuLudwigPRB,Roy2009,Roy20093D,ryu2010topological,Hassan2010,Qi2011,slager2013space,Chiu2016}  

For three-dimensional (3D) insulators, a $4$-component unit vector $\bs{\hat{d}}_j(\bs{k})$ can be embedded in $\hat{P}_j(\bs{k})$, which wraps around the Brillouin zone (BZ) three-torus. Such instanton or tunneling configurations of $\bs{\hat{d}}_j(\bs{k})$ can be classified by the third spherical homotopy group $\pi_3(S^3)=\mathbb{Z}$, leading to the 3D winding number $n_{3j} \in \mathbb{Z}$. When $n_{3j} \neq 0$, $U(2)$ Berry connection $\bs{A}_j(\bs{k})$ inherits 3D tunneling configurations. Therefore, Wilson loop calculations can facilitate identification of $n_{3,j}$.

Exploiting rotation and mirror symmetries, $U(2)$ redundancy of Bloch wave functions can be reduced to $U(1) \times U(1)$ (or a smaller discrete sub-group). If such gauge-fixing procedure is properly implemented, the 3D winding number can be related to the Chern-Simons coefficient~\cite{ZhangFT,ryu2010topological,EssinMagnetoelectric,Essin2010,malashevich2010theory,Coh2011,varnava2020}
\begin{eqnarray}\label{A11}
\mathcal{CS}_j &=&\frac{1}{8\pi^2} \;  \int d^{3}k  \; \epsilon^{abc} \; \text{Tr}[A_{a,j}\partial_{b}A_{c,j}+\frac{2i}{3}A_{a,j}A_{b,j} \nn \\ && A_{c,j}]
=\frac{n_{3,j}}{2},
\end{eqnarray} 
and the magneto-electric coefficient or axion angle
\begin{equation}
\theta_j = 2\pi \mathcal{CS}_j = \pi n_{3,j}.
\end{equation} 
The primary goal of this work is to identify $|n_{3,j}|$ from symmetry analysis and the gauge-invariant spectrum of $SU(2)$ Wilson loops. 

For concreteness, we will focus on materials, possessing $\mathcal{P}$ and $\mathcal{T}$ symmetries. For such systems, 
the numerical cost for Wilson loop calculations can be substantially reduced by symmetry analysis. The main idea is to first perform \emph{a coarse classification} of bulk winding numbers, with fictitious ``order parameter" type quantities, defined in momentum space on Miller hyper-cube, which are known as symmetry-indicators (SI). 

The application of SIs for $\mathcal{P}$- and $\mathcal{T}$- symmetric topological insulators (TIs) was pioneered by Fu, Kane and Mele.~\cite{FuKaneMele2007,FuKane} They identified the strong, $\mathbb{Z}_2$ topological index (STI)
$(-1)^{\nu_{0,j}}=(-1)^{n_{3,j}}$
from the product of parity eigenvalues at time-reversal-invariant-momentum (TRIM) points. 
The STI of a ground state, with $m$ occupied bands is given by $
\nu_{0,GS}= \sum_{j=1}^{m} \; \nu_{0,j} \; \text{mod} \; 2$. When an odd (even) number of $\mathbb{Z}_2$-non-trivial bands are occupied, $\nu_{0,GS}$ identifies the ground state as a non-trivial (trivial) insulator.

By construction, $\nu_{0,j}$ cannot distinguish (i) between $n_{3,j}=0$, and $n_{3,j}=2s \neq 0$, and (ii) between different odd integers. Since the ground states of topological crystalline insulators (TCIs) support a combination of $\mathbb{Z}_2$-trivial bands and an even number of $\mathbb{Z}_2$-non-trivial bands, their analysis requires new methods.~\cite{slager2013space,Chiu2016} This led to generalization of SIs,~\cite{Kruthoff2017,bradlyn2017topological,po2017symmetry,KhalafSymm,CanoBuildingBlocks2018,vergniory2019complete,zhang2019catalogue,tang2019efficient,tang2019comprehensive, vergniory2021all,xu2020high,elcoro2020magnetic,Bouhon2021,Lange2021} $K$-theory analysis,~\cite{freed2013twisted,Okuma2019} and the analysis of Wilson loop spectrum or Wannier charge centers (WCC).~\cite{yu2011equivalent,alexandradinata2014wilson,Taherinejad2014,Z2pack,Soluyanov2011,bouhon2019wilson,bradlyn2019disconnected} However, we are not aware of any direct method for computing $n_{3,j}$ beyond the scope of $\mathbb{Z}_2$-classification scheme~\cite{ZhangFT,ryu2010topological,EssinMagnetoelectric,Essin2010,malashevich2010theory,Coh2011,varnava2020}. Therefore, it is difficult to predict whether TCIs can support quantized magneto-electric response with $\theta=2s \pi$ and $s \neq 0$.

\begin{figure}
\centering
\includegraphics[scale=0.25]{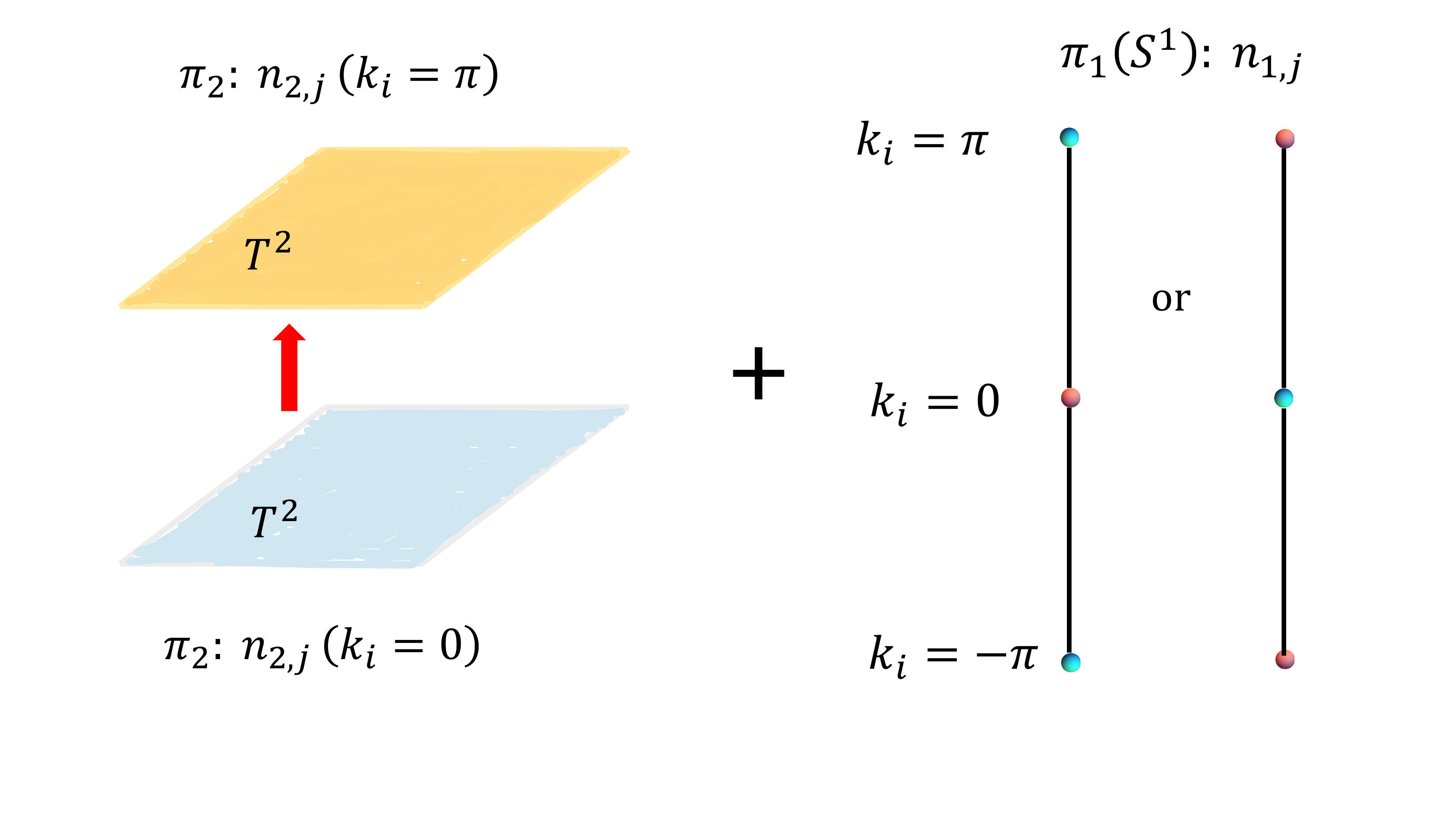} 
   \caption{Schematic of $3$-dimensional tunneling configurations. The winding number $n_{3,j}$ corresponds to the change of $2$-dimensional winding number $n_{2,j}$ of high-symmetry planes, which are equivalent to $T^2$. $2$-dimensional winding number identifies the presence of quantized, non-Abelian Berry flux for high-symmetry planes, and $|n_{3,j}|=|n_{2,j}(k_i=\pi)-n_{2,j}(k_i=0)|$. The change of Berry flux must be accompanied by non-trivial $1$-dimensional winding number ($n_{1,j}$) for one or more high-symmetry axes, along the tunneling direction. The converse is not true. While $n_{1,j}$ and $n_{2,j}$ can be obtained from Wilson loop calculations for Berry connection, this process can be guided by the staggered symmetry-indicator $\kappa_{AF,j}$. }
 \label{fig:instanton}
\end{figure}

In this work, we will develop a comprehensive theoretical framework for computing $n_{3,j}$. We will introduce a staggered SI $\kappa_{AF,j} \in \mathbb{Z}$ for recognizing patterns of parity (and rotation) eigenvalues that lead to $n_{3,j} \neq 0$. Using $\kappa_{AF,j}$, the uniform $\mathbb{Z}_4$ index $\kappa_{1,j}$~\cite{po2017symmetry,KhalafSymm}, and weak $\mathbb{Z}_2$-indices $(\nu_{1,j},\nu_{2,j},\nu_{3,j})$~\cite{FuKaneMele2007,FuKane}, following three classes of \emph{non-trivial} band topology will be identified: 
\begin{eqnarray}
&&  (i) \; \text{class A:}  \; \text{3D} \; \mathbb{Z}_2\text{-topology}, \; \text{and} \;  n_{3,j}=(2s_j+1); \nn  \\
&&  (ii) \; \text{class B:} \; \text{weak/2D} \; \mathbb{Z}_2\text{-topology}, \; \text{and} \;  \; n_{3,j}=0; \nn \\
&& (iii) \; \text{class C:} \;  \; \mathbb{Z}_2\text{-trivial, 3D topology}, \; \text{and} \; n_{3,j}=2s_j .\nn
\end{eqnarray}  
\emph{For these three classes of bands, $\kappa_{AF,j}$ respectively displays odd integer, zero, and even integer values}. After performing coarse-classification with SIs, we will show that $n_{3,j}$ can be calculated from tunneling configurations of $SU(2)$ Berry flux (see illustration of   Fig.~\ref{fig:instanton} ). This will be accomplished with a joint analysis of WCC for high-symmetry axes and in-plane Wilson loops for high-symmetry planes. 

The manuscript is organized as follows. In Sec.~\ref{symmetryindicators}, we introduce $\kappa_{AF,j}$ and discuss its relationship with $D$-dimensional winding number. In Sec.~\ref{Cubic}, we analyze $O_h$-symmetry-protected, tunneling configurations, using an analytically tractable 4-band model. Contrasting properties of Wilson loops and surface Dirac fermions for three classes A, B, and C are demonstrated. In Sec.~\ref{Kuramoto}, we describe essential features of $D_{3d}$-symmetry-protected instantons by considering a 4-band model of rhombohedral systems. In Sec.~\ref{abinitio}, we compute $|n_{3,j}|$ for \emph{ab initio} band structure of Bi. In Sec.~\ref{summ}, we conclude with a brief discussion of our results. 
In Appendix~\ref{schindler}, we present calculate of signed winding numbers of an $8$-band tight-binding model of Bi. 

\section{Staggered index and homotopy classification}\label{symmetryindicators}
We begin with a physical perspective on SIs of parity eigenvalues for $D$-dimensional, simple cubic systems.
The TRIM points of $D$-dimensional BZ (vertices of Miller hyper-cube) can be written as
\begin{equation}
  \bs{Q}^i= \frac{1}{2} \sum_{a=1}^{D}  l^i_a \; \bs{b}_a, \; \text{with} \;  i=1,..,2^D,
\end{equation}
where $\bs{b}_a$ are reciprocal vectors, and $l^i_a=0,1$. The parity eigenvalues of $j$-th band are Ising variables $\delta^i_{j}=\pm 1$, located on the vertices of Miller-cube. Topological information encoded in $2N \times 2N$ diagonal matrices 
\begin{equation}\label{flagcenter}
\mathcal{P}^i= \text{diag} [\delta^i_1 \sigma_0,..,\delta^i_N \sigma_0], 
\end{equation}
can be extracted by using matrix-valued ``order parameters" or SIs. 

\begin{figure}
 \includegraphics[scale=0.26]{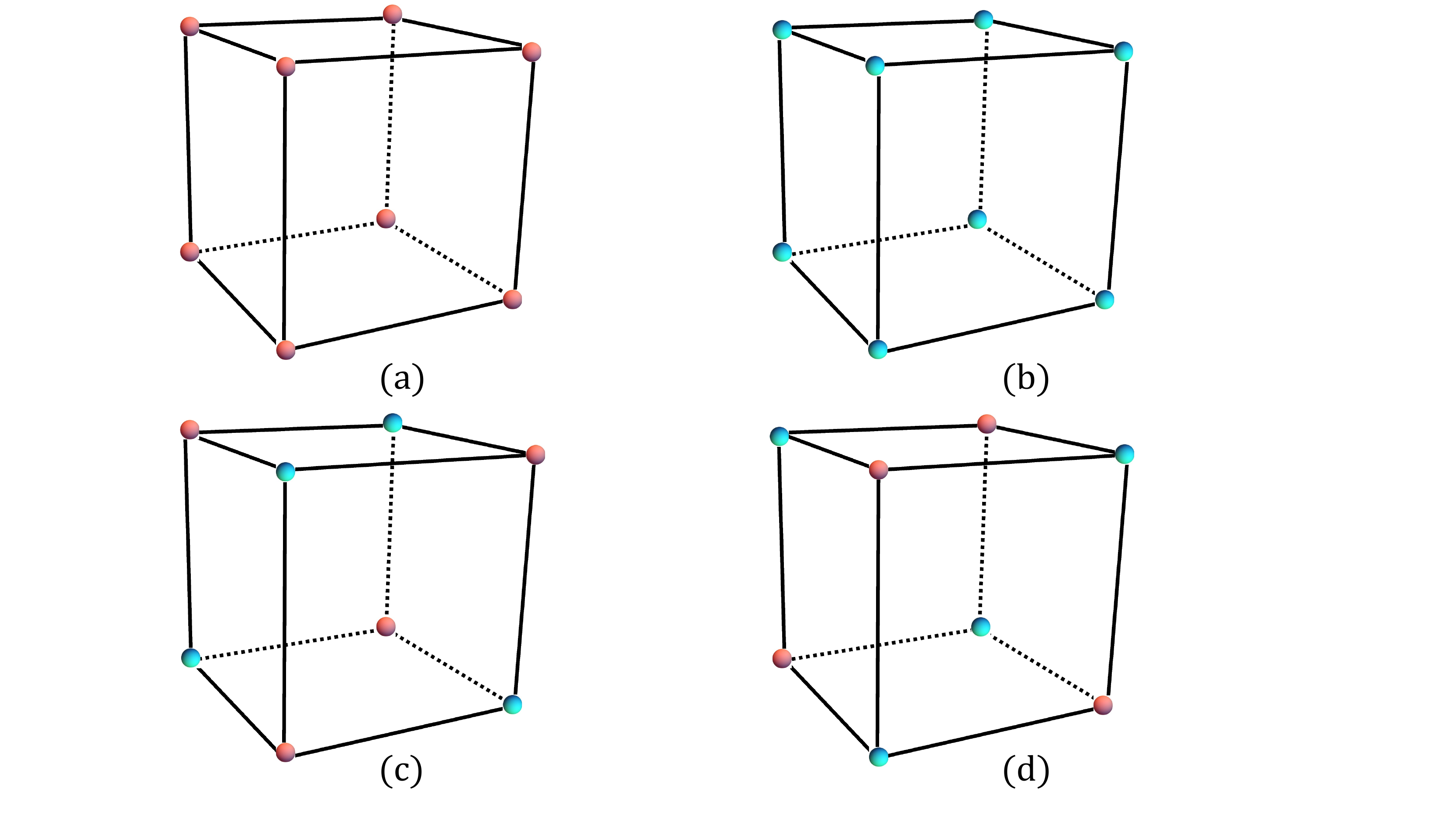} 
    \caption{(a)-(b) Ferromagnetic configurations of parity eigenvalues for topologically trivial bands. (c)-(d) Maximally staggered, N\'{e}el configurations of parity eigenvalues of topologically non-trivial bands. Eigenvalue $+1$ ($-1$) is denoted by light-red (cyan) dot.  } 
    \label{fig:FMAFM}
\end{figure}

The $\mathbb{Z}_2$ STIs of constituent bands are given by
 \begin{eqnarray}
&& \prod_{i} \mathcal{P}^i = \text{diag} [(-1)^{\nu_{0,1}},..,(-1)^{\nu_{0,N}}], \\\
&& (-1)^{\nu_{0,j}} = \prod_{i=1}^{2^D} \; \delta^i_j.
 \end{eqnarray}
The uniform or ferromagnetic indices~\cite{po2017symmetry,KhalafSymm} are defined as
\begin{eqnarray}
&&\boldsymbol \kappa_1 = \frac{1}{2} \sum_{i=1}^{2^D} \; P^i = \text{diag} [ \kappa_{1,1},.., \kappa_{1,N}], \\
&&\kappa_{1,j}=\frac{1}{2} \sum_{i=1}^{2^D} \; \delta^i_{j},
\end{eqnarray}
and $\kappa_{1,j}$ can acquire $(2^D +1)$ values
\begin{equation}
\kappa_{1,j}= 0, \pm 1, \pm 2,.., \pm 2^{D-1}.
\end{equation}
Due to the lack of band inversion, perfect ferromagnetic configurations [see Figs.~\ref{fig:FMAFM}(a) and \ref{fig:FMAFM}(b) ], describe topologically trivial bands, with $\kappa_{1,j}=\pm 2^{D-1}$. The uniform index of a ground state with $m$ occupied bands is defined by
 \begin{equation}
 \kappa_{1,GS}=\sum_{j=1}^{m} \kappa_{1,j}   \; \text{mod} \; 2^{D-1}, \; \text{when} \; D>1,
 \end{equation} 
as it can be shifted by adding topologically trivial bands. Thus, $\kappa_{1,GS}=0$, and $\kappa_{1,GS} = 2^{D-1}  \times l $ with $l \in \mathbb{Z}$ correspond to topologically equivalent, trivial states, leading to the $\mathbb{Z}_{2^{(D-1)}}$-classification scheme for the ground state.

There exist 
\begin{equation}
N_0=\frac{2^D!}{[2^{(D-1)}!]^2}
\end{equation}
Ising configurations, with $2^{D-1}$ positive, and $2^{D-1}$ negative parity eigenvalues, leading to $\kappa_{1,j}=0$. We need new indicators to classify them. \emph{Notably, both topologically non-trivial configurations at $D=1$ possess $\kappa_{1,j}=0$}. 
By focusing on maximally staggered, N\'{e}el configurations (see Fig.~\ref{fig:FMAFM}(c) and \ref{fig:FMAFM}(d)), let us define 
\begin{eqnarray}
&&\boldsymbol \kappa_{AF} = \frac{1}{2} \; \sum_{i=1}^{2^{D}} \; (-1)^{l^i_1+l^i_2+...+l^i_D} \; \mathcal{P}^i \nn \\ 
&&=\text{diag} [ \kappa_{AF,1},.., \kappa_{AF,N}], \\
&&\kappa_{AF,j}= \frac{1}{2} \sum_{i=1}^{2^{D}} \; (-1)^{l^i_1+l^i_2+...+l^i_D} \; \delta^{i}_{j}, \\
&& \kappa_{AF,GS} = \sum_{j=1}^{m} \; \kappa_{AF, j}.
\end{eqnarray}
Akin to $\kappa_{1,j}$, $\kappa_{AF,j}$ can also acquire $(2^D+1)$ distinct values 
\begin{equation}
\kappa_{AF,j}=0, \pm 1, \pm 2,.., \pm 2^{D-1},
\end{equation}
As trivial bands with perfect ferromagnetic configurations lead to $\kappa_{AF, j}=0$, $\kappa_{AF, GS} \neq 0$ cannot be deformed to $0$ by adding topologically trivial bands. Therefore, the staggered index is a stable, $\mathbb{Z}$-valued SI, which can be used for all inversion-symmetric systems. By construction, $(-1)^{\kappa_{1,j}}=(-1)^{\kappa_{AF,j}}=(-1)^{\nu_{0,j}}$, and $(-1)^{\kappa_{1,GS}}=(-1)^{\kappa_{AF,GS}}=(-1)^{\nu_{0,GS}}$.

\emph{If our primary goal is to understand which configurations are capable of producing $D$-dimensional winding numbers, we can ignore $N_0$ configurations with} $\kappa_{AF,j}=0$. This can be seen from the explicit homotopy classification
of minimal model 
\begin{eqnarray}\label{genD}
&&\hat{H}(\bs{k})=\sum_{j=1}^{D+1} d_j(\bs{k}) \Gamma_j 
=t_p \sum_{j=1}^{D} \;  \sin k_j \Gamma_j \nn \\ &&+ t_s [M-\Delta_1 \sum_{j=1}^{D} \cos k_j ] \Gamma_{D+1}.
\end{eqnarray} 
of $D$-dimensional cubic topological insulators. Here $t_p$ and $t_s$ are hopping parameters, $(M, \Delta_1)$ are dimensionless tuning parameters, and $\Gamma_j$'s are $2^l \times 2^l$ mutually anti-commuting matrices, with $l \geq \left[\frac{D+1}{2} \right]$. The operation of $\mathcal{P}$ is implemented as $\Gamma_{D+1} H(-\bs{k}) \Gamma_{D+1} = H(\bs{k})$. Non-trivial $D$-dimensional band topology arises from instanton configurations of $O(D+1)$ unit vector $\bs{\hat{d}}(\bs{k})=\bs{d}(\bs{k})/|\bs{d}(\bs{k})|$, which are classified by the $D$-th spherical homotopy group $\pi_{D}(S^D)=\mathbb{Z}$. The corresponding winding number 
\begin{eqnarray}
n_D= \frac{\Gamma(\frac{D+1}{2})}{2 \pi^{\frac{D+1}{2}}} \; \int_{T^D} \; d^Dk \; \epsilon_{i_1...i_{D+1}} \hat{d}_{i_1} \partial_1 \hat{d}_{i_2}...\partial_D \hat{d}_{i_{D+1}}, \nn \\
\end{eqnarray}
counts how many times the BZ $D$-torus $T^D$ wraps around the unit-sphere $S^D$, and $\partial_{a}= \frac{\partial}{\partial k_{a}}$.

\begin{table}
\def\arraystretch{1.5}
	\begin{tabular}{|c|c|c|}
		\hline
		\begin{tabular}{c}
			 Parity eigenvalues \\ $(\delta_\Gamma, \delta_R, \delta_X, \delta_M)$
		\end{tabular} & \begin{tabular}{c}
		Symmetry  indicators\\ $(\kappa_1; \kappa_{AF}; \nu_{1}, \nu_2, \nu_3)$
		\end{tabular} & Class\\
		\hline 
		$1. \; (+1, +1,+1,+1)$ &$(+4;0;0,0,0)$& Trivial\\
		\hline
		$2. \; (-1, -1,-1,-1)$ &$(-4;0;0,0,0)$& Trivial\\
		\hline
		$3. \; (-1, +1,+1,+1)$ &$(+3;-1;0,0,0)$& A\\
		\hline
		$4. \; (+1, -1,-1,-1)$ &$(-3;+1;0,0,0)$& A\\
		\hline
		$5. \; (+1, -1,+1,+1)$ &$(+3;+1;1,1,1)$& A\\
		\hline
		$6. \; (-1, +1,-1,-1)$ &$(-3;-1;1,1,1)$& A\\
		\hline
		$7. \; (+1, +1,-1,+1)$ &$(+1;+3;1,1,1)$& A\\
		\hline
		$8. \; (-1, -1,+1,-1)$ &$(-1;-3;1,1,1)$& A\\
		\hline
		$9. \; (+1, +1,+1,-1)$ &$(+1;-3;0,0,0)$& A\\
		\hline
		$10. \; (-1, -1,-1,+1)$ &$(-1;+3;0,0,0)$& A\\
		\hline
		$11. \; (-1, -1,+1,+1)$ &$(+2;0;1,1,1)$& B\\
		\hline
		$12. \; (+1, +1,-1,-1)$ &$(-2;0;1,1,1)$& B\\
		\hline
		$13. \; (-1, +1,-1,+1)$ &$(0;+2;1,1,1)$& C\\
		\hline
		$14. \; (+1, -1,+1,-1)$ &$(0;-2;1,1,1)$& C\\
		\hline
		$15. \; (-1, +1,+1,-1)$ &$(0;-4;0,0,0)$& C\\
		\hline
		$16. \; (+1, -1,-1,+1)$ &$(0;+4;0,0,0)$& C\\
		\hline
\end{tabular}
\caption{Patterns of parity eigenvalues and symmetry indicators [see Eq.~\ref{SISC}] for simple cubic systems, with $\Gamma=(0,0,0)$, $R=(1,1,1)$, $X=\{(1,0,0), (0,1,0), (0,0,1)\}$, and $M=(1,1,0)$, $(0,1,1)$, and $(1,0,1)$. In Sec.~\ref{Cubic}, we show that $O_h$-symmetry-protected tunneling configurations for simple cubic systems can be completely understood by considering the change of $SU( 2)$ Berry flux along $4$-fold or $3$-fold axes. The staggered index tracks the bulk winding number, the number of normalizable surface Dirac cones, and the surface Hall conductivity of $(001)$ and $(111)$ surfaces. 
} \label{tab1}
\end{table}

\begin{table}
\def\arraystretch{1.5}
	\begin{tabular}{|c|c|c|}
		\hline
		\begin{tabular}{c}
			 Parity eigenvalues \\ $(\delta_\Gamma, \delta_X, \delta_L)$
		\end{tabular} & \begin{tabular}{c}
		Symmetry  indicators\\ $(\kappa_1; \kappa_{AF}; \nu_{1}, \nu_2, \nu_3)$
		\end{tabular} & Class\\
		\hline 
		$1. \; (+1, +1,+1)$ &$(+4;0;0,0,0)$& Trivial\\
		\hline
		$2. \; (-1, -1,-1)$ &$(-4;0;0,0,0)$& Trivial\\
		\hline
		$3. \; (-1, +1,+1)$ &$(+3;-1;0,0,0)$& A\\
		\hline
		$4. \; (+1, -1,-1)$ &$(-3;+1;0,0,0)$& A\\
		\hline
		$5. \; (+1, -1,+1)$ &$(+1;-3;0,0,0)$& A\\
		\hline
		$6. \; (-1, +1,-1)$ &$(-1;+3;0,0,0)$& A\\
		\hline
		$7. \; (+1, +1,-1)$ &$(0;+4;0,0,0)$& C\\
		\hline
		$8. \; (-1, -1,+1)$ &$(0;-4;0,0,0)$& C\\
		\hline
		\end{tabular}
\caption{Patterns of parity eigenvalues and symmetry indicators [see Eq.~\ref{SIFCC}] for FCC crystals, with $\Gamma=(0,0,0)$, $X=\{(1,1,0), (0,1,1), (1,0,1)\}$, and $L=\{(1,0,0), (0,1,0), (0,0,1), (1,1,1) \}$. Notice that FCC crystals only support class A and class C bands, and class C bands exhibit maximally staggered, configurations of parity eigenvalues. Class C bands are important for describing topology of SnTe and PbTe. In contrast to simple cubic systems, tunneling of Berry flux for FCC systems occurs along $(111)$ axis. Detailed analysis of tight-binding model and \emph{ab initio} band structures of SnTe and PbTe will be presented in a separate work. 
} \label{tab2}
\end{table}

\begin{table}
\def\arraystretch{1.5}
	\begin{tabular}{|c|c|c|}
		\hline
		\begin{tabular}{c}
			 Parity eigenvalues \\ $(\delta_\Gamma, \delta_T, \delta_L, \delta_X)$
		\end{tabular} & \begin{tabular}{c}
		Symmetry  indicators\\ $(\kappa_1; \kappa_{AF}; \nu_{1}, \nu_2, \nu_3)$
		\end{tabular} & Class\\
		\hline 
		$1. \; (+1, +1,+1,+1)$ &$(+4;0;0,0,0)$& Trivial\\
		\hline
		$2. \; (-1, -1,-1,-1)$ &$(-4;0;0,0,0)$& Trivial\\
		\hline
		$3. \; (-1, +1,+1,+1)$ &$(+3;-1;0,0,0)$& A\\
		\hline
		$4. \; (+1, -1,-1,-1)$ &$(-3;+1;0,0,0)$& A\\
		\hline
		$5. \; (+1, -1,+1,+1)$ &$(+3;+1;1,1,1)$& A\\
		\hline
		$6. \; (-1, +1,-1,-1)$ &$(-3;-1;1,1,1)$& A\\
		\hline
		$7. \; (+1, +1,-1,+1)$ &$(+1;+3;1,1,1)$& A\\
		\hline
		$8. \; (-1, -1,+1,-1)$ &$(-1;-3;1,1,1)$& A\\
		\hline
		$9. \; (+1, +1,+1,-1)$ &$(+1;-3;0,0,0)$& A\\
		\hline
		$10. \; (-1, -1,-1,+1)$ &$(-1;+3;0,0,0)$& A\\
		\hline
		$11. \; (-1, -1,+1,+1)$ &$(+2;0;1,1,1)$& B\\
		\hline
		$12. \; (+1, +1,-1,-1)$ &$(-2;0;1,1,1)$& B\\
		\hline
		$13. \; (-1, +1,-1,+1)$ &$(0;+2;1,1,1)$& C\\
		\hline
		$14. \; (+1, -1,+1,-1)$ &$(0;-2;1,1,1)$& C\\
		\hline
		$15. \; (-1, +1,+1,-1)$ &$(0;-4;0,0,0)$& C\\
		\hline
		$16. \; (+1, -1,-1,+1)$ &$(0;+4;0,0,0)$& C\\
		\hline
\end{tabular}
\caption{Patterns of parity eigenvalues and symmetry indicators [see Eq.~\ref{SIR3m}] for rhombohedral systems, with $\Gamma=(0,0,0)$, $T=(1,1,1)$, $X=\{(1,1,0), (0,1,1), (1,0,1)\}$, and $L=(1,0,0)$, $(0,1,0)$, and $(0,0,1)$. In Sec.~\ref{Kuramoto}, we show that $D_{3d}$-symmetry-protected tunneling configurations can be understood by considering the change of $SU(2)$ Berry flux along $3$-fold axis. The staggered index tracks the bulk winding number, the number of normalizable surface Dirac cones, and the surface Hall conductivity of $(111)$ surface. While addressing topology of \emph{ab initio} band structure of Bi, we will encounter bands with $3$-fold rotation eigenvalues $e^{\pm i \pi/3}$, and $e^{\pm i \pi}$. All SIs for $e^{\pm i \pi/3}$ will remain unchanged. For bands carrying $e^{\pm i \pi}$, $\kappa_{AF}$ must be modified according to Eq.~\ref{modified-st}. } \label{tab3}
\end{table}

The TRIM points support parity eigenvalues $\delta^i_\pm = \pm \text{sgn}[d_{(D+1)} (\bs{Q}^i)]$ for $2^{(l-1)}$-fold degenerate conduction ($+$) and valence ($-$) bands. When $t_s=0$ and $t_p \neq 0$, they serve as hedgehogs of $O(D)$ unit vector, which can also be understood as merons of $O(D+1)$ unit vector, with hedgehog charge 
\begin{equation}
n^i_{h} = (\text{sgn}(t_p))^D (-1)^{l^i_1+l^i_2+...+l^i_D}.
\end{equation}
By combining the hedgehog charge and parity eigenvalues, we arrive at 
\begin{eqnarray}
n_D = \mp \; (\text{sgn}(t_p))^D \kappa_{AF, \pm} .
 \end{eqnarray}
Therefore, the homotopy analysis of intra-band Berry connection provides information about 
\begin{eqnarray}
n_{D, \pm} = (\text{sgn}(t_p))^D \kappa_{AF, \pm} = \mp n_D.
\end{eqnarray} 
Furthermore, we can rewrite $n_D$ as 
\begin{equation}\label{gentunneling}
n_D = n_{D-1}(k_i = \pi) -n_{D-1}(k_i=0),
\end{equation}
which describes the change of $(D-1)$-dimensional winding number along $i$-th high-symmetry direction, supporting band inversion. This scheme of dimensional reduction provides an intuitive way to think about instanton configurations of vector fields and non-Abelian Berry connection [see Fig.~\ref{fig:instanton}]. The staggered index precisely keeps track of such tunneling configurations. 

The dimensional reduction for non-Abelian Berry connection can be performed by Wilson loop along $j$-th axis 
\begin{equation}
    W_{j,-}(\bs{k}_\perp)= \text{P} \; \text{exp}\left[i\int_{-\pi}^{\pi} A_{j,-}(k_j, \bs{k}_\perp)dk_{j}\right],
\end{equation}
where $\text{P}$ indicates path-ordering, and WCC are given by
\begin{equation}
\bar{j}_{-}(\bs{k}_\perp) = \frac{1}{2 \pi} \; \text{Im} \left[ \ln (W_{j,-} (\bs{k}_\perp) ) \right] .
\end{equation}  When two TRIM points with identical (opposite) parity eigenvalues are joined by Wilson loop, $W_{j,-} \to \mathbb{1}$ ($-\mathbb{1}$) which are the center elements of gauge group (spin groups which are double covers of special orthogonal groups). The element $-\mathbb{1}$ corresponds to $\pi$ Berry phase or time-reversal polarization, and WCC describe interpolation between center elements as a function of $(D-1)$-dimensional, transverse wave vector $\bs{k}_\perp$. In the following sections, we consider explicit examples of 3D simple cubic and rhombohedral models to elucidate the relationship between staggered index, bulk invariant and Wilson loops. To set the stage for such analysis, we provide simplified expressions of relevant SIs for some representative crystalline systems.

\subsection{Staggered index of selected 3D systems}

At $D=3$, there are total $2^8=256$ configurations of parity eigenvalues. The perfect ferromagnetic (trivial bands) and N\'{e}el configurations (bands with maximal winding numbers) are respectively characterized by 
\begin{eqnarray}
( \kappa_{1,j}; \kappa_{AF,j}, \nu_{1,j}, \nu_{2,j}, \nu_{3,j})=( \pm 4; 0; 0,0,0), \label{FM1} \nn \\ \\
(\kappa_{1,j}; \kappa_{AF,j}, \nu_{1,j}, \nu_{2,j}, \nu_{3,j})=(0; \pm 4; 0,0,0), \label{AFM1} \nn \\ 
\end{eqnarray}
and the weak $\mathbb{Z}_2$ indices $(\nu_{1,j}, \nu_{2,j}, \nu_{3,j})$ identify odd vs. even integer distinction of 2D winding numbers for $(100)$, $(010)$, and $(001)$ planes, passing through the high-symmetry point $(l_1, l_2, l_3)=(1,1,1)$. For example, \begin{eqnarray}
 && (-1)^{\nu_{1,j}}= \delta^{(1,1,1)}_j \delta^{(1,1,0)}_j \delta^{(1,0,0)}_j \delta_{j}^{(1,0,1)}, \\
 && \nu_{1,GS}=\sum_{j=1}^{m} \nu_{1,j} \; \text{mod} \; 2.
 \end{eqnarray}

Other $(2^8-4)=252$ configurations of parity eigenvalues display imperfect ferromagnetic and staggered moments. Not all configurations are allowed by underlying crystal symmetries. For simple cubic systems (space groups 221 to 224) three $X$ points ($M$) points support identical parity eigenvalue $\delta_{X}$ ($\delta_M$). Therefore, only 16 configurations can be realized, with SIs
\begin{eqnarray}\label{SISC}
&&\kappa_{1,j}= \frac{1}{2}(\delta_{\Gamma,j}+\delta_{R,j}+3\delta_{X,j}+3\delta_{M,j}), \nn \\ 
&&\kappa_{AF,j}= \frac{1}{2}(\delta_{\Gamma,j}-\delta_{R,j}-3\delta_{X,j}+3\delta_{M,j}), \nn \\
&&\nu_{1,j}=\nu_{2,j}=\nu_{3,j}=\frac{1}{2}(1-\delta_R \delta_X ).
\end{eqnarray} Using $(\kappa_{1,j}; \kappa_{AF,j}; \nu_{1,j}, \nu_{2,j}, \nu_{3,j})$, we arrive at the coarse classification of Kramers-degenerate bands, listed in Table.~\ref{tab1}. The SIs for primitive tetragonal and orthorhomic systems are easily obtained by distinguishing different $X$ and $M$ points. Consequently, additional configurations can be allowed. But the main idea of tracking 3D winding numbers with $\kappa_{AF,j}$ remains unaffected.

For space groups 225-228, underlying FCC crystals lead to three $X$ points and four $L$ points. Therefore, only $8$ configurations are allowed, which are listed in Table~\ref{tab2}, with SIs 
\begin{eqnarray}\label{SIFCC}
&&\kappa_{1,j}= \frac{1}{2}(\delta_{\Gamma,j}+3\delta_{X,j}+4\delta_{L,j}), \nn \\ 
&&\kappa_{AF,j}= \frac{1}{2}(\delta_{\Gamma,j}+3\delta_{X,j}-4\delta_{L,j}), \nn \\
&&\nu_{1,j}=\nu_{2,j}=\nu_{3,j}=0.
\end{eqnarray} 
Importantly, FCC crystals do not support class B bands. 

Rhombohedral systems are related to distorted FCC lattice. Due to rhombohedral distortion, $(111)$ L point becomes inequivalent with other three $L$ points, and is commonly known as the $T$ point. Thus, rhombohedral systems allow $16$ configurations of parity eigenvalues. The SIs are given by
\begin{eqnarray}\label{SIR3m}
&&\kappa_{1,j}= (\delta_{\Gamma,j}+\delta_{T,j}+3\delta_{X,j}+3\delta_{L,j}), \nn \\ 
&&\kappa_{AF,j}= (\delta_{\Gamma,j}-\delta_{T,j}+3\delta_{X,j}-3\delta_{L,j}), \nn \\
&&\nu_{1,j}=\nu_{2,j}=\nu_{3,j}=\frac{1}{2}(1-\delta_T \delta_L) .
\end{eqnarray}
and are listed in Table~\ref{tab3}. These SIs can be directly applied for analyzing \emph{ab initio} band structures of materials like Bi, Sb, and Bi$_2$Se$_3$, when bands possess $3$-fold rotation eigenvalues $e^{ \pm i \pi/3}$. 

For bands with rotation eigenvalues $e^{ \pm i \pi}$, $\Gamma$ and $T$ points support $n^{\Gamma}_h= \pm 3$, $n^{T}_h=\mp 3$, $n^X_h=\mp 1$, $n^L_h=\pm 1$ as hedgehog charge. Therefore, the staggered index of such bands is given by
\begin{eqnarray}\label{modified-st}
 \kappa_{AF,j}^{\pm \pi}=\frac{3}{2}(\delta_{\Gamma,j}-\delta_{T,j}+\delta_{L,j}-\delta_{X,j})
 \end{eqnarray}
Consequently, the staggered index of class A configurations $3$-$10$ of Table~\ref{tab3} will be modified as
\begin{eqnarray}
\kappa_{AF,j}^{\pm \pi}=-3,+3,+3,-3,-3,+3,+3,-3,
\end{eqnarray}
respectively. Class C configurations $13-16$ support
\begin{eqnarray}
\kappa_{AF}^{\pm \pi}=-6,+6,0,0
\end{eqnarray}
Hence, maximally staggered configurations with rotation eigenvalues $e^{\pm i \pi}$ do not lead to 3D winding number. If the rotation data is not taken into account, Wilson loop calculations for $3$-fold planes would guarantee that the correct magnitude of winding number is obtained.

For primitive hexagonal crystals, bands carrying rotation eigenvalues $e^{\pm i \frac{\pi}{6}}$ and $e^{\pm i \frac{\pi}{2}}$, the staggered index can be defined as
  \begin{eqnarray}
 &&\kappa^{\pi/6}_{AF,j}= \frac{1}{2}(\delta_{\Gamma,j}-\delta_{A,j}-3\delta_{X,j}+3\delta_{L,j}), \nn \\ 
&&\kappa^{\pi/2}_{AF,j}= \frac{3}{2}(\delta_{\Gamma,j}-\delta_{A,j}-\delta_{X,j}+\delta_{L,j}).
 \end{eqnarray} 
For simple toy models of bands with $e^{\pm i \frac{\pi}{6}}$, $K$ and $H$ points can also participate in band inversion, and $\kappa^{\pi/6}_{AF,j}$ should be modified by adding $(\delta_K-\delta_H)$. Such examples can be found in Appendix~\ref{schindler}. Akin to rhombohedral systems, primitive hexagonal systems also support maximum staggered index $\pm 6$.

\section{Simple cubic systems and $\bs{O_h}$ instantons}\label{Cubic}

To understand topology of $O_h$ instanton configurations of Table~\ref{tab1} and $SU(2)$ Wilson loops, 
we consider a tight-binding model of two Kramers-degenerate bands, described by 
\begin{equation} \label{Sp4}
H(\mathbf{k})=\sum_{j=1}^{5}d_{j}(\mathbf{k})\Gamma_{j}.
\end{equation} Here $\Gamma_{j}$'s are $4\times 4$ anti-commuting matrices, given explicitly as $\Gamma_{j=1,2,3}=\tau_{1}\otimes \sigma_{j}$, $\Gamma_{4}=\tau_{2}\otimes \sigma_{0}$, and $\Gamma_5=\tau_3 \otimes \sigma_0$, where $\sigma_{0,1,2,3}$($\tau_{0,1,2,3}$) are $2\times 2$ identity matrix and three Pauli matrices, operating on spin (orbital) index, respectively. 
Using $T_{1u}$ and $A_{1g}$ harmonics of $O_h$ point group, we define the following map 
\begin{eqnarray}\label{eq:cubic}
  &&  d_{j}(\mathbf{k})=t_{p}\sin k_{j}, \; \text{with} \; j=1,2,3, \nn \\ && d_4(\bs{k}) = M^\prime, \nn \\ &&d_{5}(\mathbf{k})=t_{s}\bigg(M-\Delta_{1}\sum_{j=1}^{3}\cos k_{j} -\Delta_{2}\sum_{i<j=1}^{3}\cos k_{i}\cos k_{j}  \nn \\ && -\Delta_{3}\prod_{j=1}^{3}\cos k_{j}\bigg),
\end{eqnarray}
where $t_{p,s}$ are hopping parameters with units of energy, and $M, \Delta_{1}, \Delta_{2}, \Delta_{3}$ are dimensionless tuning parameters. For simplicity, the lattice constant has been set to unity. Parity and time-reversal symmetries are implemented as $\mathcal{P}^{\dagger}H(-\mathbf{k})\mathcal{P}=H(\mathbf{k})$, $\mathcal{T}^\dagger H^\ast(-\bs{k}) \mathcal{T}=H(\bs{k})$, with $\mathcal{P}=\Gamma_5$, $\mathcal{T}=i\Gamma_{31}=i\tau_0 \otimes \sigma_2$, respectively, and $\Gamma_{ab}=[\Gamma_a, \Gamma_b]/(2i)$. 

The pseudo-scalar mass $M^\prime \neq 0$ breaks $\mathcal{P}$ and $\mathcal{T}$ symmetries, but preserves the combined $\mathcal{PT}$ symmetry, $\Gamma_{24} H^\ast(\bs{k}) \Gamma_{24} = H(\bs{k})$, and Kramers-degeneracy. When $M^\prime \neq 0$, the 4-band model describes generic magneto-electric insulators, and $\mathcal{P}$ and $\mathcal{T}$ symmetric topological insulators are obtained for $M^\prime =0 $. While computing Chern-Simons coefficient, it is convenient to use $M^\prime \to 0^+$, as a suitable regulator of Dirac string singularities at TRIM locations.

\begin{figure}
    \centering
   \includegraphics[scale=0.35]{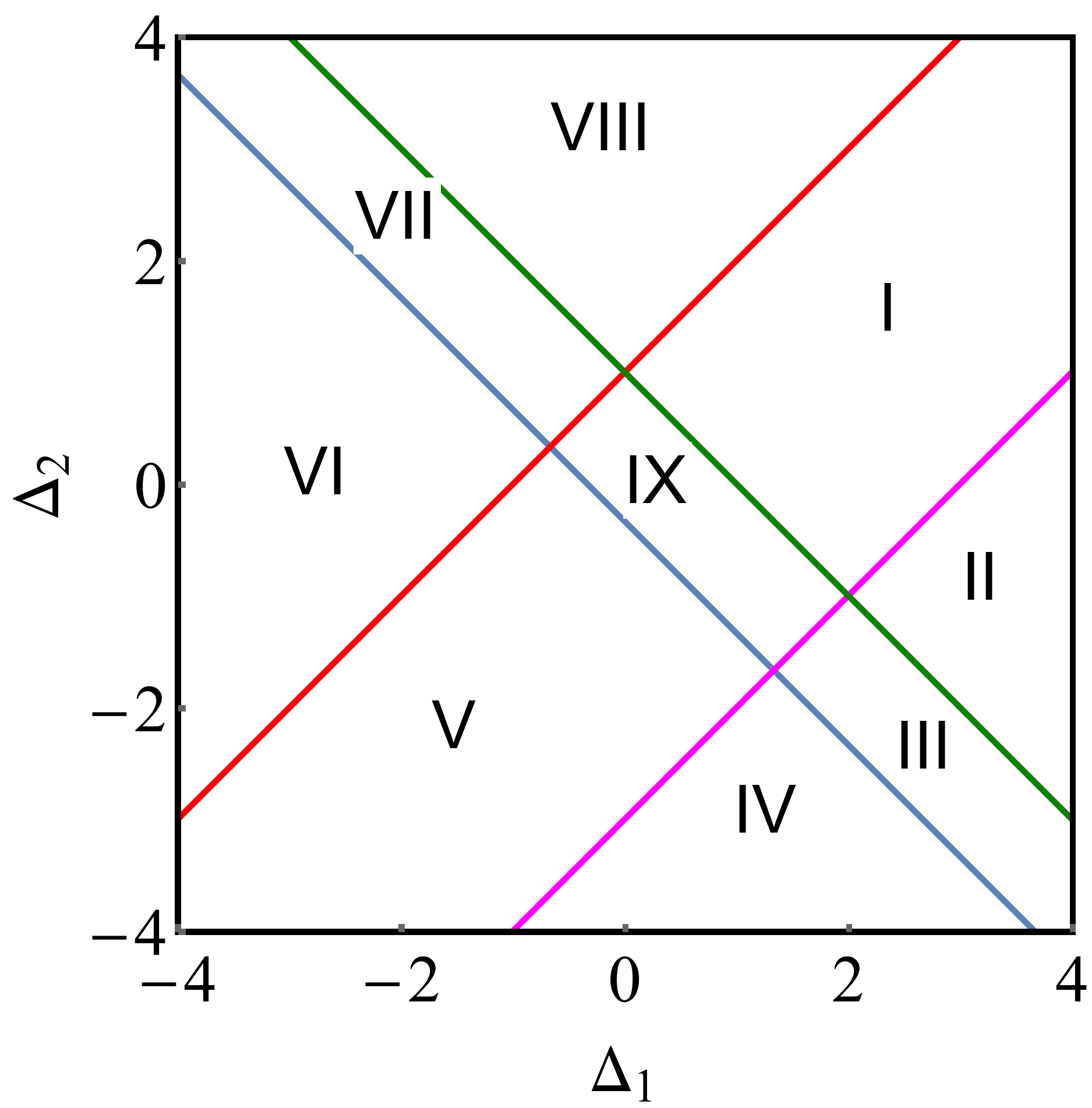} 
        \caption{Phase diagram of cubic model [Eq.~\ref{eq:cubic} ], when $M=1$ and $\Delta_{3}=2$. The configurations of parity eigenvalues, symmetry-indicators, and bulk winding numbers are summarized in Fig.~\ref{fig:cubic}. Along blue, green, red, and magenta colored lines, band gap can vanish at $\Gamma$, $M$, $R$, and $X$ points, respectively. These lines are described by $d_5(\bs{Q}^i )=0$.}
    \label{fig:Phases}
\end{figure}

\begin{figure*}
    \centering
           \includegraphics[scale=0.58]{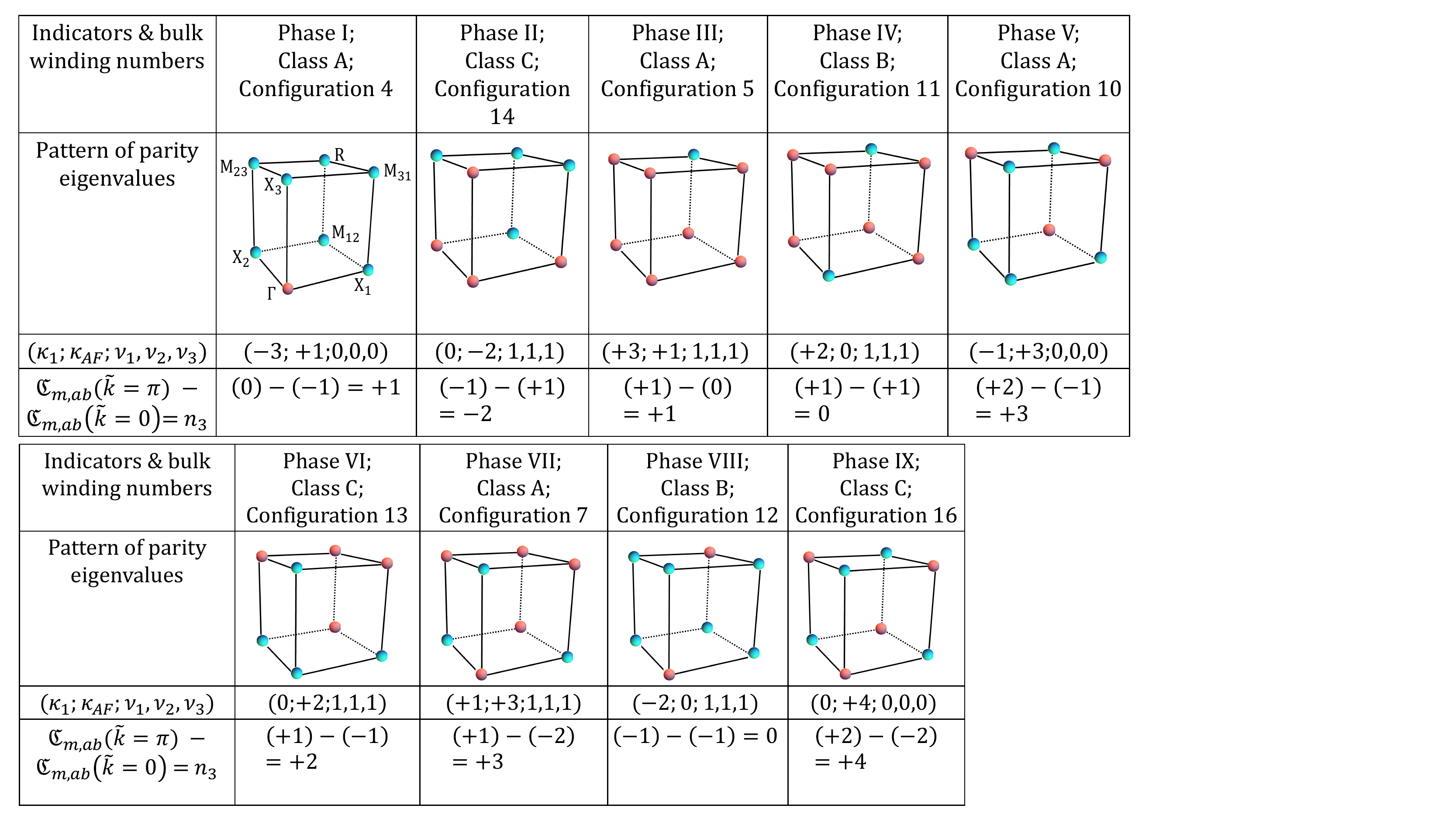} 
           \caption{Summary of symmetry-indicators and bulk winding numbers for nine homotopically distinct phases from Fig.~\ref{fig:Phases}. Under Phase I, time-reversal-invariant-momentum points $\Gamma=(0,0,0)$, $X_1=(1,0,0)$, $X_2=(0,1,0)$, $X_3=(0,0,1)$, $M_{12}=(1,1,0)$, $M_{23}=(0,1,1)$, $M_{31}=(1,0,1)$, $R=(1,1,1)$ are labeled. Parity eigenvalue $+1$ $(-1)$ of valence bands is indicated by red (cyan) dot. Three-dimensional winding number obeys Eq.~\ref{mirrortunneling}, describing tunneling configurations of mirror Chern numbers along $4$-fold axes. Therefore, for \emph{ab initio} band structures of many simple cubic topological insulators, signed 3-dimensional winding numbers of constituent bands and ground state can be easily obtained from Wilson loop spectrum for $D_{4h}$ symmetric mirror planes. This computational scheme is also applicable for other crystalline systems, when the tunneling is protected by $C_{nh}$ and $D_{nh}$ point group symmetries.} 
    \label{fig:cubic}
\end{figure*}

All salient properties of topological insulators follow from the $O(4)$ vector $(d_1,d_2,d_3,d_5)$, and the $SU(2)$ matrix 
\begin{equation}
u(\bs{k})=\hat{d}_5(\bs{k}) \sigma_0 + i \sum_{j=1}^{3} \; \hat{d}_j(\bs{k}) \sigma_j.
\end{equation}
At TRIM points parity eigenvalues of conduction ($+$) and valence ($-$) bands are given by $\pm \text{sgn}(d_5(\bs{Q}^i))$ and $u(\bs{k})$ maps to $SU(2)$ center elements $\pm \sigma_0$. The 3D winding number is determined by
\begin{eqnarray}\label{eq: o4wind}
 &&  n_{3}=\frac{1}{2\pi^2}\int_{T^3}d^{3}k \; \epsilon^{abcd} \; \hat{d}_{a}\partial_{k_{x}}\hat{d}_{b}\partial_{k_{y}}\hat{d}_{c}\partial_{k_{z}}\hat{d}_{d}, \nn \\
    &&=\frac{1}{24 \pi^2} \int_{T^3} d^3k \; \epsilon^{jlm} \; [(u^\dagger \partial_{j} u) u^\dagger \partial_l u) (u^\dagger \partial_m u) ],  \nn \\
\end{eqnarray}
and the present model can realize
\begin{equation}
n_3= \text{sgn}(t_p) \; \kappa_{AF, -} = \pm 1, \pm 2, \pm 3, \pm 4.
\end{equation}
A representative phase diagram is shown in Fig.~\ref{fig:Phases} for $M=+1$, $\Delta_3=+2$. In Fig.~\ref{fig:cubic}, we display configurations of parity eigenvalues, SIs, and winding numbers for these phases.

\begin{figure*}
    \centering
           \includegraphics[scale=0.42]{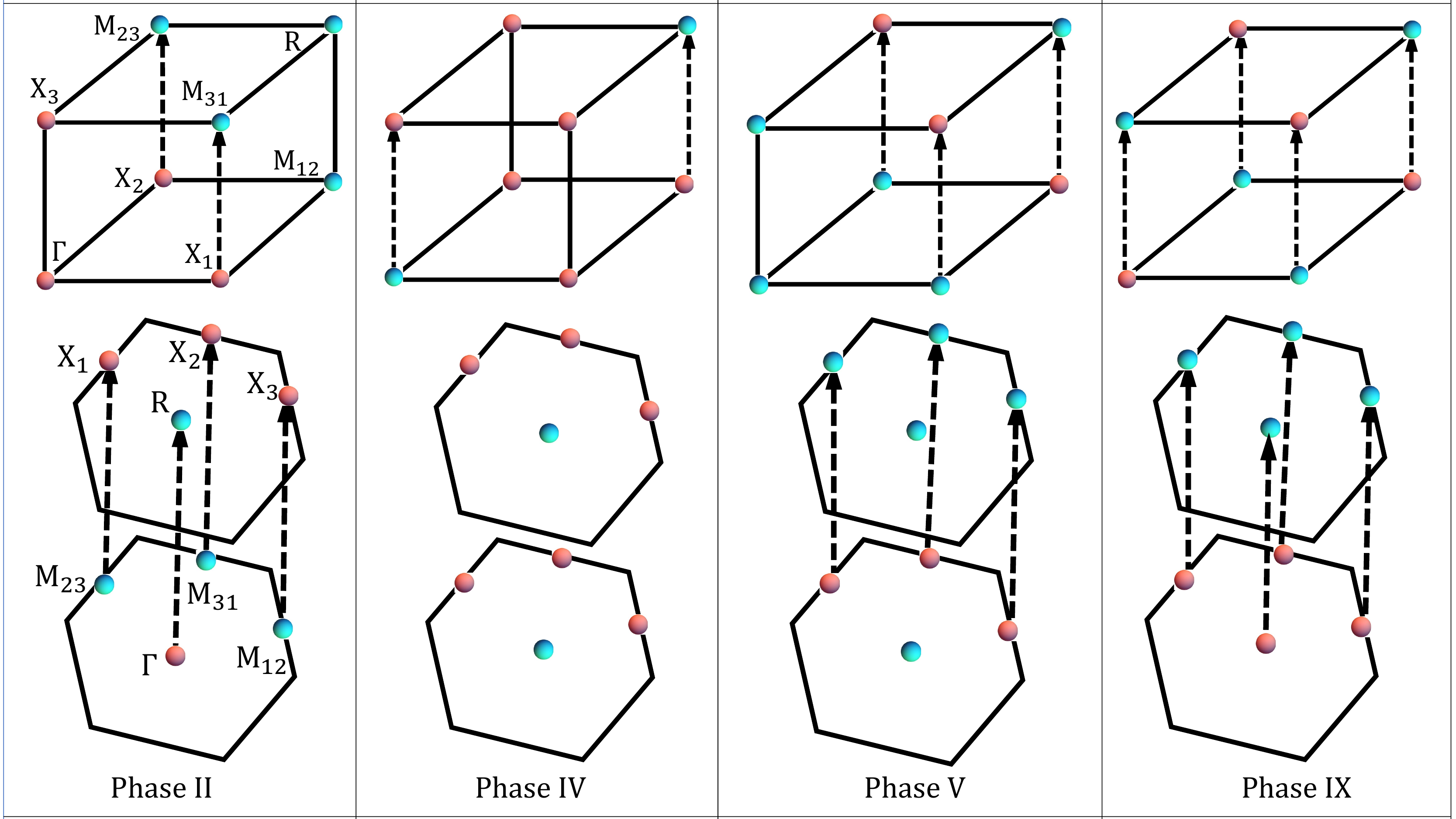} 
           \caption{Contrasting pictures of cubic-symmetry-protected instantons, when respectively viewed along $4$-fold axis $(001)$, and $3$-fold axis $(111)$. The dashed lines carry non-trivial one-dimensional winding numbers [see Eq.~\ref{1Ddirected} and Eq.~\ref{1Dstaggered} ]. Wilson lines along these axes support $\pi$ Berry phase, and the number of dashed lines counts the total number of 2-component, massless, Dirac fermions under open boundary conditions. For class B, weak topological insulator (Phase IV), the net one-dimensional winding number along $z$-axis vanishes, signifying the absence of tunneling. Consequently, anomalous Hall conductivity of $[001]$ surface vanishes for Phase IV. The absence of tunneling is more directly seen from the perspective of $3$-fold axis. Being a $\mathbb{Z}_2$ topological insulator, Phase V supports three non-trivial Wilson lines along $[001]$ and $[111]$ directions, and three gapless, Dirac fermions, and surface Hall conductivity $\pm \frac{3}{2} e^2/h$ for both surfaces. The number of surface Dirac fermions for Phase II for $(001)$ ($(111)$) surface is $2$ ($4$). But both surfaces possess Hall conductivity $\pm e^2/h$. Finally, phase IX displays $4$ non-trivial Wilson lines along both directions, and supports four gapless Dirac fermions, and surface Hall conductivity $\pm 2 e^2/h$.  } 
    \label{fig:4vs3}
\end{figure*}
As a consequence of cubic symmetry, all $4$-fold symmetric planes exhibit $D_{4h}$ symmetry. Consequently, they manifest as crystal-symmetry-enforced defects of Bloch map, and the $O(4)$ vector reduces to $O(3)$ vector. Topology of such planes can be classified by the second homotopy group $\pi_2(S^2) = \mathbb{Z}$, and 2D winding numbers   \begin{eqnarray}
    && \mathfrak{C}_{m,ab}\left(\frac{1}{2}\epsilon_{jab}k_{j}=0,\pi \right) = \nn \\
    && \frac{1}{4\pi} \int_{T^2} \; dk_adk_b \; \tilde{\mathbf{d}}(\mathbf{k})\cdot \left(\partial_{k_{a}}\tilde{\mathbf{d}}(\mathbf{k}) \times \partial_{k_{b}}\tilde{\mathbf{d}}(\mathbf{k})\right)  ,
\end{eqnarray}
correspond to mirror Chern numbers, describing quantized, non-Abelian Berry flux through high-symmetry plane,  and 
\begin{multline}
   \tilde{\mathbf{d}}(\mathbf{k})=\{d_{a}(\mathbf{k}),d_{b}(\mathbf{k}),d_{5}(\mathbf{k})\}/ \\\sqrt{d_{a}(\mathbf{k})^2+d_{b}(\mathbf{k})^2+d_{5}(\mathbf{k})^2}. 
\end{multline}
Phases with 3D winding numbers $n_{3,-} \neq 0$, support distinct values of quantized, non-Abelian Berry flux $2\pi \mathfrak{C}_{R,ab}$ for $k_{j}=0,\pi$ planes. As a consequence of cubic symmetry, $n_{3,-}$ is precisely related to the tunneling configuration of non-Abelian Berry flux, along three principal $4$-fold axes
\begin{equation}\label{mirrortunneling}
n_{3,-}=\mathfrak{C}_{m,ab}(\frac{1}{2} \epsilon_{abj} k_j=\pi) - \mathfrak{C}_{m,ab}(\frac{1}{2} \epsilon_{abj} k_j=0).
\end{equation}
For the present model, we can express mirror Chern numbers as
\begin{eqnarray}
 \mathfrak{C}_{m,ab}(\frac{1}{2} \epsilon_{abj} k_j=0)=\frac{1}{2}(2\delta_X-\delta_{\Gamma}-\delta_M), \nn \\
  \mathfrak{C}_{m,ab}(\frac{1}{2} \epsilon_{abj} k_j=\pi)=\frac{1}{2}(2\delta_M-\delta_R-\delta_X),
  \end{eqnarray}
 leading to the exact relationship $n_{3,-}=\kappa_{AF,-}$.
  
As weak topological insulators (Phase IV and Phase VIII) exhibit identical mirror Chern numbers for both $k_j=0, \pi$ planes, they do not support 3D tunneling configurations. 
In contrast to this, Phase II and Phase VI, which are commonly denoted as weak topological insulators, carry 3D winding numbers $n_{3,-} = \pm 2$. Since weak $\mathbb{Z}_2$ indices do not carry information regarding sign of mirror Chern numbers, they cannot address the presence or absence of even integer winding numbers. 

Finally, Phase IX is of particular interest, as it supports tunneling configurations of even integer valued mirror Chern numbers for all $4$-fold mirror planes. Also note that even integer mirror Chern numbers occur for $k_j = \pi$ planes of Phase V and $k_j=0$ planes of Phase VII, leading to strong topological insulators with higher winding number $n_{3,-} =+3$.

Along any high-symmetry axis, joining $\bs{Q}^i$ and $\bs{Q}^j$, the $O(4)$ vector $(d_1,d_2,d_3,d_5)$ reduces to $O(2)$ vector, which can be classified by the fundamental group of circle $\pi_1(S^1)=\mathbb{Z}$. Let us consider four high-symmetry lines parallel to the $\hat{z}$ axis, passing through $(k_x,k_y)=(0,0),(\pi,0),(0,\pi), (\pi,\pi)$. As these points correspond to TRIM locations of $(001)$ surface BZ, we will denote them as $\bar{\bs{Q}}=\bar{\Gamma}, \bar{X}, \bar{Y}, \bar{M}$, respectively. The signed 1D winding numbers for high-symmetry axes are given by
\begin{eqnarray}\label{1Ddirected}
&& n_{1}(\bar{\Gamma})=\frac{\text{sgn}(t_p)}{2}(\delta_X-\delta_\Gamma), \nn \\ 
&& n_{1}(\bar{X})=
n_{1}(\bar{Y}) =  \frac{\text{sgn}(t_p)}{2}(\delta_M  -\delta_X), \nn \\
&& n_{1}(\bar{M})=\frac{\text{sgn}(t_p)}{2}(\delta_R-\delta_M), 
\end{eqnarray} 
and these can be combined to write
\begin{eqnarray}\label{1Dstaggered}
&& n_{3,-}=[-n_1(\bar{\Gamma})-n_1(\bar{M})+ n_1(\bar{X}) + n_{1}(\bar{Y})] \nn \\
&& =\text{sgn}(t_p) \kappa_{AF,-}.
\end{eqnarray}
When a high-symmetry axis supports non-trivial 1D winding number, it leads to normalizable 2-component, gapless Dirac fermions, under open boundary conditions along $(001)$ direction. In the presence of infinitesimal regulator $M^\prime \to 0^+$, the surface Hamiltonians in the vicinity of TRIM locations are given by
\begin{eqnarray}
&& H(\bar{\bs{Q}}+ \delta \bs{k}) \approx \text{sgn}[n_{1D}(\bar{\bs{Q}})] \; [\cos(\bar{Q}_y) \delta k_y  \sigma_1 \nn \\ && - \cos(\bar{Q}_x) \delta k_x \sigma_2] -\sigma_3 M^\prime.
\end{eqnarray}
Therefore, the chirality of Dirac cone and the surface Hall conductivity is determined by $\text{sgn}(n_{1D})$. While weak topological insulators (Phases IV and VIIII) support Dirac cones at $(k_x, k_y)=(0,0), (\pi, \pi)$, they come with opposite chirality, causing zero surface Hall conductivity. In contrast to this, Phases II and IX possess net surface Hall conductivity $\pm e^2/h$, $\pm 2 e^2/h$, respectively. Thus, the staggered index provides a precise description of bulk topology and bulk-boundary correspondence. 

The importance of staggered index can be further emphasized by considering tunneling configurations of Berry flux along the $3$-fold axis $(111)$ [see Fig.~\ref{fig:4vs3}]. Phases II, IV, V, and IX respectively lead to $4$, $0$, $3$, and $4$ Dirac cones on $(111)$ surface. But the signed 1D winding number and $\kappa_{AF,j}$ reveal that the Dirac cones for Phase II at the center ($\bar{\Gamma}$) and the boundary of surface BZ ($\bar{M}$) possess opposite chirality. Thus, the net surface Hall conductivity of Phase II remains fixed to $\pm e^2/h$, despite the presence of $4$ Dirac cones.

These collective properties of $O(4)$ vector control topology of $SU(2)$ Berry connection and the regularized Chern-Simons coefficient 
\begin{equation}
\mathcal{CS}_-(M^\prime \to 0^+)=\frac{n_{3,-}}{2}.
\end{equation}
For numerical tight-binding models of \emph{ab initio} band structure, the staggered index will provide a clear idea about the presence ($n_{3,j} \neq 0$) or absence ($n_{3,j} = 0$) of tunneling and the magnitude of winding number can be confirmed by Wilson loop calculations. Due to the $D_{4h}$ symmetry of mirror planes, bands carrying 3D winding number exhibit fully connected, gapless spectrum for $W_{j}(\bs{k}_\perp)$, with $j=x,y,z$. Moreover, the mirror Chern numbers of different planes can be obtained from winding of WCC. Therefore, tunneling configurations along $4$-fold axes of simple cubic systems can be fully characterized by gauge-invariant spectrum of $SU(2)$ Wilson loops. 

Explicit calculations on analytically controlled $4$-band model reveals the following features for Wilson loop $W_{111}$: (i) class A supports fully connected, gapless spectrum; (ii) class B shows gapped spectrum; and (iii) class C exhibits disconnected gapless spectrum. The number of gapless points are precisely counted by the number of non-trivial high-symmetry axes parallel to $(111)$, or the staggered index [see Fig.~\ref{fig:4vs3}]. The calculation of Berry flux for $3$-fold planes has some subtleties, which are explained in the following sections.

\begin{figure*}[t]
    \centering
 \subfigure[]{\includegraphics[scale=0.12]{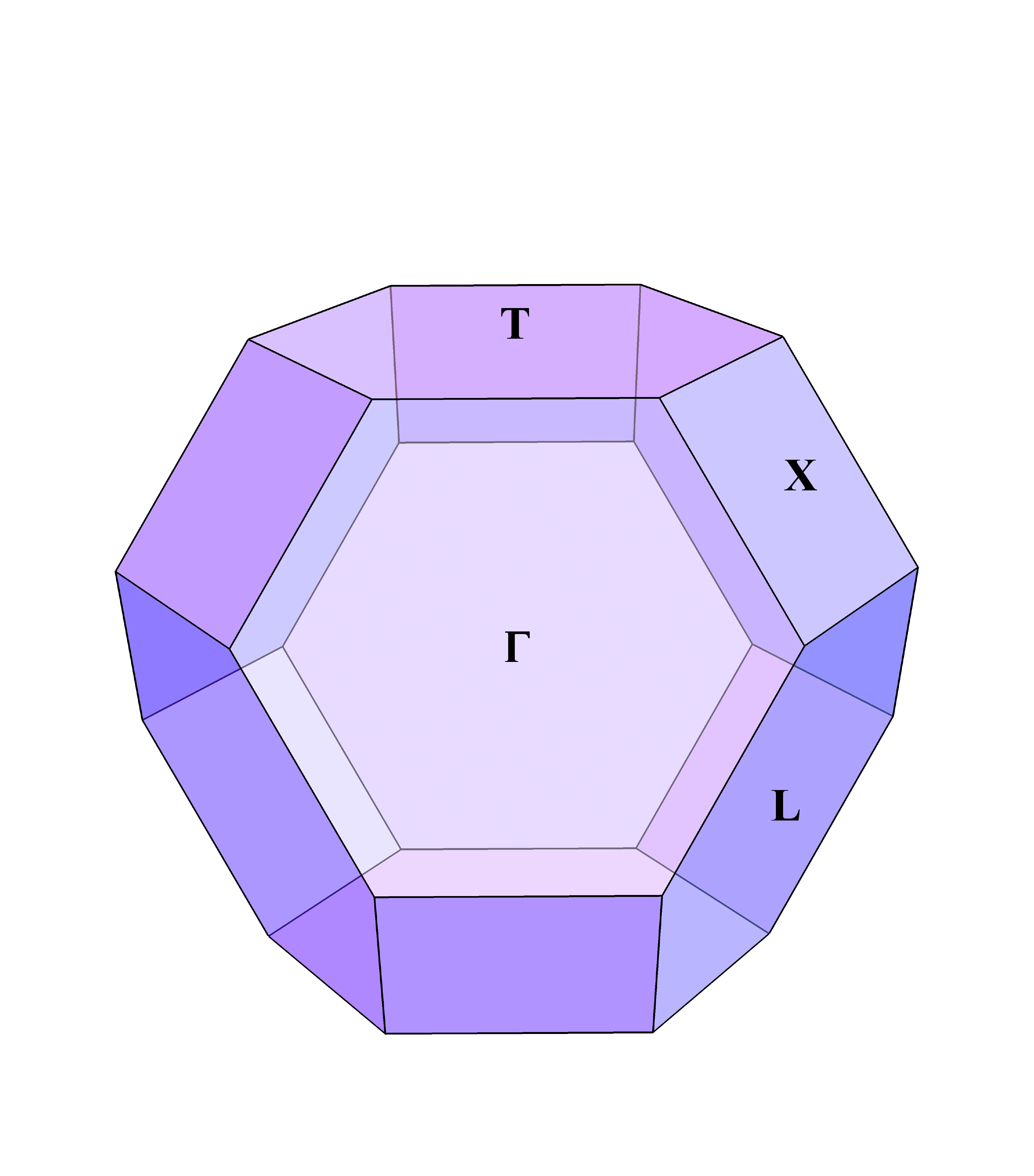} 
      \label{fig:KuramotoBZ}}
       \subfigure[]{
      \includegraphics[scale=0.14]{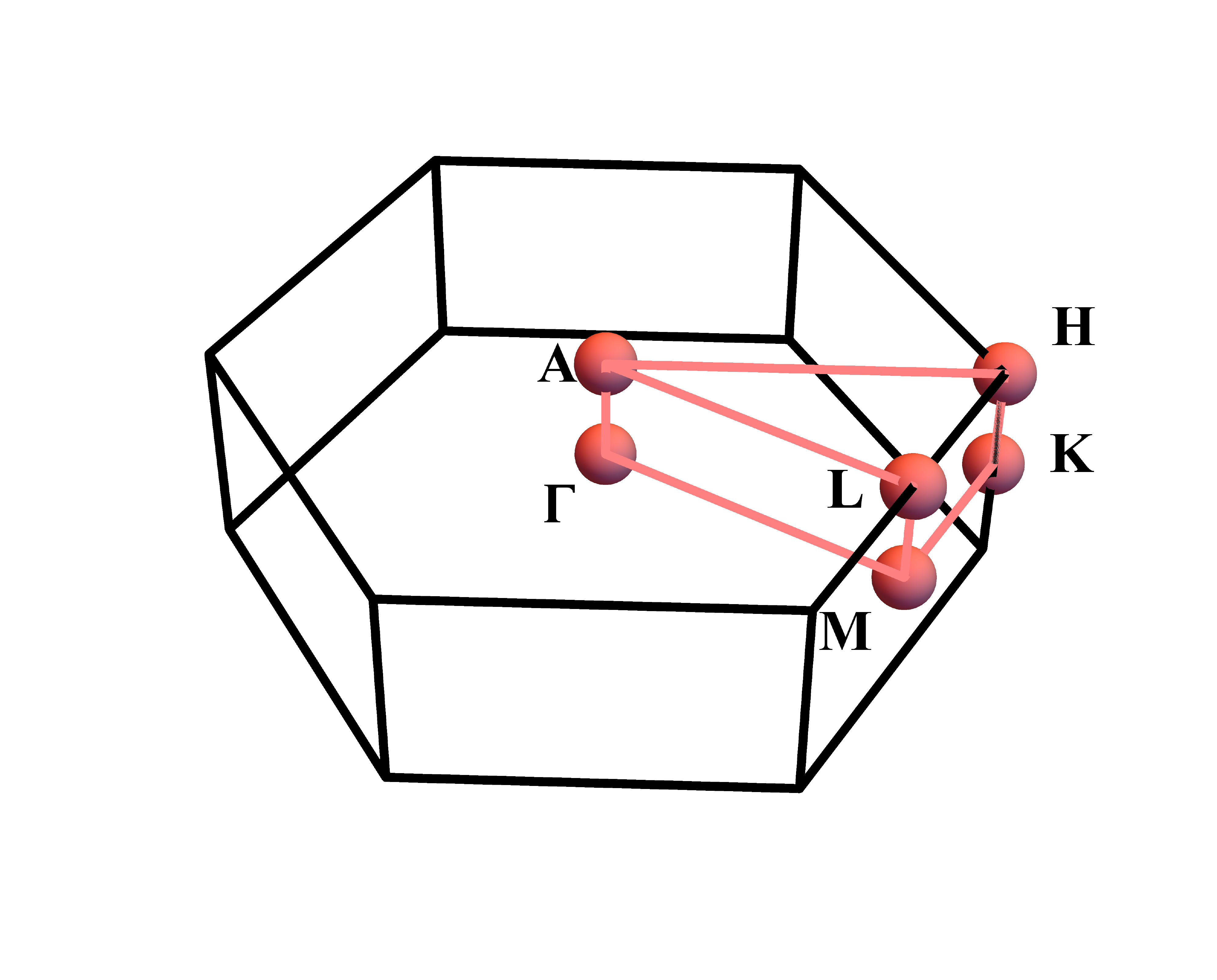} 
         \label{fig:SchBZ}}
  \subfigure[]{ \includegraphics[scale=0.25]{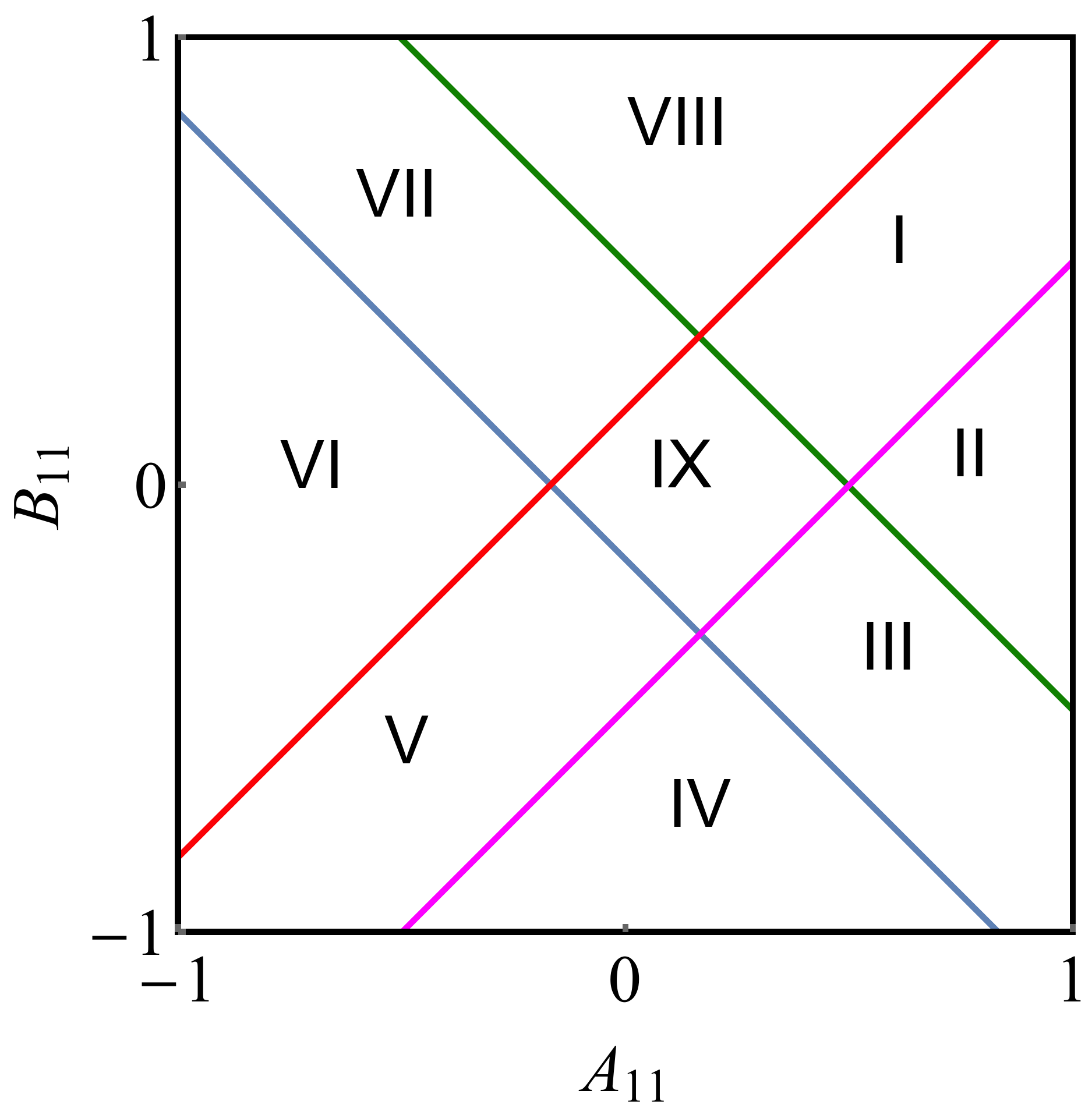}
    \label{fig:Kuramoto_phasediagram}}
      \caption{(a) For the primitive unit cell, the bulk Brillouin zone of rhombohedral systems has the shape of a truncated octahedron, with high-symmetry points $\Gamma$, $T$, $X$, and $L$.  (b) For addressing topology of $3$-fold planes, it is convenient to work with conventional unit cell, which leads to a hexagonal Brillouin zone. The volume of hexagonal Brillouin zone is three times larger than the volume of truncated octahedron. The $\Gamma$, $T$, $X$, and $L$ points of truncated octahedron respectively map to the $\Gamma$, $A$, $M$, and $L$ points of hexagonal Brillouin zone. (c) The phase diagram of rhombohedral model [ Eq.~\ref{SO5} ] when $m_{11}=+1$. Along blue, green, red, and magenta colored lines, band gap can vanish at $\Gamma$, $X$, $T$, and $L$ points, respectively. These lines are described by $d_5(\bs{Q}^i)=0$.}
   \end{figure*}

\begin{figure*}
    \centering
    \subfigure[]{
        \includegraphics[scale=0.45]{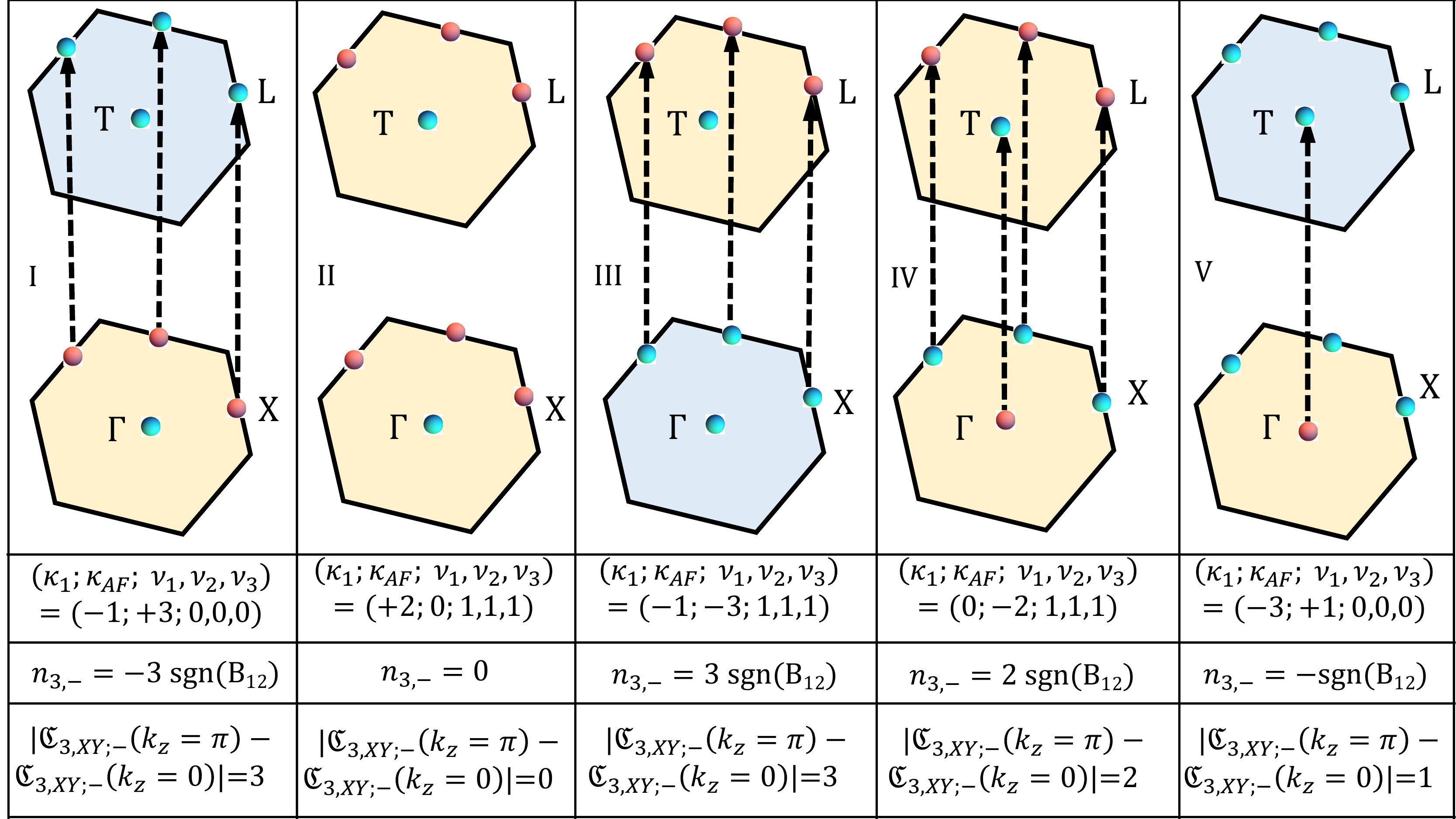}  
         \label{fig:Kuramoto_tunneling1}}
  \subfigure[]{
      \includegraphics[scale=0.45]{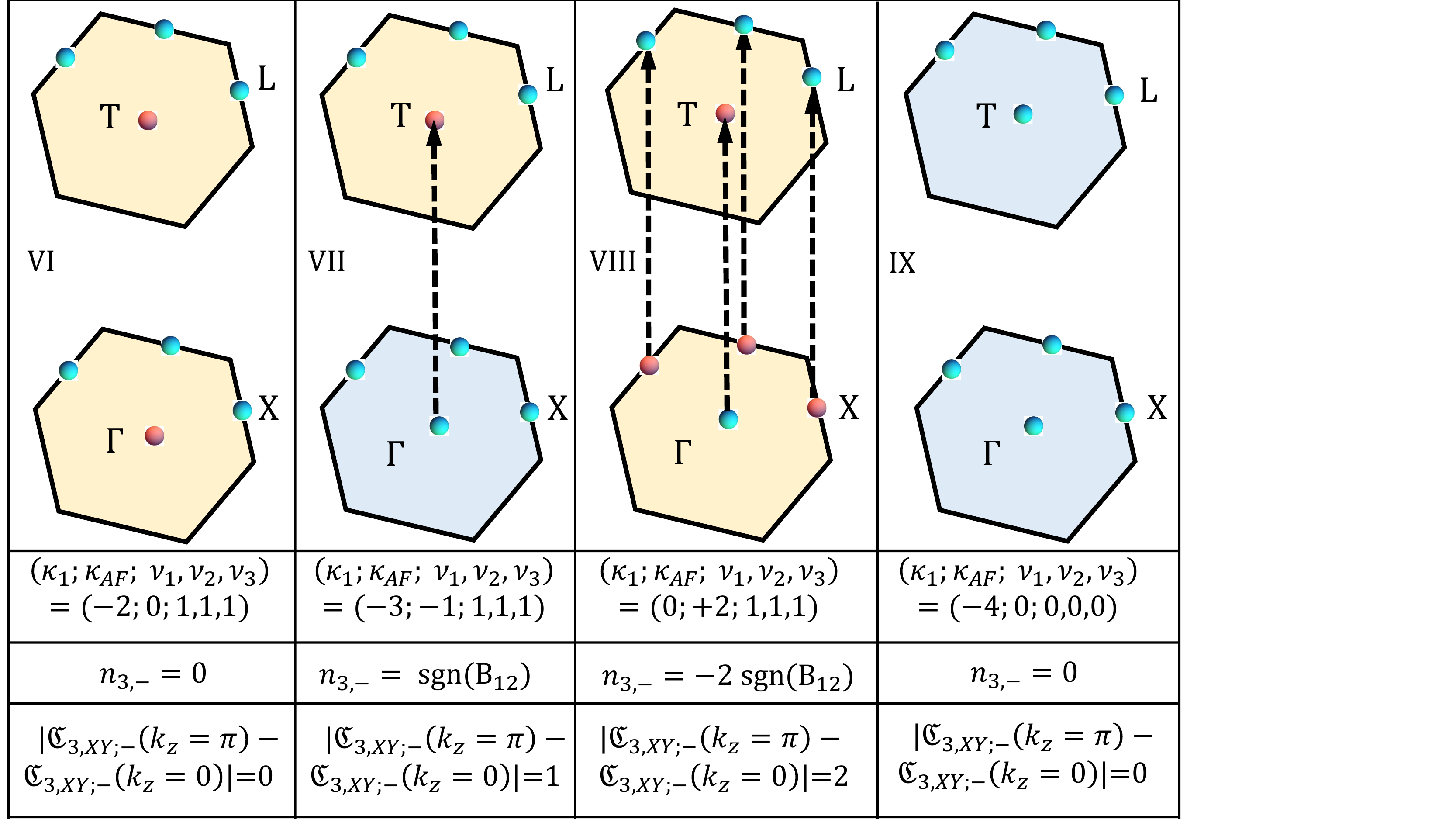} 
       \label{fig:Kuramoto_tunneling2}}
           \caption{Tunneling configurations, symmetry-indicators, and bulk winding numbers for nine homotopically distinct, rhombohedral phases of Fig.~\ref{fig:Kuramoto_phasediagram}. Parity eigenvalue $+1$ $(-1)$ of valence bands is indicated by red (cyan) dot. All three types of band topology can be succinctly understood in terms of tunneling configurations of $C_3$-symmetry protected Berry flux through $XY$ planes. The $\mathbb{Z}_2$ trivial (non-trivial) planes are colored blue (yellow).  } 
    \label{fig:Kuramoto_tunneling}
\end{figure*}

\section{Rhombohedral systems and $D_{3d}$ instantons}\label{Kuramoto}

In Ref.~\onlinecite{Mao2011}, an elegant four-band, tight-binding model was proposed by Mao \emph{et. al} for describing Bi$_2$Se$_3$. 
The 3D bulk Brillouin zone has the shape of a truncated octahedron, as shown in Fig.~\ref{fig:KuramotoBZ}. The primitive reciprocal lattice vectors are given by
\begin{eqnarray}\label{R3mvectors}
  &&  \mathbf{b}_{1}=(-1,-\sqrt{3}/3,b)g,\; 
    \mathbf{b}_{2}=(1,-\sqrt{3}/3,b)g,\; \nn \\
  && \mathbf{b}_{3}=(0,2\sqrt{3}/3,b)g,
\end{eqnarray}
where $b=1/3$ and $g=2\pi$, and the TRIM points are labeled by 
\begin{eqnarray}
    \Gamma=(0,0,0),\;L=\{(1,0,0),(0,1,0),(0,0,1)\}, \nn \\ T=(1,1,1),\; X=\{(1,1,0),(0,1,1),(1,0,1)\}. 
\end{eqnarray}
and the SIs follow from Eq.~\ref{SIR3m}. The underlying point group corresponds to $D_{3d}$ and primary crystalline symmetries are: (i) 3-fold rotation about the $[111]$ axis ($C_{3z}$); (ii) 2-fold rotations about $[1\bar{1}0]$, $[10\bar{1}]$, and $[01\bar{1}]$ axes ($C_{2}$);  (iii) mirror symmetries for 2-fold planes $\mathcal{M}_{1\bar{1}0}$, $\mathcal{M}_{10\bar{1}}$, and $\mathcal{M}_{01\bar{1}}$; (iv) space-inversion symmetry ($\mathcal{P}$).

\begin{figure*}[t]
    \centering
     \centering
     \subfigure[]{
              \includegraphics[scale=0.37]{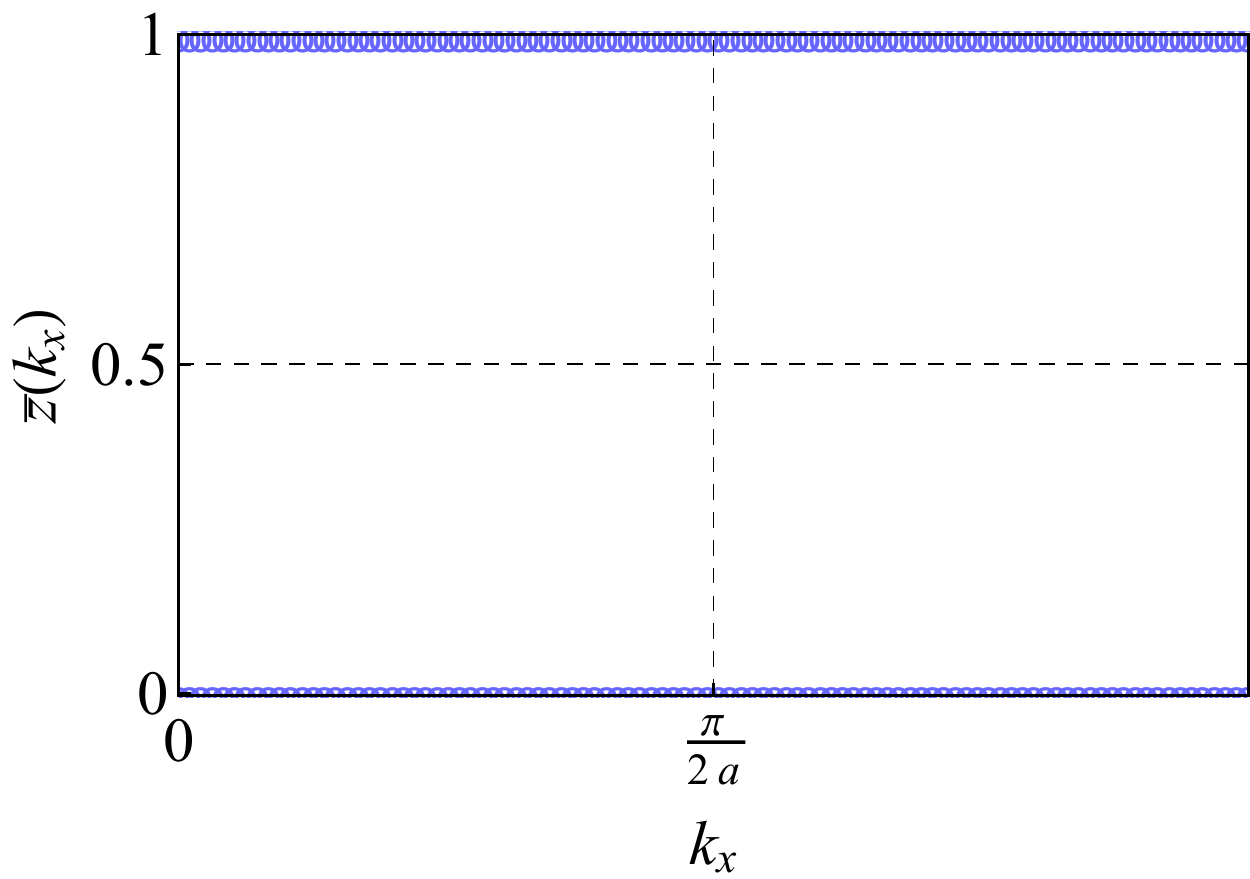} 
      \label{fig:KurBWL}}
          \subfigure[]{
        \includegraphics[scale=0.37]{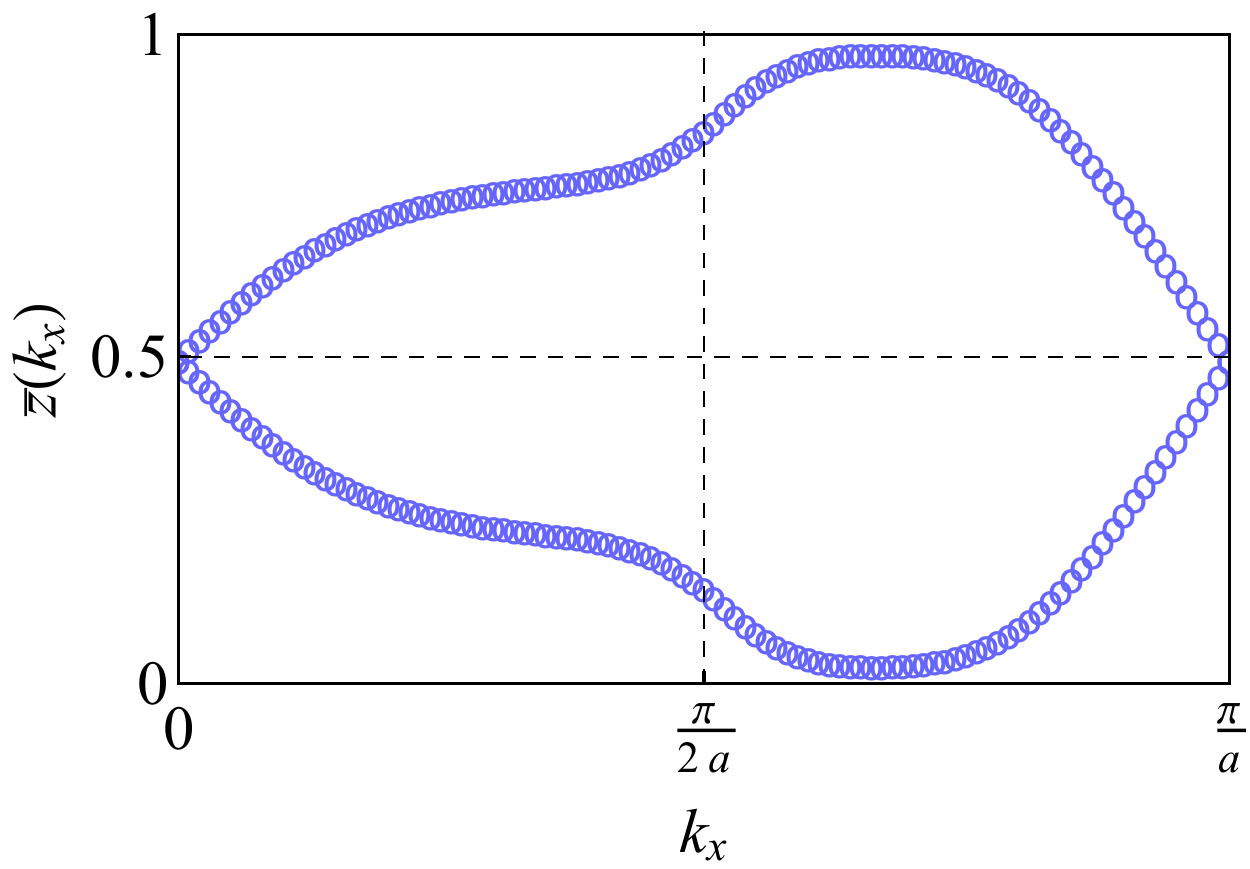} 
        \label{fig:KurCWL}}
    \subfigure[]{
        \includegraphics[scale=0.37]{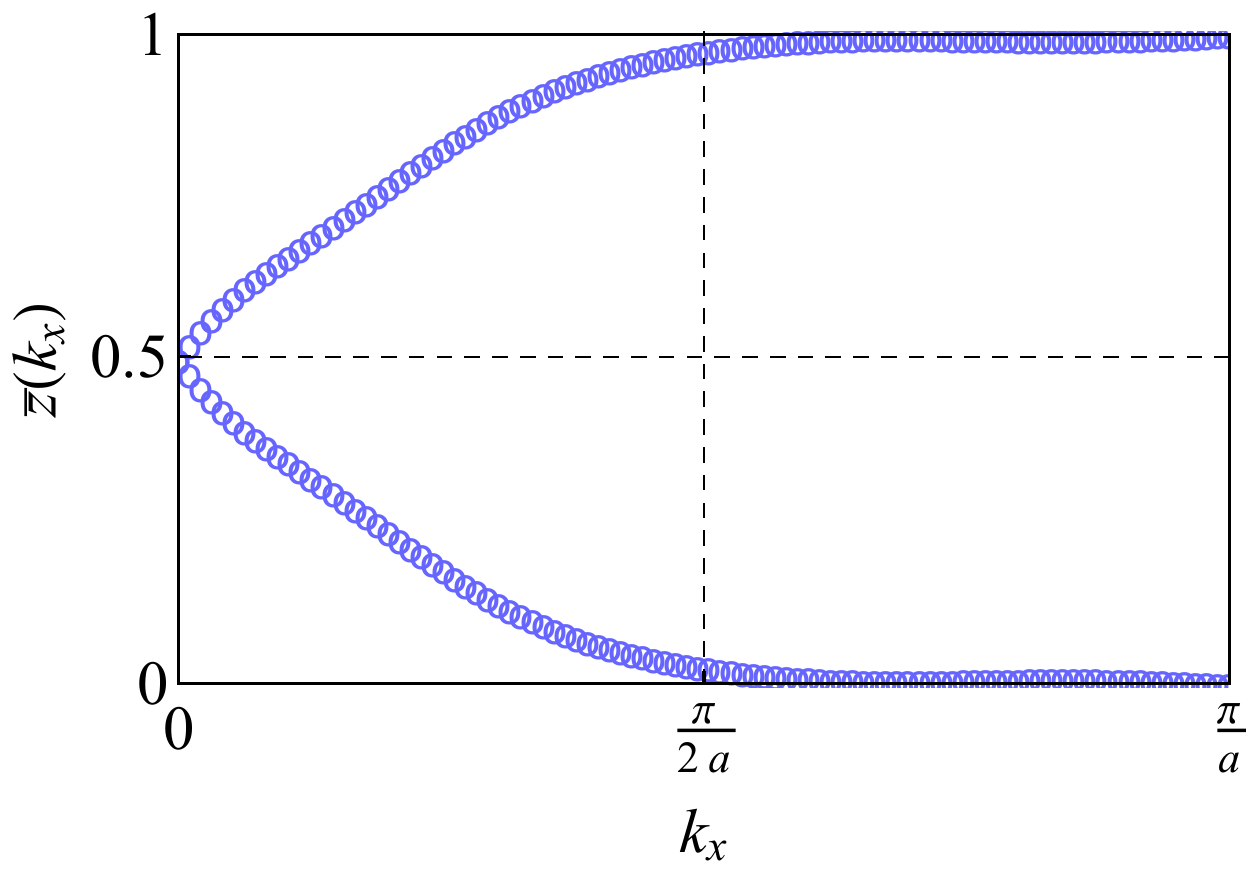} 
        \label{fig:KurAWL}}
  \subfigure[]{
            \includegraphics[scale=0.8]{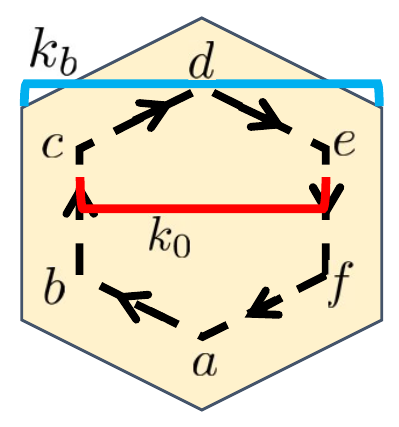} 
 \label{fig:PWLschem}}
  \subfigure[]{ \includegraphics[scale=0.4]{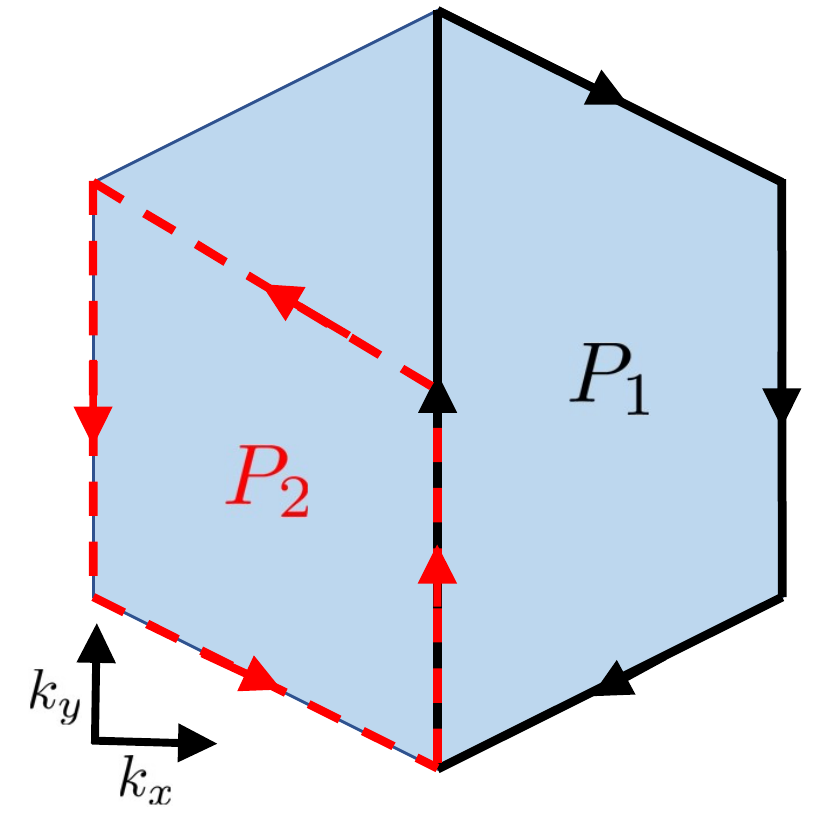}  \label{fig:PWLschemP1P2} }
       \subfigure[]{ \includegraphics[scale=0.2]{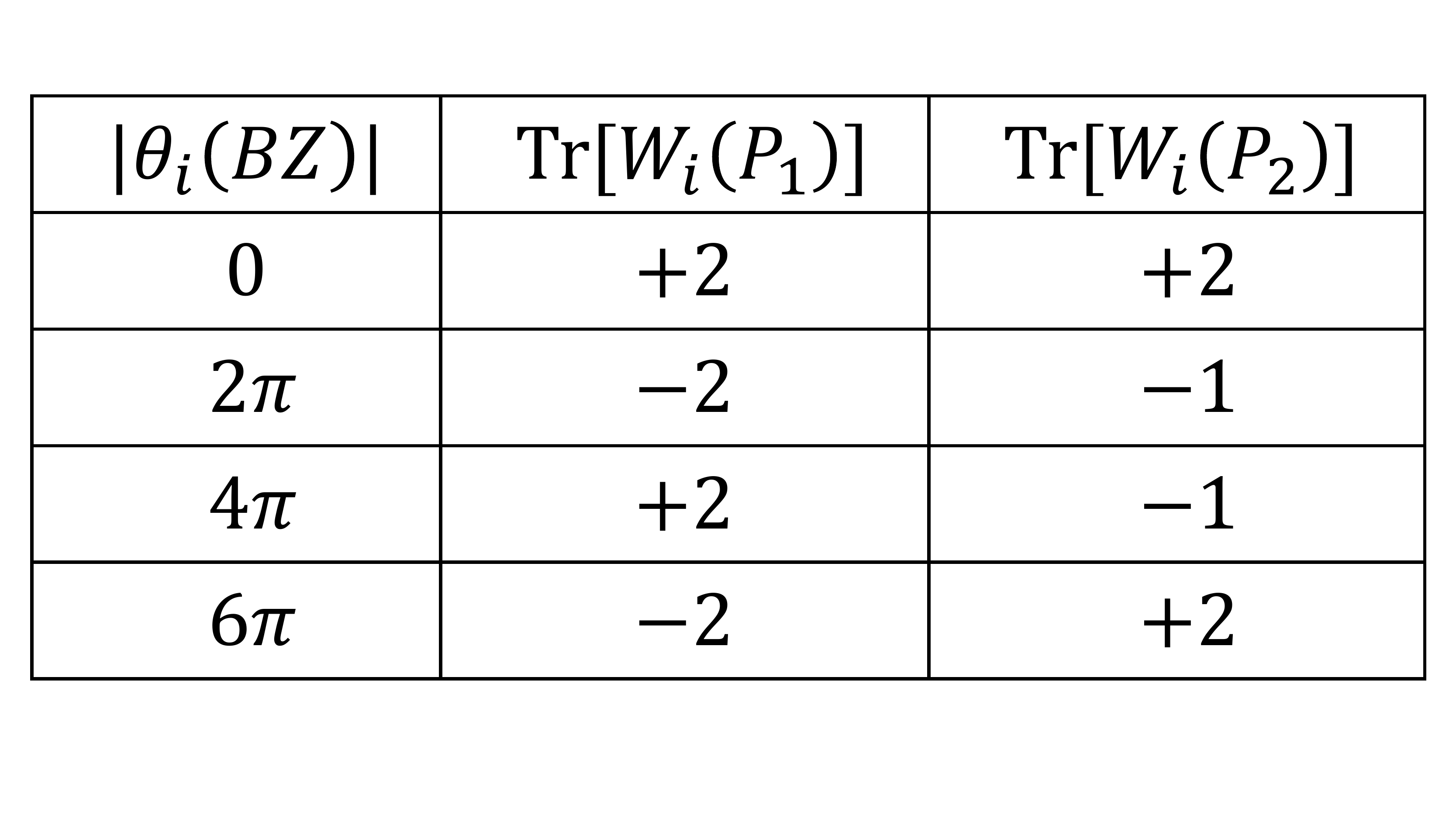} \label{tablePWL}}
  \subfigure[]{
     \includegraphics[scale=0.55]{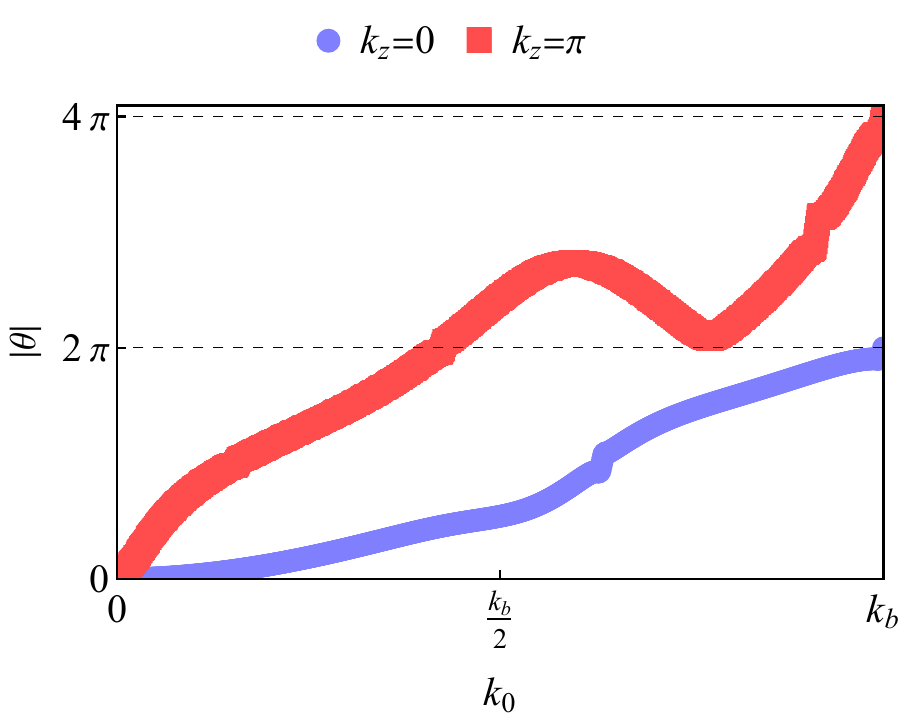} 
       \label{fig:KurPWLI}}
      \subfigure[]{
        \includegraphics[scale=0.55]{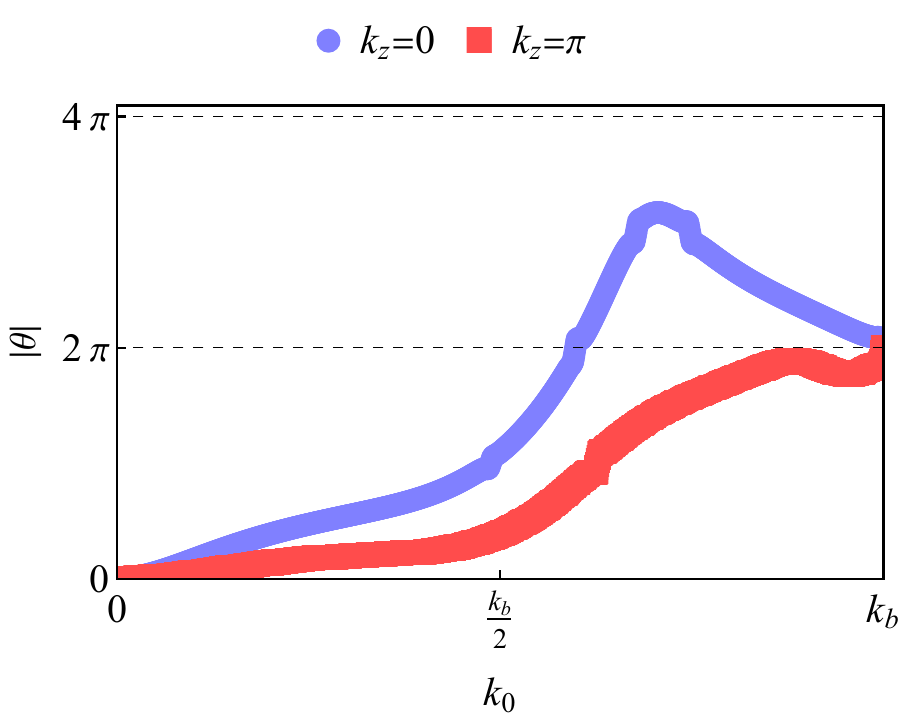} 
        \label{fig:KurPWLIV}}
        \subfigure[]{
        \includegraphics[scale=0.55]{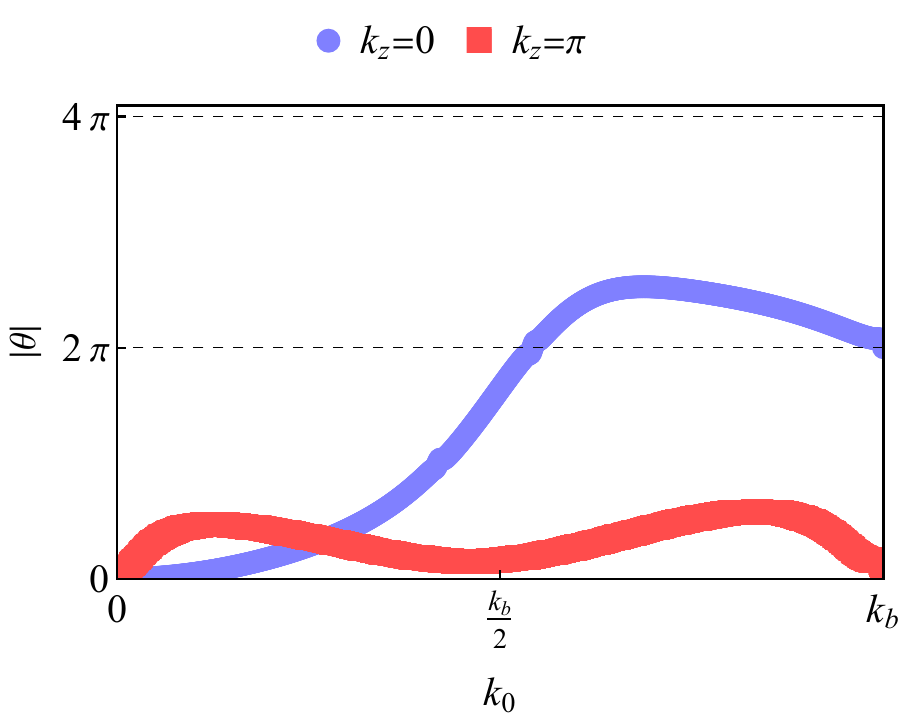} 
        \label{fig:KurPWLV}}
           \caption{Summary of Wilson loop analysis. a) Phase II (class B bands with $n_{3,j}=0$) exhibits gapped spectrum. (b) Phase IV (class C bands with $n_{3,j}=\pm2$) supports disconnected, gapless spectrum. (c) Phase V (class A bands with $n_{3,j}=\pm 1$) exhibits fully connected, gapless spectrum. (d) Schematic of $3$-fold symmetric contour, around which Wilson loop is calculated for $k_z=0, \pi$ planes. The yellow region represents the first Brillouin zone of $xy$ planes. The size of contour is parameterized by $k_{0}$. When $k_{0}=k_{b}$, the path encloses the full Brillouin zone. (e) To reduce computational cost, we can further exploit symmetry of $3$-fold planes and compute in-plane Wilson loops along high-symmetry paths $P_1$ and $P_2$. (c) If a Kramers pair supports quantized flux $|\theta_i|=2 l_i \pi$ through the hexagonal Brillouin zone, $P_1$ and $P_2$ respectively enclose $l_i \pi$, and $\frac{2}{3}  l_i \pi$. Hence, from the trace of Wilson loops, we can distinguish between bands, possessing two-dimensional winding numbers $|\mathfrak{C}_{3,XY;i}|= 0, 1, 2, 3$ for different hexagonal planes. (g) For Phase I  $k_z=0$ and $\pi$ planes respectively possess $|\mathfrak{C}_{3,XY;i}|=1$, $|\mathfrak{C}_{3,XY;i}|=2$. (h) For Phase IV both planes possess $|\mathfrak{C}_{3,XY;i}|=1$. (i) For phase V, $k_z=0,\pi$ planes carry $|\mathfrak{C}_{3,XY;i}|=1,0$, respectively. By applying an infinitesimal $\mathcal{T}$-breaking, training field we obtain signed $\mathfrak{C}_{3,XY;i}$.  }
    \label{fig:PWL_Path1}
\end{figure*}

The Bloch Hamiltonian has the form of Eq.~\ref{Sp4} and $\mathbf{d}(\mathbf{k})$ 
is given by 
\begin{eqnarray}\label{SO5}
       d_{1}(\bs{k}) &=&-2A_{14}\sin w (\sin k_{a2}-\sin k_{a3}) \nn \\ && +2B_{14}[\sin k_{g1}+\cos w(\sin k_{g2} +\sin k_{g3})], \nn \\
       d_{2}(\bs{k})&=&-2B_{14}\sin w (\sin k_{g2} -\sin k_{g3}) \nn \\ && -2A_{14}[\sin k_{a1}+\cos w(\sin k_{a2}+\sin k_{a3})], \nn \\
     d_{3}(\bs{k})&=&2A_{12}\sum_{i=1}^{3}\sin k_{ai}, \nn \\
   d_{4}(\bs{k})&=&-2B_{12}\sum_{i=1}^{3}\sin k_{gi}, \nn  \\
  d_{5}(\bs{k})&=&2A_{11}\sum_{i=1}^{3}\cos k_{ai}+2B_{11}\sum_{i=1}^{3}\cos k_{gi}+m_{11}, \nn \\
\end{eqnarray}
where $w=-2\pi/3$, $k_{ai}=\mathbf{k}\cdot \mathbf{a}_{i}$ and $k_{gi}=\mathbf{k}\cdot \mathbf{g}_{i}$, $a_{1}=(a,0,0)$, $a_{2}=(-\frac{a}{2},\frac{\sqrt{3}a}{2},0)$, $a_{3}=(-\frac{a}{2},-\frac{\sqrt{3}a}{2},0)$ $g_{1}=(0,\frac{a}{\sqrt{3}},c)$, $g_{2}=(-\frac{a}{2},-\frac{\sqrt{3}a}{6},c)$, and $g_{3}=(\frac{a}{2},-\frac{\sqrt{3}a}{6},c)$, and $c/a=b=1/3$.

Under symmetry operations of D$_{3d}$ point group, $(d_1(\bs{k}),d_2(\bs{k}))$, $d_{3}(\mathbf{k})$, $d_4(\bs{k})$, and $d_5(\bs{k})$ respectively transform as $E_u$-doublet, $A_{2u}$-singlet, $A_{1u}$-singlet, and $A_{1g}$-singlet. The operations of $\mathcal{P}$, $\mathcal{T}$, $C_{3z}$, $C_{2x}$, and $M_{YZ}$ symmetries are respectively implemented with $$\Gamma_5, \; i \Gamma_{31}, \; e^{i \pi/3 \Gamma_{12}}, \; i \Gamma_{14}, \; \text{and} \;  \Gamma_5 \Gamma_{14}=\Gamma_{23}.$$ The presence of $A_{2u}$ harmonic is a natural consequence of crystalline symmetry. It does not affect symmetry-indicators and bulk winding numbers, and universal topological properties are captured by $4$-component vector $(d_1, d_2, d_4, d_5)$, and the $SU(2)$ matrix $u(\bs{k})=\hat{d}_5(k) \sigma_0 + i \hat{d}_1(\bs{k}) \sigma_1 + i \hat{d}_2 (\bs{k})\sigma_2 + i \hat{d}_4 (\bs{k}) \sigma_3$. Thus, Eq.~\ref{SO5} is a non-trivial example of 4-band model, where a homotopically non-trivial $O(4)$ vector remains embedded within $O(5)$ unit vector.

The current model is sufficient for capturing $14$ out of $16$ configurations (except $\kappa_{AF}=\pm 4$) of Table~\ref{tab3} and 3D winding numbers
\begin{equation}
n_{3,-} = -\text{sgn}(B_{12}) \kappa_{AF,-}=\pm 1, \pm 2, \pm 3,
\end{equation}
for $u(\bs{k})$. After regulating $$d_3(\bs{k}) \to d_3^\prime(\bs{k}) = M^\prime + d_3(\bs{k}),$$ the regularized Chern-Simons coefficient of valence bands is given by $\mathcal{CS}_-(M^\prime \to 0^+ )=\frac{n_{3,-}}{2}$. A representative phase diagram involving nine phases are shown in Fig.~\ref{fig:Kuramoto_phasediagram}. A summary of SIs, bulk winding numbers, and Wilson loop analysis are presented in Fig.~\ref{fig:Kuramoto_tunneling}.

\subsection{$SU(2)$ Wilson loops}
Along all high-symmetry axes joining two TRIM points, the $O(5)$ vector reduces to different $O(2)$ vectors. For understanding the presence or absence of tunneling, we first consider 1D winding numbers for high-symmetry axes, which are parallel to $\hat{z}$. Using hexagonal BZ, these lines can be identified as $\Gamma T \equiv \Gamma A$ passing through $(k_x,k_y)=(0,0)$, and three $XL \equiv ML$ lines, respectively passing through $(2\pi, 0)$, $(\pi, \sqrt{3} \pi)$, and $(-\pi, \sqrt{3} \pi)$. The signed 1D winding numbers and 3D winding number can be related as
\begin{eqnarray}
&& n_{1}(\Gamma T) = \text{sgn}(B_{12}) \frac{1}{2} (\delta_{T,-}-\delta_{\Gamma,-}), \nn \\ 
&& n_{1}(XL)=\text{sgn}(B_{12}) \frac{1}{2}(\delta_{L,-}-\delta_{X,-}), \nn \\
&&n_1(\Gamma T)+3n_1(XL)=n_{3,-}=-\text{sgn}(B_{12}) \kappa_{AF,-}. \nn \\
\end{eqnarray}
When $n_1$ is non-trivial, $SU(2)$ Wilson loop $W_{z,-}(k_x, k_y)$ displays gapless spectrum. The WCC spectrum $\bar{z}(k_x, k_y)$ for classes A (Phase V), B (Phase II), and C (Phase IV) are shown in Figs.~\ref{fig:KurBWL}-\ref{fig:KurAWL}. As $3$-fold planes lack mirror symmetry, class C bands exhibit disconnected, gapless spectrum (as emphasized for simple cubic systems). Therefore, the full connectivity of WCC is not an essential criterion to determine 3D winding number.

For $3$-fold $XY$ planes, $O(5)$ vector does not reduce to $O(3)$ vector. Thus, 2D winding numbers must be found from in-plane Wilson loops, defined as  \begin{equation}
    W_{j}(C)=\text{P} \; \text{exp}\left[i\oint \sum_{a=1}^{2}A_{a,j}(\mathbf{k}(l))\frac{dk_{a}}{dl}dl\right].
\end{equation}
It describes $SU(2)$ Berry phase accrued by the $j$-th Kramers-degenerate band, when parallel-transported along a closed, non-intersecting curve $C$, parameterized by $\mathbf{k}(l)$. As an element of $SU(2)$ group, $W_{j}$ can be written as, 
\begin{equation}\label{inplaneWL}
W_{j}(C)=\text{exp}\left[i\theta_{j} \hat{\mathbf{\Omega}}_j \cdot \boldsymbol \sigma \right].
\end{equation}
By employing non-Abelian Stokes theorem, the gauge invariant angle $\theta_{j}$ can be related to the magnitude of $SU(2)$ Berry flux enclosed by the loop $C$.~\cite{tyner2020topology,tyner2021quantized} The in-plane loop will be calculated with $3$-fold symmetry preserving contours [see Fig.~\ref{fig:PWLschem} ], and the area of the loop will be gradually increased from $0$ to the area of hexagonal BZ. The magnitude of relative Chern number $|\mathfrak{C}_{3,XY; j}|$ is found from
\begin{equation}
|\mathfrak{C}_{3,XY; j}| = \frac{1}{2\pi} |\theta_j(k_0=k_b) - \theta_j(k_0=0)|.
\end{equation}
The computation of 2D winding numbers can be further simplified by taking advantage of crystalline symmetry, as explained in Figs.~\ref{fig:PWLschemP1P2} and ~\ref{tablePWL}. The results for in-plane Wilson loops for Phases I, IV, and V are shown in Figs.~\ref{fig:KurPWLIV}-\ref{fig:KurPWLV}. 

To further elaborate on important differences with simple cubic systems, we study $YZ$ mirror planes, passing through $k_x=0, 2\pi$. For these planes $d_2=d_3=0$, and we can compute mirror Chern numbers from the $O(3)$ vector $(d_1, d_4, d_5)$. While class A bands support mirror Chern numbers 
\begin{eqnarray}
\mathfrak{C}_{m,YZ}(k_x=0)=\mathfrak{C}_{m,YZ}(k_x=2\pi)=\pm 1, \nn \\
\end{eqnarray}
no tunneling occurs along $2$-fold axis. In contrast to this, Class B and Class C bands do not possess any mirror Chern numbers. Thus, $n_{3,-}$ cannot be identified from $\mathfrak{C}_{m,YZ}$. With analytical and numerical insights gained for $R\bar{3}m$ instantons, in the following section we address topology of \emph{ab initio} band structure of Bi.

\begin{figure*}[t]
    \centering
\subfigure[]{ \includegraphics[scale=0.45]{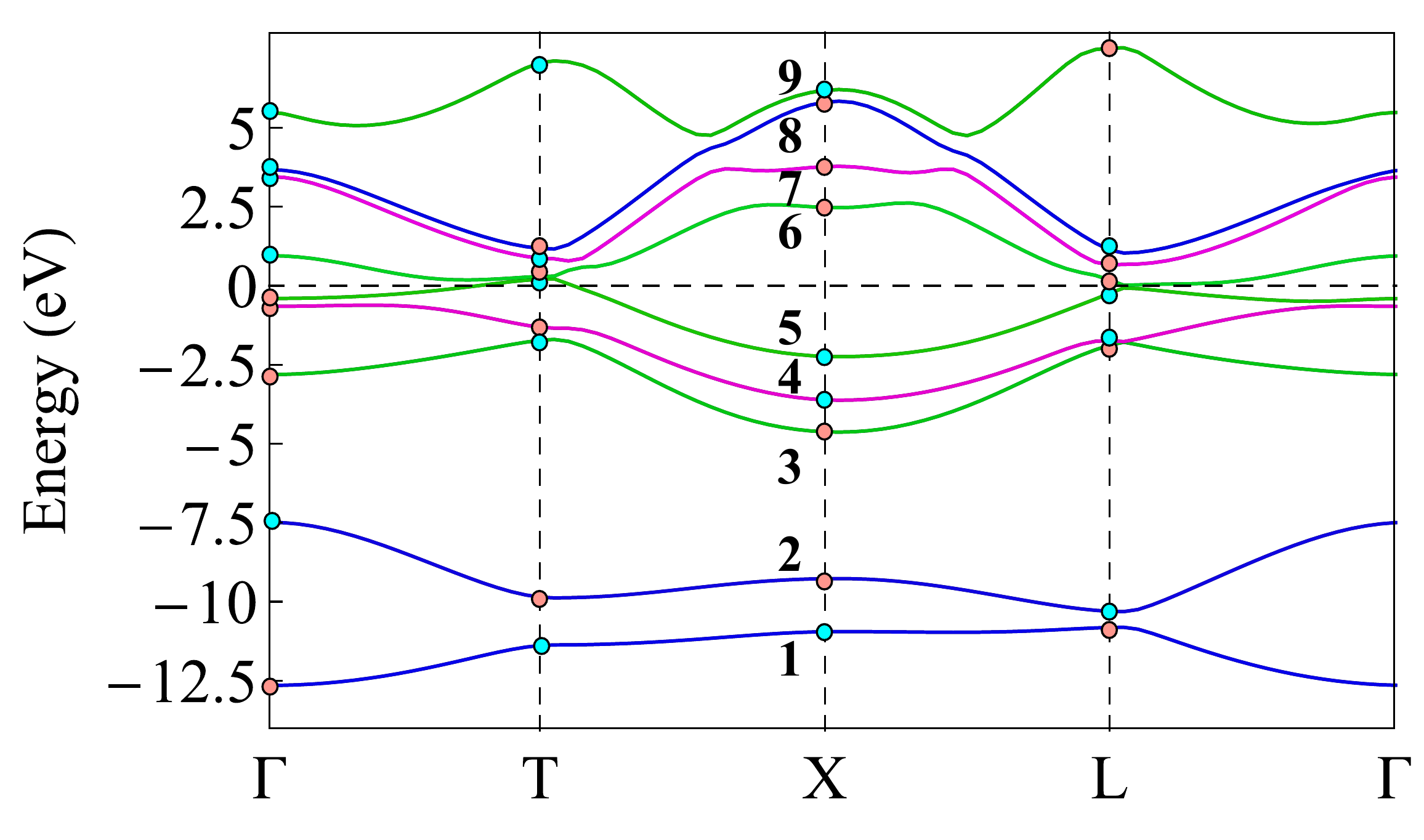}  \label{fig:Pan0}}
 \subfigure[]{ \includegraphics[scale=0.42]{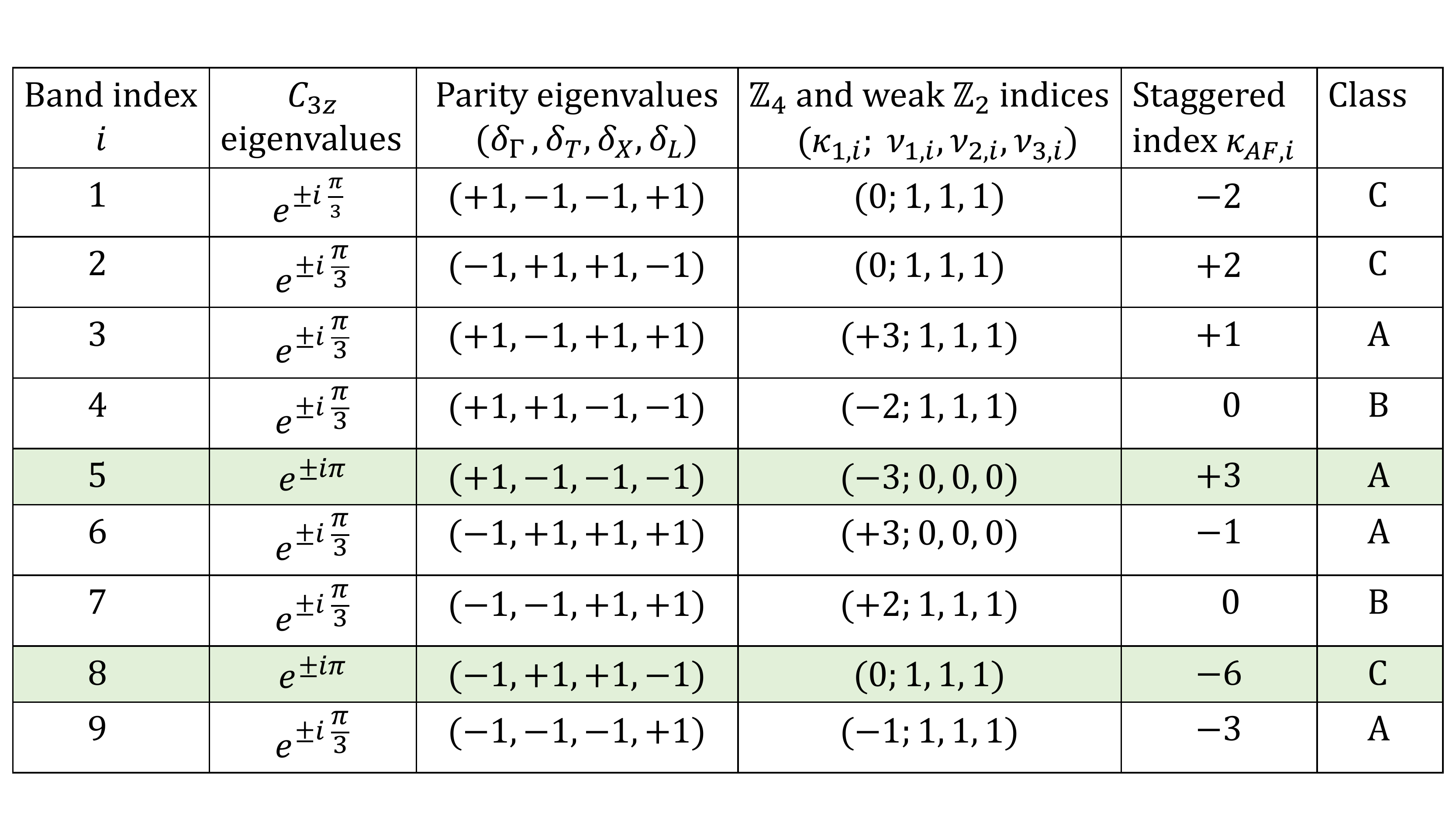} \label{fig:Pan1}}
            \caption{(a) Bulk band structure of Bi along high symmetry paths of primitive Brillouin zone (Fig.~\ref{fig:KuramotoBZ} ) and light-red (cyan) dots denote parity eigenvalues +1 (-1) at time-reversal-invariant momentum points. Class A, Class B and Class C bands are respectively colored as green, purple, and blue. (b) Summary of symmetry data and indicators for various Kramers-degenerate bands of bismuth. According to Table~\ref{tab3} bands $i=1$ through $9$ have parity eigenvalue configuration numbers $14$, $13$, $5$, $12$, $4$, $3$, $11$, $13$, and $8$, respectively. We have used Eq.~\ref{SIR3m}, and ~\ref{modified-st} to account for the rotation eigenvalues. The hypothetical insulator with fully occupied bands $1$ through $5$ is a higher-order topological insulator, with ground state indicators given by  Eq.~\ref{BismuthGDST}. To understand these indicators one must consider the combined effects of bands $3$, $4$, and $5$. If the chemical potential is placed between bands $1$ and $2$, the resulting insulator will exhibit class C topology. In contrast to this, all ground state indicators would vanish for the insulator obtained by placing the chemical potential between bands $2$ and $3$. }
  \end{figure*}

\begin{figure*}
    \centering
    \subfigure[]{
   \includegraphics[scale=0.54]{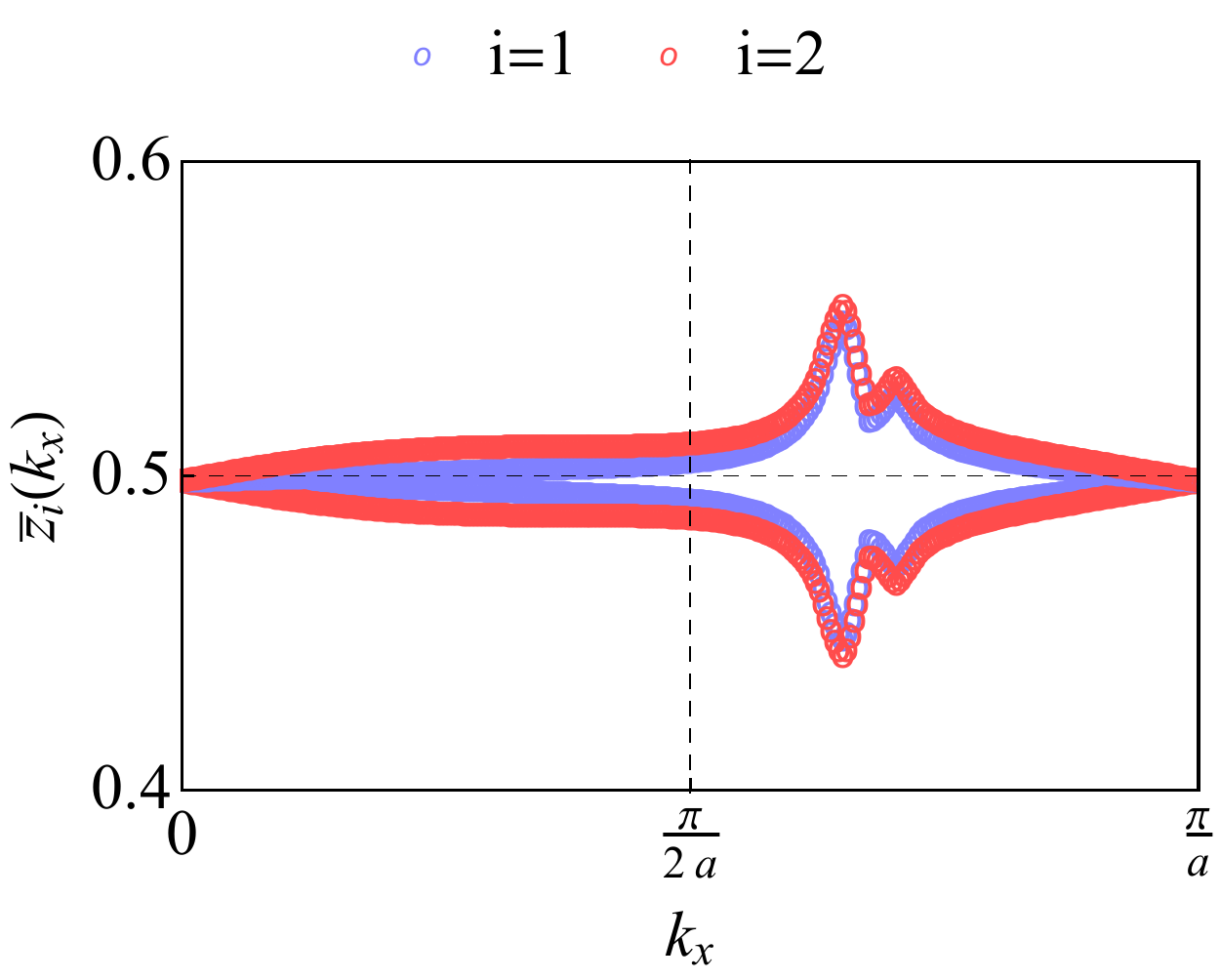} 
         \label{fig:2a}}
    \subfigure[]{
        \includegraphics[scale=0.54]{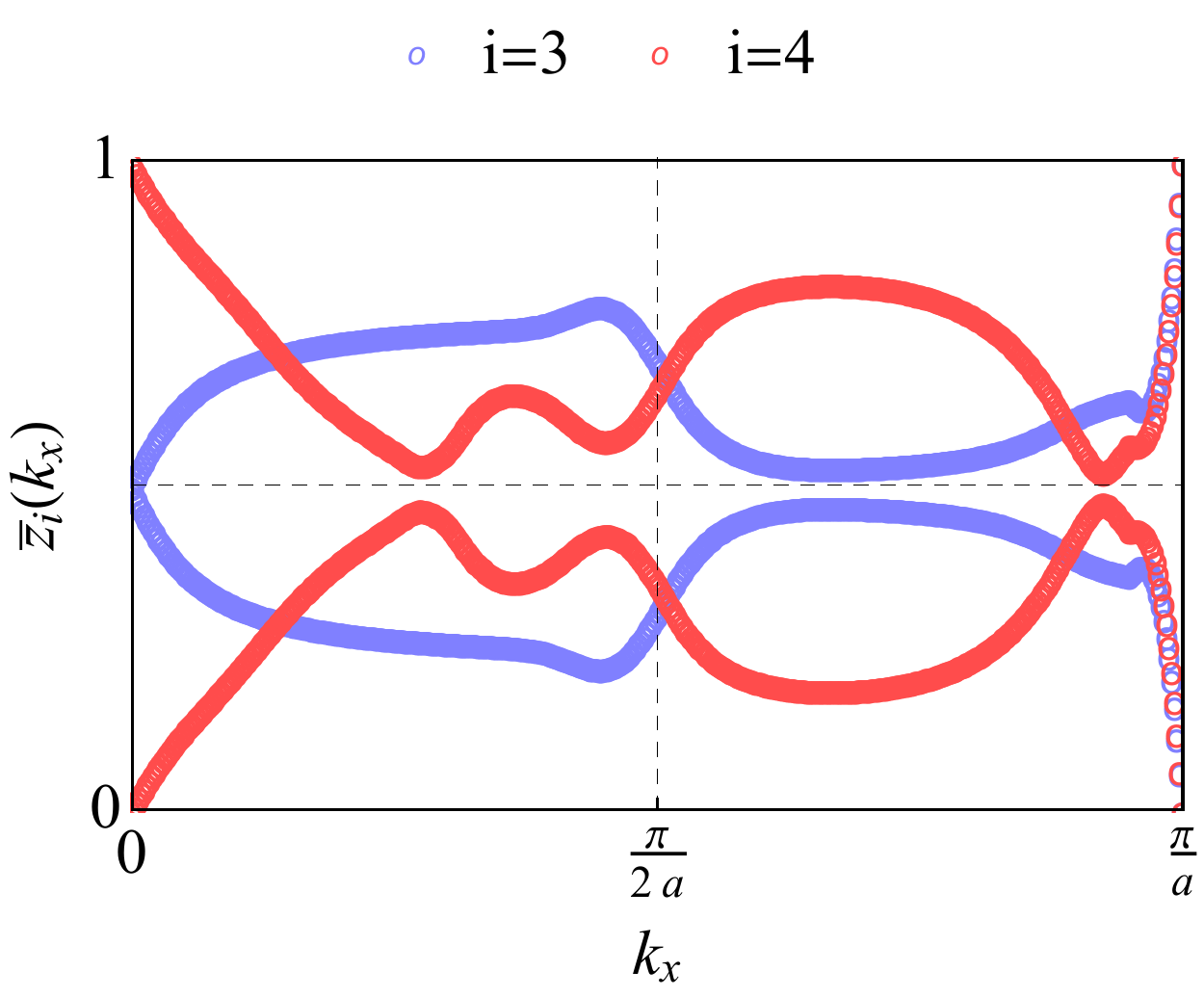} 
         \label{fig:2b}}
    \subfigure[]{
        \includegraphics[scale=0.54]{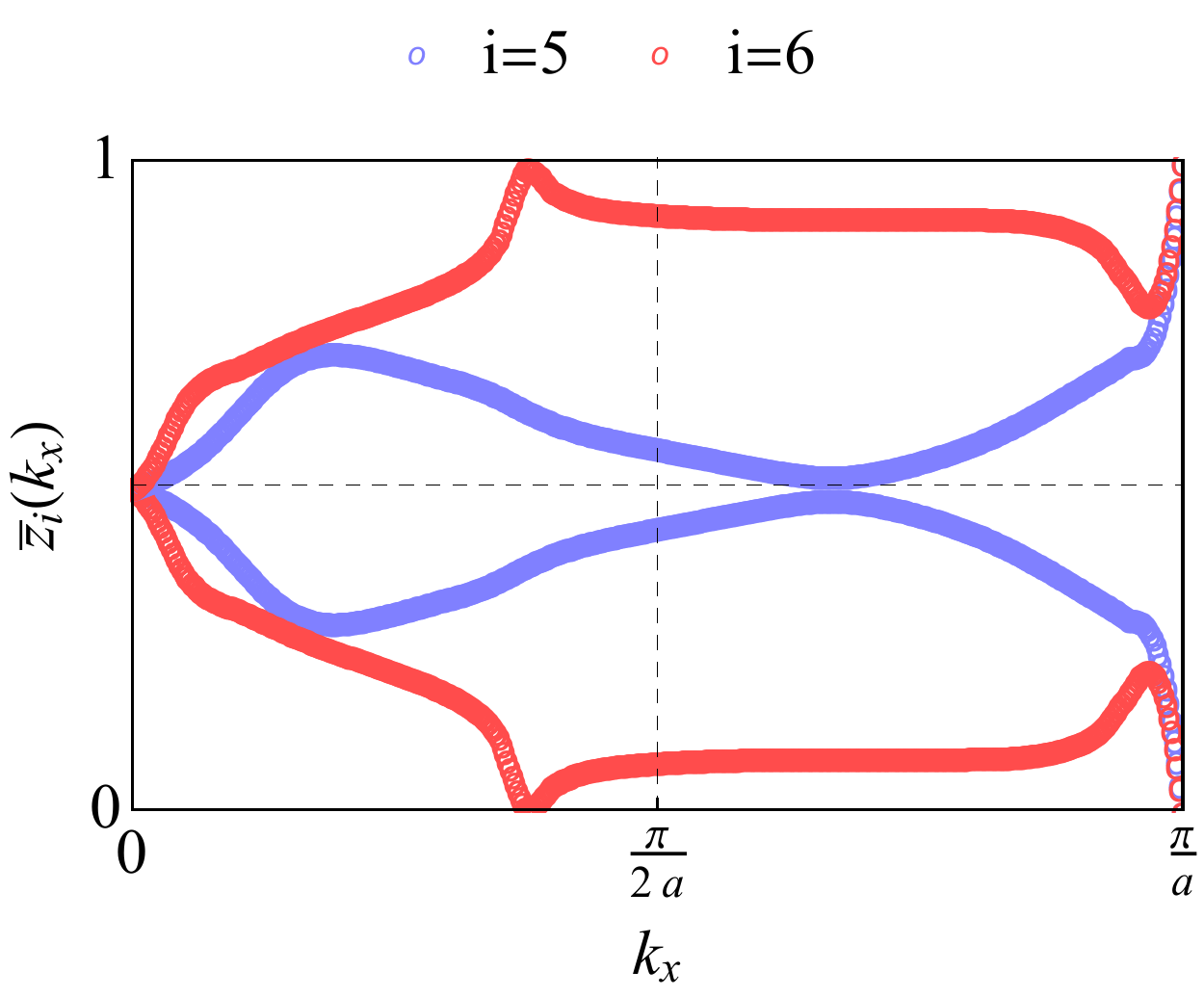} 
       \label{fig:2c}}
        \subfigure[]{ \includegraphics[scale=0.54]{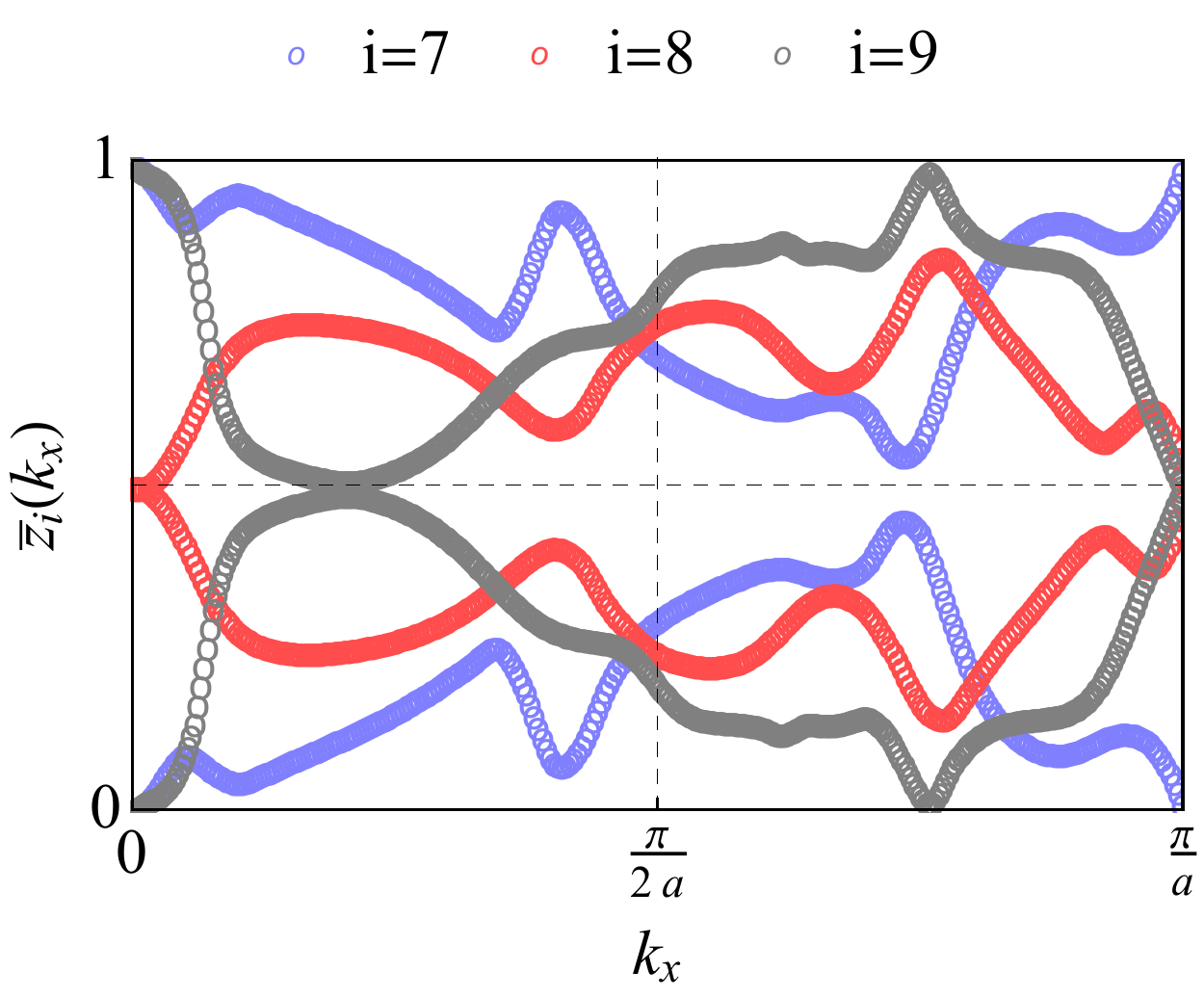} 
         \label{fig:2d}}
    \caption{Spectra of $SU(2)$ Wilson line $W_{z,i}(k_x,k_y=0)$ for different bands. In precise agreement with results from previous section, (i) Class A bands ($3$, $5$, $6$, and $9$) exhibit fully connected, gapless spectrum; (ii) Class B bands ($4$ and $7$) possesses gapped spectrum; (iii) Class C bands ($1$, $2$, and $8$) display disconnected, gapless spectrum. The number of gapless points for bands $3$, $5$ and $6$ [$9$] is one [$3$] and located at the $\bar{\Gamma}$ point [three $\bar{M}$ points] of surface Brillouin zone. The total number of gapless points for class C bands is $4$ ($\bar{\Gamma}$, and three $\bar{M}$ points of surface Brillouin zone). As indicated by the staggered index, class B bands lack tunneling of Berry flux.}
    \label{fig:WCC}
\end{figure*}

\section{\emph{Ab initio} band structure of bismuth}\label{abinitio}

Though originally considered to be a topologically trivial system, refined symmetry-indicators show that the ground state admits both higher-order and rotational-symmetry-protected crystalline topology.~\cite{schindler2018higher,Rudenko2017,kim2016topological,zhu2019evidence,bieniek2017stability,hsu2019topology,hofmann2006surfaces} Does this imply the existence of even integer 3D winding number? We will affirmatively answer this question with a combined analysis of $\kappa_{AF}$ and $C_3$-symmetry-protected tunneling configurations of non-Abelian Berry flux. This tunneling configuration also underpins the diversity of topological phases that can be realized by varying the number of buckled honeycomb layers and the strength of bucking.~\cite{MurakamiBiFilm,rasche2013stacked,drozdov2014one,Chen2013bilayer,nayak2019resolving,TakayamBi,lei2016electronic,chang2019band,BiFilmIto,BiFilmTB} We will directly analyze \emph{ab initio} data, as the sixteen band Liu-Allen model~\cite{liu1995electronic} does not faithfully capture topological properties.~\cite{teo2008surface}

Since Bi is a rhombohedral system with space group $R\bar{3}m$\cite{golin}, the BZ for primitive unit cell has the shape of truncated octahedron [see Fig.~\ref{fig:KuramotoBZ}]. The primitive reciprocal lattice vectors are given by Eq.~\ref{R3mvectors}, with $b=0.384919$ and $g=1.36307 \r{A}^{-1}$.~\cite{Jain2013} 
All density-functional theory (DFT) are carried out with Quantum Espresso software package.~\cite{QE-2009,QE-2017,QE-2020} Exchange-correlation potentials employ Perdew-Burke-Ernzerhof (PBE) parametrization of generalized gradient approximation (GGA).~\cite{Perdew1996} All topological analysis are performed with Wannier90 and Z2pack software packages.~\cite{Pizzi2020,Z2pack,Soluyanov2011}

The bulk band structure for primitive unit cell and coarse topological classification are respectively shown in Fig.~\ref{fig:Pan0} and Fig.~\ref{fig:Pan1}. Bismuth is a compensated semimetal with band $5$ ($6$) producing hole pocket around $T$ point (electron pockets around $L$ points). This does not affect topology of constituent bands. The hypothetical gapped, ground state, involving occupied bands $1$ through $5$ is a higher-order, TCI, with ground state indicators 
\begin{eqnarray}\label{BismuthGDST}
&&(\nu_{0,GS}; \kappa_{1,GS}; \kappa_{AF,GS}; \nu_{1,GS}, \nu_{2,GS}, \nu_{3,GS}) \nn \\ && =(0; -2; +4; 0, 0, 0).
\end{eqnarray}
Without considering rotation eigenvalues $e^{\pm i \pi}$ of band $5$, we would have found $\kappa_{AF, GS}=+2$.

To guarantee the absence of non-trivial Wilson lines through generic locations of hexagonal planes, we have computed WCCs ($\bar{z}_i(k_{x},k_y)$) for different bands, which are displayed in Fig.~\ref{fig:WCC}. The results are in direct correspondence with those presented in Fig.~\ref{fig:KurBWL}-\ref{fig:KurAWL}. Hence, we can conclude that class A, B, and C bands of Bi respectively support odd, zero, and even integer values of flux tunneling along $C_3$ axis. 
The calculation of mirror Chern numbers of occupied bands leads to
\begin{eqnarray}
\mathfrak{C}_{m,YZ} = \text{diag} (0, 0, +1, 0, -1).
\end{eqnarray}
Again in full agreement with results of $4$-band model, only class A bands are found to possess mirror Chern numbers. As bands $3$ and $5$ carry opposite mirror Chern numbers, the net mirror Chern number for the ground state vanishes. 

\begin{figure}
    \centering
       \includegraphics[scale=0.38]{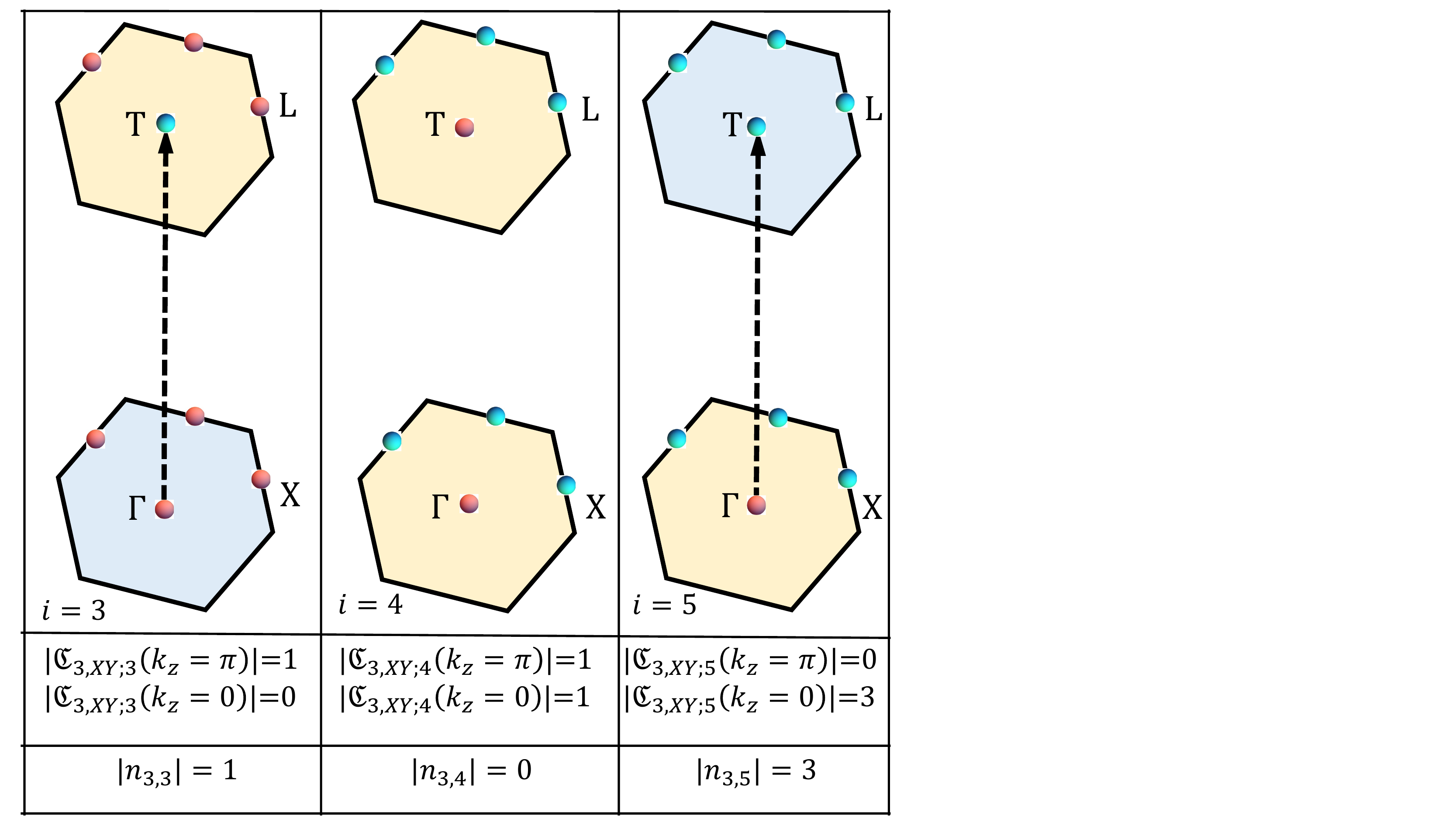} \caption{Tunneling configurations of occupied bands $i=3,4,5$, using conventional unit cell. The magnitudes of quantized Berry flux through $3$-fold planes are found from in-plane Wilson loop calculations. The $\mathbb{Z}_2$ trivial (non-trivial) planes are colored light-blue (light yellow).}
    \label{fig:BiTunneling}
\end{figure}

The in-plane Wilson loop calculations for different bands also support classification based on $\kappa_{AF}$. Since the staggered index of bands $1$ and $2$ cancel each other, we only show the results for occupied bands $3$, $4$, and $5$ in Fig.~\ref{fig:BiTunneling}. Therefore, the ground state can carry net winding number $\pm 2$ or $\pm 4$. This uncertainty can be resolved by implementing detailed gauge fixing process for Berry connection. As Bi is ultimately a semimetal, we do not pursue such numerically expensive analysis for \emph{ab initio} band structure. However, in Appendix~\ref{schindler}, we address signed winding number of an 8-band tight-binding model of Bi~\cite{schindler2018higher}, which can support ground state winding number $4$. 

\section{Conclusions}\label{summ}

\par 
Our analysis for cubic model demonstrates the power of staggered index and Wilson loop for identifying signed 3D winding number for constituent Kramers degenerate bands and ground state. Similar analysis of tunneling of mirror Chern number can be carried out for tetragonal systems with space groups 83-88 ($C_{4h}$ instantons) and 123-142 ($D_{4h}$ instantons); hexagonal systems with space groups 174 ($C_{3h}$ instantons), 175-176 ($C_{6h}$ instantons), 187-190 ($D_{3h}$ instantons), 191-194 ($D_{6h}$ instantons). 
When high-symmetry planes lack mirror symmetry, the gauge-invariant magnitudes of relative Chern number and 3D winding number can be determined from Wilson loops, without detailed knowledge of underlying basis. Therefore, a combined analysis of staggered index, in-plane Wilson loop, and straight Wilson lines are sufficient to perform $\mathbb{N}$ classification of 3D winding numbers for all $\mathcal{P}$ and $\mathcal{T}$ symmetric systems. While the application of staggered index requires $\mathcal{P}$ symmetry, Wilson loop calculations can be applied for addressing topology of $\mathcal{PT}$-preserving, magneto-electric systems.  
\par 
Our work shows that states with $\kappa_{AF, GS} \neq 0$ can support $\sum_{j=1}^{m} n_{3,j} \neq 0$. Thus, we expect $\mathbb{Z}_2$-trivial, topological crystalline insulators with $\kappa_{AF, GS} \neq 0$ to possess quantized, magneto-electric coefficient $\theta = 2 s \pi$, with $s \in \mathbb{Z}$. The tight-binding model as well as \emph{ab initio} band structure of bismuth support such conclusions. Can quantized topological response of such states be detected? To affirmatively answer this question, in Part II, we will probe topological response with magnetic monopole and vortices. 
 
 \appendix 

\section{8-band model of bismuth}  \label{schindler}
\begin{figure}[t]
    \centering
        \subfigure[]{  \includegraphics[scale=0.25]{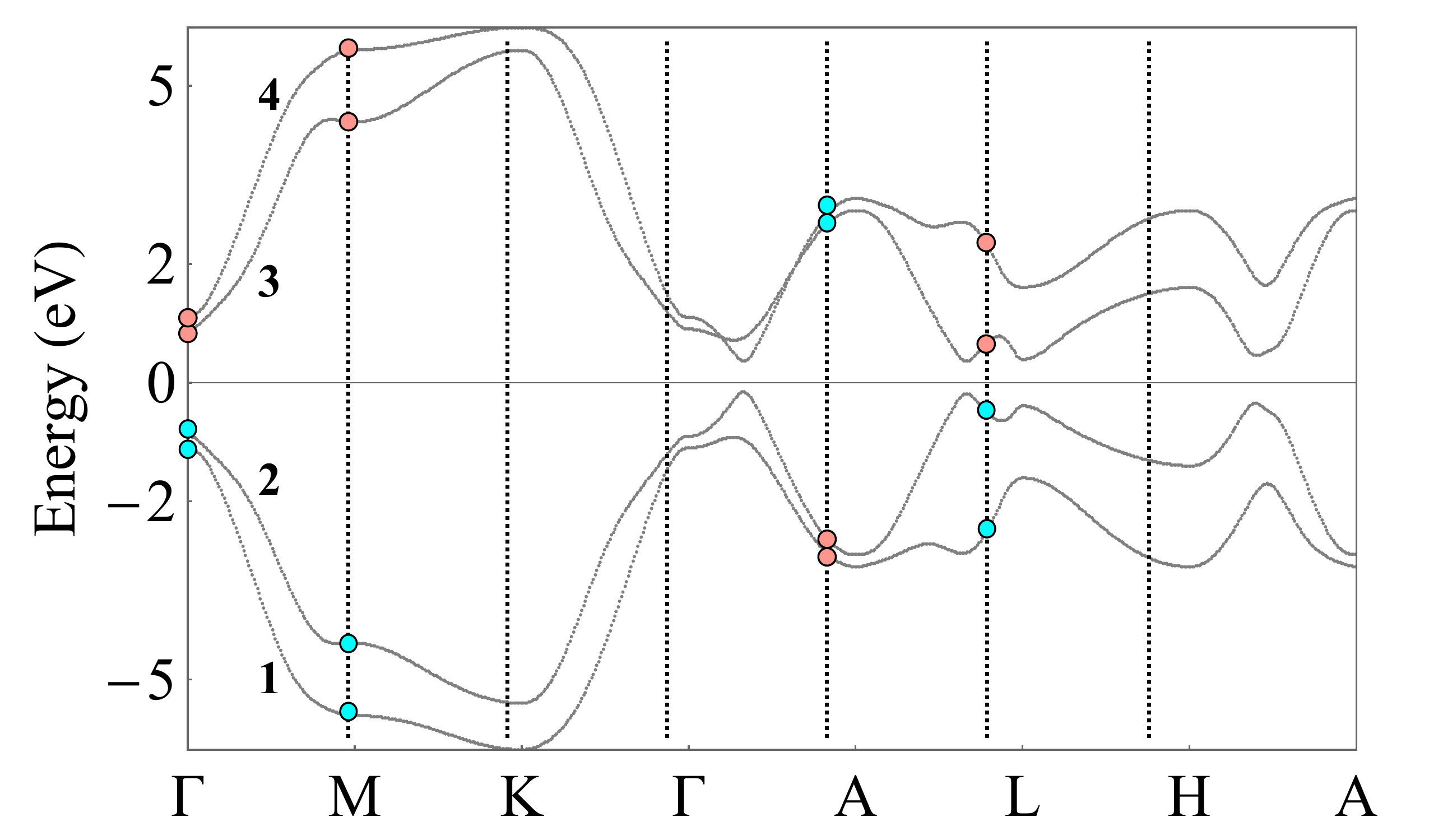}  
        \label{fig:SchBands}}
        \subfigure[]{
                   \includegraphics[scale=0.45]{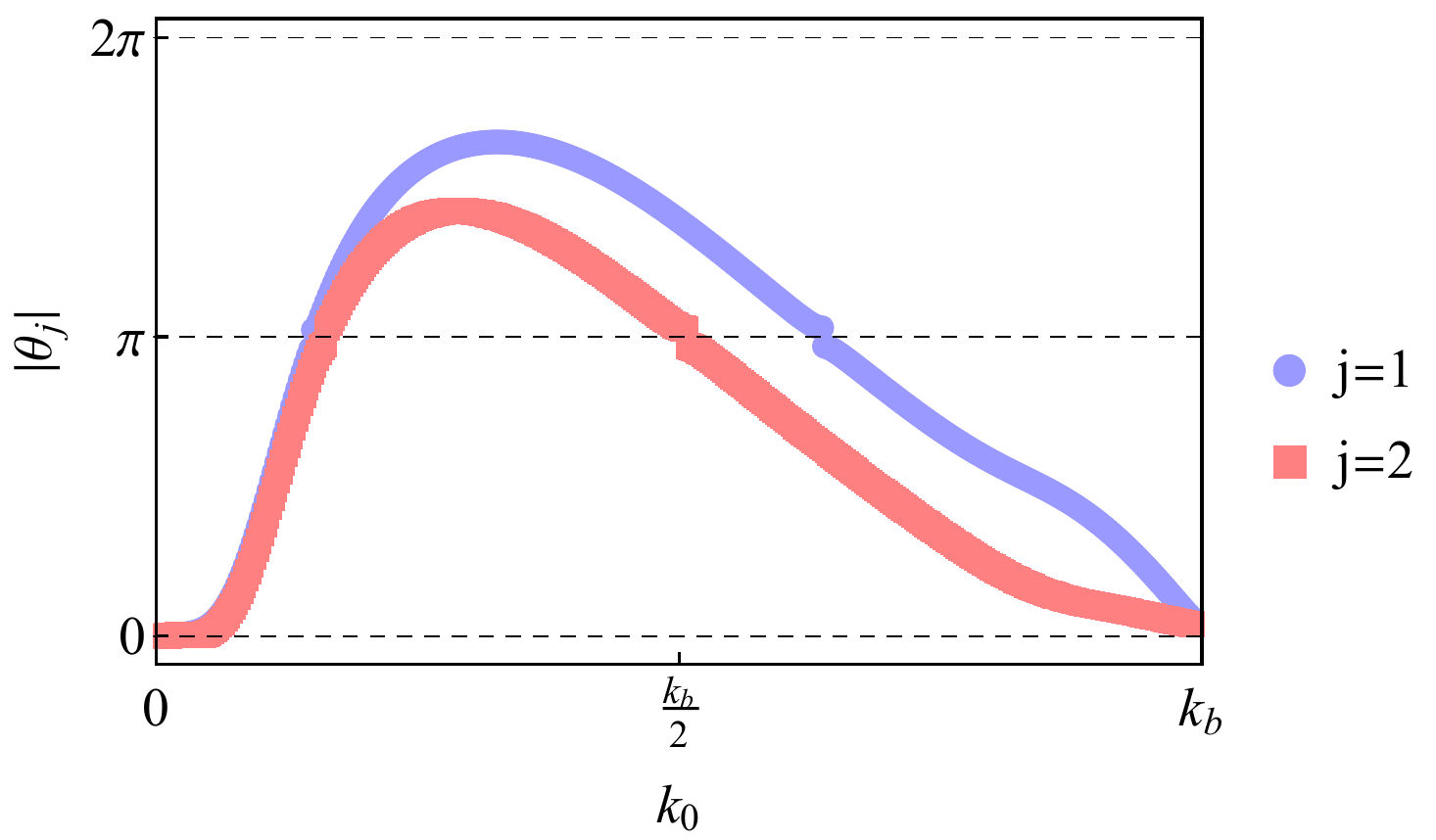} 
    \label{fig:PWLGamma}}
    \subfigure[]{
        \includegraphics[scale=0.45]{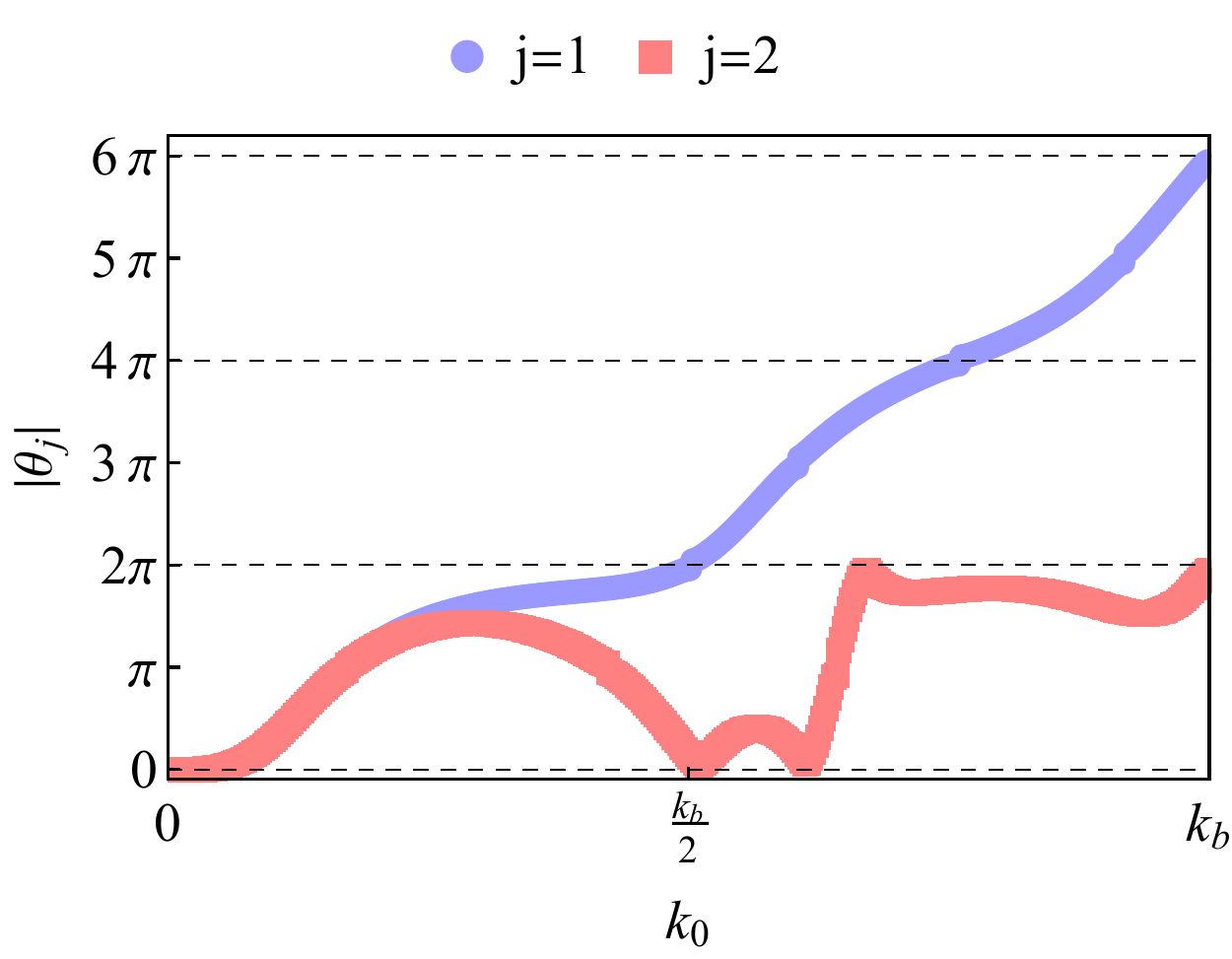} 
     \label{fig:PWLA}}
  \caption{(a) Band structure of 8-band tight-binding model, using hexagonal Brillouin zone for conventional unit cell. Parity eigenvalues +1 (-1) are labeled with red (cyan) dots. Bands are labeled by their energy at the $\Gamma$ point, i.e. $E_{1}(0)<E_{2}(0)<E_{3}(0)<E_{4}(0)$. Results for $|\theta_j|$ (b) $k_z=0$ and (c) $k_z=\pi$ planes, respectively.}
      \label{fig:SchBandtopology}
\end{figure}

The model is written using conventional unit cell, with BZ shown in. The Bloch Hamiltonian has the form
\begin{equation}\label{eq: TB1}
    H(\mathbf{k})=\begin{bmatrix}
    H_{TB,I}(\mathbf{k})+\epsilon & \delta M_{TB}(\mathbf{k})\\
    \delta M_{TB}(\mathbf{k})^{\dagger} & H_{TB,II}(\mathbf{k})-\epsilon
    \end{bmatrix},
\end{equation} 
where $H_{TB, I/II}(\mathbf{k})$ describe two 4-band, strong $\mathbb{Z}_2$ topological insulators, coupled by the hybridization matrix $M_{TB}(\mathbf{k})$. The $\mathcal{P}$ and $C_{3z}$ for $H(\mathbf{k})$ follow as $\mathcal{P}=\mathcal{P}_{I}\oplus \mathcal{P}_{II}$ and $C_{3z}=C_{3z,I}\oplus C_{3z,II}$. For details of model parameters and explicit representations of symmetry operators, please consult the supplementary information of Ref.~\onlinecite{schindler2018higher}. Along the $3$-fold axis all elements of $M_{TB}$ vanish as a consequence of $C_{3z}$ symmetry. The band structure and SIs are respectively shown in Fig.~\ref{fig:SchBands}.

We will perform direct analysis of tunneling configurations of $SU(2) \times SU(2)$ Berry connection for occupied valence bands. 
From numerical results shown in Fig.~\ref{fig:PWLGamma}, we find
\begin{eqnarray}
&& (|\mathfrak{C}_{3,XY; 1}|, |\mathfrak{C}_{3,XY; 2}| )(k_z=0)=(0,0), \\
&& (|\mathfrak{C}_{3,XY; 1}|, |\mathfrak{C}_{3,XY; 2}| )(k_z=\pi)=(3, 1).
\end{eqnarray} 
 Thus. occupied bands $1$ and $2$ possess tunneling of Berry flux and the magnitudes of 3D winding numbers are
\begin{equation}
(|n_{3,1}|, |n_{3,2}|)=(3,1). 
\end{equation}
Hence, the ground state can exhibit net even integer winding number $\pm 4, \pm 2$.

To resolve uncertainties, we have performed explicit Abelian gauge-fixing in the following manner. We first regulate the Bloch Hamiltonian as
\begin{equation}
H(\mathbf{k})\rightarrow H(\mathbf{k})+\alpha \Gamma_{NA} \end{equation}
where the traceless diagonal matrix
\begin{equation}
\Gamma_{NA}=\sigma_{3}\otimes \sigma_{3} \otimes \sigma_3,
\end{equation}
 is a generator of Cartan sub-algebra for the coset space, which commutes with $C_{3,z}$. This separates Kramers-pairs by $|2\alpha|$ at the BZ center. Appealing to Abelian Stokes theorem, the signed Berry flux for non-degenerate bands can be calculated with Abelian in-plane Wilson loops or TKNNY formula for Chern number~\cite{tknn1982} 
 \begin{equation}
    \mathfrak{C}_{3,XY; j}=  \lim_{\alpha \to 0} \; \frac{1}{2\pi} \; \int_{T^2} d^2k F_j(\bs{k}) \end{equation}
 where \begin{equation} F_j(\bs{k})=\sum_{j \neq l} 2 \text{Im}\frac{\bra{\psi_{l,\mathbf{k}}}\partial_x \hat{H} \ket{\psi_{j,\mathbf{k}}}\bra{\psi_{j,\mathbf{k}}}\partial_y \hat{H}\ket{\psi_{l,\mathbf{k}}}}{(E_{j,\mathbf{k}}-E_{l,\mathbf{k}})^2}. \nn \\
\end{equation}
By implementing this calculation, we find signed winding numbers
\begin{eqnarray}
&& (\mathfrak{C}_{3,XY; 1}, \mathfrak{C}_{3,XY; 2} )(k_z=0) =(0, 0),  \\
&& (\mathfrak{C}_{3,XY; 1}, \mathfrak{C}_{3,XY; 2} )(k_z=\pi) =(+3, +1), \\
&& (n_{3,1}, n_{3,2})= (+3, +1).\label{signed}
\end{eqnarray} 
Therefore, the ground state of the $8$-band model carries net 3D winding number $+4$. 

If the hybridization matrix is switched off, the signed Berry flux and $\mathcal{N}_3$ for 4-band models $ H_{TB,I}(\mathbf{k})$ and $H_{TB,II}(\mathbf{k})$ can be calculated following Secs.~\ref{Cubic} and \ref{Kuramoto}. The decoupled model also leads to Eq.~\ref{signed} for constituent occupied bands. This demonstrates the stability of third homotopy classification determined from the tunneling configurations of non-Abelian Berry flux.

\acknowledgements{
This work was supported by the National Science Foundation MRSEC program (DMR-1720139) at the Materials Research Center of Northwestern University, and the start up funds of P. G. provided by the Northwestern University. A part of this work was performed at the Aspen Center for Physics, which is supported by National Science Foundation grant PHY-1607611.}

\bibliographystyle{apsrev4-1}
\nocite{apsrev41Control}
\bibliography{ref.bib}

\begin{thebibliography}{72}%
\makeatletter
\providecommand \@ifxundefined [1]{%
 \@ifx{#1\undefined}
}%
\providecommand \@ifnum [1]{%
 \ifnum #1\expandafter \@firstoftwo
 \else \expandafter \@secondoftwo
 \fi
}%
\providecommand \@ifx [1]{%
 \ifx #1\expandafter \@firstoftwo
 \else \expandafter \@secondoftwo
 \fi
}%
\providecommand \natexlab [1]{#1}%
\providecommand \enquote  [1]{``#1''}%
\providecommand \bibnamefont  [1]{#1}%
\providecommand \bibfnamefont [1]{#1}%
\providecommand \citenamefont [1]{#1}%
\providecommand \href@noop [0]{\@secondoftwo}%
\providecommand \href [0]{\begingroup \@sanitize@url \@href}%
\providecommand \@href[1]{\@@startlink{#1}\@@href}%
\providecommand \@@href[1]{\endgroup#1\@@endlink}%
\providecommand \@sanitize@url [0]{\catcode `\\12\catcode `\$12\catcode
  `\&12\catcode `\#12\catcode `\^12\catcode `\_12\catcode `\%12\relax}%
\providecommand \@@startlink[1]{}%
\providecommand \@@endlink[0]{}%
\providecommand \url  [0]{\begingroup\@sanitize@url \@url }%
\providecommand \@url [1]{\endgroup\@href {#1}{\urlprefix }}%
\providecommand \urlprefix  [0]{URL }%
\providecommand \Eprint [0]{\href }%
\providecommand \doibase [0]{http://dx.doi.org/}%
\providecommand \selectlanguage [0]{\@gobble}%
\providecommand \bibinfo  [0]{\@secondoftwo}%
\providecommand \bibfield  [0]{\@secondoftwo}%
\providecommand \translation [1]{[#1]}%
\providecommand \BibitemOpen [0]{}%
\providecommand \bibitemStop [0]{}%
\providecommand \bibitemNoStop [0]{.\EOS\space}%
\providecommand \EOS [0]{\spacefactor3000\relax}%
\providecommand \BibitemShut  [1]{\csname bibitem#1\endcsname}%
\let\auto@bib@innerbib\@empty
\bibitem [{\citenamefont {Kane}\ and\ \citenamefont {Mele}(2005)}]{Kane2005}%
  \BibitemOpen
  \bibfield  {author} {\bibinfo {author} {\bibfnamefont {C.~L.}\ \bibnamefont
  {Kane}}\ and\ \bibinfo {author} {\bibfnamefont {E.~J.}\ \bibnamefont
  {Mele}},\ }\bibfield  {title} {\enquote {\bibinfo {title} {$\mathbb{Z}_{2}$
  topological order and the quantum spin {H}all effect},}\ }\href {\doibase
  10.1103/PhysRevLett.95.146802} {\bibfield  {journal} {\bibinfo  {journal}
  {Phys. Rev. Lett.}\ }\textbf {\bibinfo {volume} {95}},\ \bibinfo {pages}
  {146802} (\bibinfo {year} {2005})}\BibitemShut {NoStop}%
\bibitem [{\citenamefont {Bernevig}\ \emph {et~al.}(2006)\citenamefont
  {Bernevig}, \citenamefont {Hughes},\ and\ \citenamefont
  {Zhang}}]{bernevig2006quantum}%
  \BibitemOpen
  \bibfield  {author} {\bibinfo {author} {\bibfnamefont {B.~A.}\ \bibnamefont
  {Bernevig}}, \bibinfo {author} {\bibfnamefont {T.~L.}\ \bibnamefont
  {Hughes}}, \ and\ \bibinfo {author} {\bibfnamefont {S.-C.}\ \bibnamefont
  {Zhang}},\ }\bibfield  {title} {\enquote {\bibinfo {title} {Quantum spin
  {H}all effect and topological phase transition in {H}g{T}e quantum wells},}\
  }\href {\doibase 10.1126/science.1133734} {\bibfield  {journal} {\bibinfo
  {journal} {Science}\ }\textbf {\bibinfo {volume} {314}},\ \bibinfo {pages}
  {1757--1761} (\bibinfo {year} {2006})}\BibitemShut {NoStop}%
\bibitem [{\citenamefont {Fu}\ \emph {et~al.}(2007)\citenamefont {Fu},
  \citenamefont {Kane},\ and\ \citenamefont {Mele}}]{FuKaneMele2007}%
  \BibitemOpen
  \bibfield  {author} {\bibinfo {author} {\bibfnamefont {L.}~\bibnamefont
  {Fu}}, \bibinfo {author} {\bibfnamefont {C.~L.}\ \bibnamefont {Kane}}, \ and\
  \bibinfo {author} {\bibfnamefont {E.~J.}\ \bibnamefont {Mele}},\ }\bibfield
  {title} {\enquote {\bibinfo {title} {Topological insulators in three
  dimensions},}\ }\href {\doibase 10.1103/PhysRevLett.98.106803} {\bibfield
  {journal} {\bibinfo  {journal} {Phys. Rev. Lett.}\ }\textbf {\bibinfo
  {volume} {98}},\ \bibinfo {pages} {106803} (\bibinfo {year}
  {2007})}\BibitemShut {NoStop}%
\bibitem [{\citenamefont {Fu}\ and\ \citenamefont {Kane}(2007)}]{FuKane}%
  \BibitemOpen
  \bibfield  {author} {\bibinfo {author} {\bibfnamefont {Liang}\ \bibnamefont
  {Fu}}\ and\ \bibinfo {author} {\bibfnamefont {C.~L.}\ \bibnamefont {Kane}},\
  }\bibfield  {title} {\enquote {\bibinfo {title} {Topological insulators with
  inversion symmetry},}\ }\href {\doibase 10.1103/PhysRevB.76.045302}
  {\bibfield  {journal} {\bibinfo  {journal} {Phys. Rev. B}\ }\textbf {\bibinfo
  {volume} {76}},\ \bibinfo {pages} {045302} (\bibinfo {year}
  {2007})}\BibitemShut {NoStop}%
\bibitem [{\citenamefont {Moore}\ and\ \citenamefont
  {Balents}(2007)}]{Moore2007}%
  \BibitemOpen
  \bibfield  {author} {\bibinfo {author} {\bibfnamefont {J.~E.}\ \bibnamefont
  {Moore}}\ and\ \bibinfo {author} {\bibfnamefont {L.}~\bibnamefont
  {Balents}},\ }\bibfield  {title} {\enquote {\bibinfo {title} {Topological
  invariants of time-reversal-invariant band structures},}\ }\href {\doibase
  10.1103/PhysRevB.75.121306} {\bibfield  {journal} {\bibinfo  {journal} {Phys.
  Rev. B}\ }\textbf {\bibinfo {volume} {75}},\ \bibinfo {pages} {121306}
  (\bibinfo {year} {2007})}\BibitemShut {NoStop}%
\bibitem [{\citenamefont {Qi}\ \emph {et~al.}(2008)\citenamefont {Qi},
  \citenamefont {Hughes},\ and\ \citenamefont {Zhang}}]{ZhangFT}%
  \BibitemOpen
  \bibfield  {author} {\bibinfo {author} {\bibfnamefont {X.-L.}\ \bibnamefont
  {Qi}}, \bibinfo {author} {\bibfnamefont {T.~L.}\ \bibnamefont {Hughes}}, \
  and\ \bibinfo {author} {\bibfnamefont {S.-C.}\ \bibnamefont {Zhang}},\
  }\bibfield  {title} {\enquote {\bibinfo {title} {Topological field theory of
  time-reversal invariant insulators},}\ }\href {\doibase
  10.1103/PhysRevB.78.195424} {\bibfield  {journal} {\bibinfo  {journal} {Phys.
  Rev. B}\ }\textbf {\bibinfo {volume} {78}},\ \bibinfo {pages} {195424}
  (\bibinfo {year} {2008})}\BibitemShut {NoStop}%
\bibitem [{\citenamefont {Schnyder}\ \emph {et~al.}(2008)\citenamefont
  {Schnyder}, \citenamefont {Ryu}, \citenamefont {Furusaki},\ and\
  \citenamefont {Ludwig}}]{RyuLudwigPRB}%
  \BibitemOpen
  \bibfield  {author} {\bibinfo {author} {\bibfnamefont {A.~P.}\ \bibnamefont
  {Schnyder}}, \bibinfo {author} {\bibfnamefont {S.}~\bibnamefont {Ryu}},
  \bibinfo {author} {\bibfnamefont {A.}~\bibnamefont {Furusaki}}, \ and\
  \bibinfo {author} {\bibfnamefont {A.~W.~W.}\ \bibnamefont {Ludwig}},\
  }\bibfield  {title} {\enquote {\bibinfo {title} {Classification of
  topological insulators and superconductors in three spatial dimensions},}\
  }\href {\doibase 10.1103/PhysRevB.78.195125} {\bibfield  {journal} {\bibinfo
  {journal} {Phys. Rev. B}\ }\textbf {\bibinfo {volume} {78}},\ \bibinfo
  {pages} {195125} (\bibinfo {year} {2008})}\BibitemShut {NoStop}%
\bibitem [{\citenamefont {Roy}(2009{\natexlab{a}})}]{Roy2009}%
  \BibitemOpen
  \bibfield  {author} {\bibinfo {author} {\bibfnamefont {R.}~\bibnamefont
  {Roy}},\ }\bibfield  {title} {\enquote {\bibinfo {title} {$\mathbb{Z}_{2}$
  classification of quantum spin {H}all systems: An approach using
  time-reversal invariance},}\ }\href {\doibase 10.1103/PhysRevB.79.195321}
  {\bibfield  {journal} {\bibinfo  {journal} {Phys. Rev. B}\ }\textbf {\bibinfo
  {volume} {79}},\ \bibinfo {pages} {195321} (\bibinfo {year}
  {2009}{\natexlab{a}})}\BibitemShut {NoStop}%
\bibitem [{\citenamefont {Roy}(2009{\natexlab{b}})}]{Roy20093D}%
  \BibitemOpen
  \bibfield  {author} {\bibinfo {author} {\bibfnamefont {R.}~\bibnamefont
  {Roy}},\ }\bibfield  {title} {\enquote {\bibinfo {title} {Topological phases
  and the quantum spin {H}all effect in three dimensions},}\ }\href {\doibase
  10.1103/PhysRevB.79.195322} {\bibfield  {journal} {\bibinfo  {journal} {Phys.
  Rev. B}\ }\textbf {\bibinfo {volume} {79}},\ \bibinfo {pages} {195322}
  (\bibinfo {year} {2009}{\natexlab{b}})}\BibitemShut {NoStop}%
\bibitem [{\citenamefont {Ryu}\ \emph {et~al.}(2010)\citenamefont {Ryu},
  \citenamefont {Schnyder}, \citenamefont {Furusaki},\ and\ \citenamefont
  {Ludwig}}]{ryu2010topological}%
  \BibitemOpen
  \bibfield  {author} {\bibinfo {author} {\bibfnamefont {S.}~\bibnamefont
  {Ryu}}, \bibinfo {author} {\bibfnamefont {A.~P.}\ \bibnamefont {Schnyder}},
  \bibinfo {author} {\bibfnamefont {A.}~\bibnamefont {Furusaki}}, \ and\
  \bibinfo {author} {\bibfnamefont {A.~W.~W.}\ \bibnamefont {Ludwig}},\
  }\bibfield  {title} {\enquote {\bibinfo {title} {Topological insulators and
  superconductors: tenfold way and dimensional hierarchy},}\ }\href {\doibase
  10.1088/1367-2630/12/6/065010} {\bibfield  {journal} {\bibinfo  {journal}
  {New J. Phys.}\ }\textbf {\bibinfo {volume} {12}},\ \bibinfo {pages} {065010}
  (\bibinfo {year} {2010})}\BibitemShut {NoStop}%
\bibitem [{\citenamefont {Hasan}\ and\ \citenamefont
  {Kane}(2010)}]{Hassan2010}%
  \BibitemOpen
  \bibfield  {author} {\bibinfo {author} {\bibfnamefont {M.~Z.}\ \bibnamefont
  {Hasan}}\ and\ \bibinfo {author} {\bibfnamefont {C.~L.}\ \bibnamefont
  {Kane}},\ }\bibfield  {title} {\enquote {\bibinfo {title} {Colloquium:
  Topological insulators},}\ }\href {\doibase 10.1103/RevModPhys.82.3045}
  {\bibfield  {journal} {\bibinfo  {journal} {Rev. Mod. Phys.}\ }\textbf
  {\bibinfo {volume} {82}},\ \bibinfo {pages} {3045--3067} (\bibinfo {year}
  {2010})}\BibitemShut {NoStop}%
\bibitem [{\citenamefont {Qi}\ and\ \citenamefont {Zhang}(2011)}]{Qi2011}%
  \BibitemOpen
  \bibfield  {author} {\bibinfo {author} {\bibfnamefont {X.-L.}\ \bibnamefont
  {Qi}}\ and\ \bibinfo {author} {\bibfnamefont {S.-C.}\ \bibnamefont {Zhang}},\
  }\bibfield  {title} {\enquote {\bibinfo {title} {Topological insulators and
  superconductors},}\ }\href {\doibase 10.1103/RevModPhys.83.1057} {\bibfield
  {journal} {\bibinfo  {journal} {Rev. Mod. Phys.}\ }\textbf {\bibinfo {volume}
  {83}},\ \bibinfo {pages} {1057--1110} (\bibinfo {year} {2011})}\BibitemShut
  {NoStop}%
\bibitem [{\citenamefont {Slager}\ \emph {et~al.}(2013)\citenamefont {Slager},
  \citenamefont {Mesaros}, \citenamefont {Juri{\v{c}}i{\'c}},\ and\
  \citenamefont {Zaanen}}]{slager2013space}%
  \BibitemOpen
  \bibfield  {author} {\bibinfo {author} {\bibfnamefont {R.-J.}\ \bibnamefont
  {Slager}}, \bibinfo {author} {\bibfnamefont {A.}~\bibnamefont {Mesaros}},
  \bibinfo {author} {\bibfnamefont {V.}~\bibnamefont {Juri{\v{c}}i{\'c}}}, \
  and\ \bibinfo {author} {\bibfnamefont {J.}~\bibnamefont {Zaanen}},\
  }\bibfield  {title} {\enquote {\bibinfo {title} {The space group
  classification of topological band-insulators},}\ }\href {\doibase
  10.1038/nphys2513} {\bibfield  {journal} {\bibinfo  {journal} {Nat. Phys.}\
  }\textbf {\bibinfo {volume} {9}},\ \bibinfo {pages} {98--102} (\bibinfo
  {year} {2013})}\BibitemShut {NoStop}%
\bibitem [{\citenamefont {Chiu}\ \emph {et~al.}(2016)\citenamefont {Chiu},
  \citenamefont {Teo}, \citenamefont {Schnyder},\ and\ \citenamefont
  {Ryu}}]{Chiu2016}%
  \BibitemOpen
  \bibfield  {author} {\bibinfo {author} {\bibfnamefont {C.-K.}\ \bibnamefont
  {Chiu}}, \bibinfo {author} {\bibfnamefont {J.~C.~Y.}\ \bibnamefont {Teo}},
  \bibinfo {author} {\bibfnamefont {A.~P.}\ \bibnamefont {Schnyder}}, \ and\
  \bibinfo {author} {\bibfnamefont {S.}~\bibnamefont {Ryu}},\ }\bibfield
  {title} {\enquote {\bibinfo {title} {Classification of topological quantum
  matter with symmetries},}\ }\href {\doibase 10.1103/RevModPhys.88.035005}
  {\bibfield  {journal} {\bibinfo  {journal} {Rev. Mod. Phys.}\ }\textbf
  {\bibinfo {volume} {88}},\ \bibinfo {pages} {035005} (\bibinfo {year}
  {2016})}\BibitemShut {NoStop}%
\bibitem [{\citenamefont {Essin}\ \emph {et~al.}(2009)\citenamefont {Essin},
  \citenamefont {Moore},\ and\ \citenamefont
  {Vanderbilt}}]{EssinMagnetoelectric}%
  \BibitemOpen
  \bibfield  {author} {\bibinfo {author} {\bibfnamefont {A.~M.}\ \bibnamefont
  {Essin}}, \bibinfo {author} {\bibfnamefont {J.~E.}\ \bibnamefont {Moore}}, \
  and\ \bibinfo {author} {\bibfnamefont {D.}~\bibnamefont {Vanderbilt}},\
  }\bibfield  {title} {\enquote {\bibinfo {title} {Magnetoelectric
  polarizability and axion electrodynamics in crystalline insulators},}\ }\href
  {\doibase 10.1103/PhysRevLett.102.146805} {\bibfield  {journal} {\bibinfo
  {journal} {Phys. Rev. Lett.}\ }\textbf {\bibinfo {volume} {102}},\ \bibinfo
  {pages} {146805} (\bibinfo {year} {2009})}\BibitemShut {NoStop}%
\bibitem [{\citenamefont {Essin}\ \emph {et~al.}(2010)\citenamefont {Essin},
  \citenamefont {Turner}, \citenamefont {Moore},\ and\ \citenamefont
  {Vanderbilt}}]{Essin2010}%
  \BibitemOpen
  \bibfield  {author} {\bibinfo {author} {\bibfnamefont {A.~M.}\ \bibnamefont
  {Essin}}, \bibinfo {author} {\bibfnamefont {A.~M.}\ \bibnamefont {Turner}},
  \bibinfo {author} {\bibfnamefont {J.~E.}\ \bibnamefont {Moore}}, \ and\
  \bibinfo {author} {\bibfnamefont {D.}~\bibnamefont {Vanderbilt}},\ }\bibfield
   {title} {\enquote {\bibinfo {title} {Orbital magnetoelectric coupling in
  band insulators},}\ }\href {\doibase 10.1103/PhysRevB.81.205104} {\bibfield
  {journal} {\bibinfo  {journal} {Phys. Rev. B}\ }\textbf {\bibinfo {volume}
  {81}},\ \bibinfo {pages} {205104} (\bibinfo {year} {2010})}\BibitemShut
  {NoStop}%
\bibitem [{\citenamefont {Malashevich}\ \emph {et~al.}(2010)\citenamefont
  {Malashevich}, \citenamefont {Souza}, \citenamefont {Coh},\ and\
  \citenamefont {Vanderbilt}}]{malashevich2010theory}%
  \BibitemOpen
  \bibfield  {author} {\bibinfo {author} {\bibfnamefont {A.}~\bibnamefont
  {Malashevich}}, \bibinfo {author} {\bibfnamefont {I.}~\bibnamefont {Souza}},
  \bibinfo {author} {\bibfnamefont {S.}~\bibnamefont {Coh}}, \ and\ \bibinfo
  {author} {\bibfnamefont {D.}~\bibnamefont {Vanderbilt}},\ }\bibfield  {title}
  {\enquote {\bibinfo {title} {Theory of orbital magnetoelectric response},}\
  }\href {\doibase 10.1088/1367-2630/12/5/053032} {\bibfield  {journal}
  {\bibinfo  {journal} {New J. Phys.}\ }\textbf {\bibinfo {volume} {12}},\
  \bibinfo {pages} {053032} (\bibinfo {year} {2010})}\BibitemShut {NoStop}%
\bibitem [{\citenamefont {Coh}\ \emph {et~al.}(2011)\citenamefont {Coh},
  \citenamefont {Vanderbilt}, \citenamefont {Malashevich},\ and\ \citenamefont
  {Souza}}]{Coh2011}%
  \BibitemOpen
  \bibfield  {author} {\bibinfo {author} {\bibfnamefont {S.}~\bibnamefont
  {Coh}}, \bibinfo {author} {\bibfnamefont {D.}~\bibnamefont {Vanderbilt}},
  \bibinfo {author} {\bibfnamefont {A.}~\bibnamefont {Malashevich}}, \ and\
  \bibinfo {author} {\bibfnamefont {I.}~\bibnamefont {Souza}},\ }\bibfield
  {title} {\enquote {\bibinfo {title} {Chern-simons orbital magnetoelectric
  coupling in generic insulators},}\ }\href {\doibase
  10.1103/PhysRevB.83.085108} {\bibfield  {journal} {\bibinfo  {journal} {Phys.
  Rev. B}\ }\textbf {\bibinfo {volume} {83}},\ \bibinfo {pages} {085108}
  (\bibinfo {year} {2011})}\BibitemShut {NoStop}%
\bibitem [{\citenamefont {Varnava}\ \emph {et~al.}(2020)\citenamefont
  {Varnava}, \citenamefont {Souza},\ and\ \citenamefont
  {Vanderbilt}}]{varnava2020}%
  \BibitemOpen
  \bibfield  {author} {\bibinfo {author} {\bibfnamefont {N.}~\bibnamefont
  {Varnava}}, \bibinfo {author} {\bibfnamefont {I.}~\bibnamefont {Souza}}, \
  and\ \bibinfo {author} {\bibfnamefont {D.}~\bibnamefont {Vanderbilt}},\
  }\bibfield  {title} {\enquote {\bibinfo {title} {Axion coupling in the hybrid
  wannier representation},}\ }\href {\doibase 10.1103/PhysRevB.101.155130}
  {\bibfield  {journal} {\bibinfo  {journal} {Phys. Rev. B}\ }\textbf {\bibinfo
  {volume} {101}},\ \bibinfo {pages} {155130} (\bibinfo {year}
  {2020})}\BibitemShut {NoStop}%
\bibitem [{\citenamefont {Kruthoff}\ \emph {et~al.}(2017)\citenamefont
  {Kruthoff}, \citenamefont {de~Boer}, \citenamefont {van Wezel}, \citenamefont
  {Kane},\ and\ \citenamefont {Slager}}]{Kruthoff2017}%
  \BibitemOpen
  \bibfield  {author} {\bibinfo {author} {\bibfnamefont {J.}~\bibnamefont
  {Kruthoff}}, \bibinfo {author} {\bibfnamefont {J.}~\bibnamefont {de~Boer}},
  \bibinfo {author} {\bibfnamefont {J.}~\bibnamefont {van Wezel}}, \bibinfo
  {author} {\bibfnamefont {C.~L.}\ \bibnamefont {Kane}}, \ and\ \bibinfo
  {author} {\bibfnamefont {R.-J.}\ \bibnamefont {Slager}},\ }\bibfield  {title}
  {\enquote {\bibinfo {title} {Topological classification of crystalline
  insulators through band structure combinatorics},}\ }\href {\doibase
  10.1103/PhysRevX.7.041069} {\bibfield  {journal} {\bibinfo  {journal} {Phys.
  Rev. X}\ }\textbf {\bibinfo {volume} {7}},\ \bibinfo {pages} {041069}
  (\bibinfo {year} {2017})}\BibitemShut {NoStop}%
\bibitem [{\citenamefont {Bradlyn}\ \emph {et~al.}(2017)\citenamefont {Bradlyn}
  \emph {et~al.}}]{bradlyn2017topological}%
  \BibitemOpen
  \bibfield  {author} {\bibinfo {author} {\bibfnamefont {B.}~\bibnamefont
  {Bradlyn}} \emph {et~al.},\ }\bibfield  {title} {\enquote {\bibinfo {title}
  {Topological quantum chemistry},}\ }\href {\doibase 10.1038/nature23268}
  {\bibfield  {journal} {\bibinfo  {journal} {Nature}\ }\textbf {\bibinfo
  {volume} {547}},\ \bibinfo {pages} {298--305} (\bibinfo {year}
  {2017})}\BibitemShut {NoStop}%
\bibitem [{\citenamefont {Po}\ \emph {et~al.}(2017)\citenamefont {Po},
  \citenamefont {Vishwanath},\ and\ \citenamefont {Watanabe}}]{po2017symmetry}%
  \BibitemOpen
  \bibfield  {author} {\bibinfo {author} {\bibfnamefont {H.~C.}\ \bibnamefont
  {Po}}, \bibinfo {author} {\bibfnamefont {A.}~\bibnamefont {Vishwanath}}, \
  and\ \bibinfo {author} {\bibfnamefont {H.}~\bibnamefont {Watanabe}},\
  }\bibfield  {title} {\enquote {\bibinfo {title} {Symmetry-based indicators of
  band topology in the 230 space groups},}\ }\href {\doibase
  10.1038/s41467-017-00133-2} {\bibfield  {journal} {\bibinfo  {journal} {Nat.
  Comms.}\ }\textbf {\bibinfo {volume} {8}},\ \bibinfo {pages} {1--9} (\bibinfo
  {year} {2017})}\BibitemShut {NoStop}%
\bibitem [{\citenamefont {Khalaf}\ \emph {et~al.}(2018)\citenamefont {Khalaf},
  \citenamefont {Po}, \citenamefont {Vishwanath},\ and\ \citenamefont
  {Watanabe}}]{KhalafSymm}%
  \BibitemOpen
  \bibfield  {author} {\bibinfo {author} {\bibfnamefont {E.}~\bibnamefont
  {Khalaf}}, \bibinfo {author} {\bibfnamefont {H.~C.}\ \bibnamefont {Po}},
  \bibinfo {author} {\bibfnamefont {A.}~\bibnamefont {Vishwanath}}, \ and\
  \bibinfo {author} {\bibfnamefont {H.}~\bibnamefont {Watanabe}},\ }\bibfield
  {title} {\enquote {\bibinfo {title} {Symmetry indicators and anomalous
  surface states of topological crystalline insulators},}\ }\href {\doibase
  10.1103/PhysRevX.8.031070} {\bibfield  {journal} {\bibinfo  {journal} {Phys.
  Rev. X}\ }\textbf {\bibinfo {volume} {8}},\ \bibinfo {pages} {031070}
  (\bibinfo {year} {2018})}\BibitemShut {NoStop}%
\bibitem [{\citenamefont {Cano}\ \emph {et~al.}(2018)\citenamefont {Cano},
  \citenamefont {Bradlyn}, \citenamefont {Wang}, \citenamefont {Elcoro},
  \citenamefont {Vergniory}, \citenamefont {Felser}, \citenamefont {Aroyo},\
  and\ \citenamefont {Bernevig}}]{CanoBuildingBlocks2018}%
  \BibitemOpen
  \bibfield  {author} {\bibinfo {author} {\bibfnamefont {J.}~\bibnamefont
  {Cano}}, \bibinfo {author} {\bibfnamefont {B.}~\bibnamefont {Bradlyn}},
  \bibinfo {author} {\bibfnamefont {Z.}~\bibnamefont {Wang}}, \bibinfo {author}
  {\bibfnamefont {L.}~\bibnamefont {Elcoro}}, \bibinfo {author} {\bibfnamefont
  {M.~G.}\ \bibnamefont {Vergniory}}, \bibinfo {author} {\bibfnamefont
  {C.}~\bibnamefont {Felser}}, \bibinfo {author} {\bibfnamefont {M.~I.}\
  \bibnamefont {Aroyo}}, \ and\ \bibinfo {author} {\bibfnamefont {B.~A.}\
  \bibnamefont {Bernevig}},\ }\bibfield  {title} {\enquote {\bibinfo {title}
  {Building blocks of topological quantum chemistry: Elementary band
  representations},}\ }\href {\doibase 10.1103/PhysRevB.97.035139} {\bibfield
  {journal} {\bibinfo  {journal} {Phys. Rev. B}\ }\textbf {\bibinfo {volume}
  {97}},\ \bibinfo {pages} {035139} (\bibinfo {year} {2018})}\BibitemShut
  {NoStop}%
\bibitem [{\citenamefont {Vergniory}\ \emph {et~al.}(2019)\citenamefont
  {Vergniory} \emph {et~al.}}]{vergniory2019complete}%
  \BibitemOpen
  \bibfield  {author} {\bibinfo {author} {\bibfnamefont {M.~G.}\ \bibnamefont
  {Vergniory}} \emph {et~al.},\ }\bibfield  {title} {\enquote {\bibinfo {title}
  {A complete catalogue of high-quality topological materials},}\ }\href
  {\doibase 10.1038/s41586-019-0954-4} {\bibfield  {journal} {\bibinfo
  {journal} {Nature}\ }\textbf {\bibinfo {volume} {566}},\ \bibinfo {pages}
  {480--485} (\bibinfo {year} {2019})}\BibitemShut {NoStop}%
\bibitem [{\citenamefont {Zhang}\ \emph {et~al.}(2019)\citenamefont {Zhang}
  \emph {et~al.}}]{zhang2019catalogue}%
  \BibitemOpen
  \bibfield  {author} {\bibinfo {author} {\bibfnamefont {T.}~\bibnamefont
  {Zhang}} \emph {et~al.},\ }\bibfield  {title} {\enquote {\bibinfo {title}
  {Catalogue of topological electronic materials},}\ }\href {\doibase
  10.1038/s41586-019-0944-6} {\bibfield  {journal} {\bibinfo  {journal}
  {Nature}\ }\textbf {\bibinfo {volume} {566}},\ \bibinfo {pages} {475--479}
  (\bibinfo {year} {2019})}\BibitemShut {NoStop}%
\bibitem [{\citenamefont {Tang}\ \emph
  {et~al.}(2019{\natexlab{a}})\citenamefont {Tang}, \citenamefont {Po},
  \citenamefont {Vishwanath},\ and\ \citenamefont {Wan}}]{tang2019efficient}%
  \BibitemOpen
  \bibfield  {author} {\bibinfo {author} {\bibfnamefont {F.}~\bibnamefont
  {Tang}}, \bibinfo {author} {\bibfnamefont {H.~C.}\ \bibnamefont {Po}},
  \bibinfo {author} {\bibfnamefont {A.}~\bibnamefont {Vishwanath}}, \ and\
  \bibinfo {author} {\bibfnamefont {X.~G.}\ \bibnamefont {Wan}},\ }\bibfield
  {title} {\enquote {\bibinfo {title} {Efficient topological materials
  discovery using symmetry indicators},}\ }\href {\doibase
  10.1038/s41586-019-0937-5} {\bibfield  {journal} {\bibinfo  {journal} {Nat.
  Phys.}\ }\textbf {\bibinfo {volume} {15}},\ \bibinfo {pages} {470--476}
  (\bibinfo {year} {2019}{\natexlab{a}})}\BibitemShut {NoStop}%
\bibitem [{\citenamefont {Tang}\ \emph
  {et~al.}(2019{\natexlab{b}})\citenamefont {Tang}, \citenamefont {Po},
  \citenamefont {Vishwanath},\ and\ \citenamefont
  {Wan}}]{tang2019comprehensive}%
  \BibitemOpen
  \bibfield  {author} {\bibinfo {author} {\bibfnamefont {F.}~\bibnamefont
  {Tang}}, \bibinfo {author} {\bibfnamefont {H.~C.}\ \bibnamefont {Po}},
  \bibinfo {author} {\bibfnamefont {A.}~\bibnamefont {Vishwanath}}, \ and\
  \bibinfo {author} {\bibfnamefont {X.~G.}\ \bibnamefont {Wan}},\ }\bibfield
  {title} {\enquote {\bibinfo {title} {Comprehensive search for topological
  materials using symmetry indicators},}\ }\href {\doibase
  10.1038/s41586-019-0937-5} {\bibfield  {journal} {\bibinfo  {journal}
  {Nature}\ }\textbf {\bibinfo {volume} {566}},\ \bibinfo {pages} {486--489}
  (\bibinfo {year} {2019}{\natexlab{b}})}\BibitemShut {NoStop}%
\bibitem [{\citenamefont {Vergniory}\ \emph {et~al.}(2021)\citenamefont
  {Vergniory}, \citenamefont {Wieder}, \citenamefont {Elcoro}, \citenamefont
  {Parkin}, \citenamefont {Felser}, \citenamefont {Bernevig},\ and\
  \citenamefont {Regnault}}]{vergniory2021all}%
  \BibitemOpen
  \bibfield  {author} {\bibinfo {author} {\bibfnamefont {M.~G.}\ \bibnamefont
  {Vergniory}}, \bibinfo {author} {\bibfnamefont {B.~J.}\ \bibnamefont
  {Wieder}}, \bibinfo {author} {\bibfnamefont {L.}~\bibnamefont {Elcoro}},
  \bibinfo {author} {\bibfnamefont {S.~S.~P.}\ \bibnamefont {Parkin}}, \bibinfo
  {author} {\bibfnamefont {C.}~\bibnamefont {Felser}}, \bibinfo {author}
  {\bibfnamefont {B.~A.}\ \bibnamefont {Bernevig}}, \ and\ \bibinfo {author}
  {\bibfnamefont {N.}~\bibnamefont {Regnault}},\ }\bibfield  {title} {\enquote
  {\bibinfo {title} {All topological bands of all stoichiometric materials},}\
  }\href@noop {} {\bibfield  {journal} {\bibinfo  {journal} {arXiv:2105.09954}\
  } (\bibinfo {year} {2021})}\BibitemShut {NoStop}%
\bibitem [{\citenamefont {Xu}\ \emph {et~al.}(2020)\citenamefont {Xu},
  \citenamefont {Elcoro}, \citenamefont {Song}, \citenamefont {Wieder},
  \citenamefont {Vergniory}, \citenamefont {Regnault}, \citenamefont {Chen},
  \citenamefont {Felser},\ and\ \citenamefont {Bernevig}}]{xu2020high}%
  \BibitemOpen
  \bibfield  {author} {\bibinfo {author} {\bibfnamefont {Y.~F.}\ \bibnamefont
  {Xu}}, \bibinfo {author} {\bibfnamefont {L.}~\bibnamefont {Elcoro}}, \bibinfo
  {author} {\bibfnamefont {Z.-D.}\ \bibnamefont {Song}}, \bibinfo {author}
  {\bibfnamefont {B.~J.}\ \bibnamefont {Wieder}}, \bibinfo {author}
  {\bibfnamefont {M.~G.}\ \bibnamefont {Vergniory}}, \bibinfo {author}
  {\bibfnamefont {N.}~\bibnamefont {Regnault}}, \bibinfo {author}
  {\bibfnamefont {Y.}~\bibnamefont {Chen}}, \bibinfo {author} {\bibfnamefont
  {C.}~\bibnamefont {Felser}}, \ and\ \bibinfo {author} {\bibfnamefont {B.~A.}\
  \bibnamefont {Bernevig}},\ }\bibfield  {title} {\enquote {\bibinfo {title}
  {High-throughput calculations of magnetic topological materials},}\ }\href
  {\doibase 10.1038/s41586-020-2837-0} {\bibfield  {journal} {\bibinfo
  {journal} {Nature}\ }\textbf {\bibinfo {volume} {586}},\ \bibinfo {pages}
  {702--707} (\bibinfo {year} {2020})}\BibitemShut {NoStop}%
\bibitem [{\citenamefont {Elcoro}\ \emph {et~al.}(2020)\citenamefont {Elcoro},
  \citenamefont {Wieder}, \citenamefont {Song}, \citenamefont {Xu},
  \citenamefont {Bradlyn},\ and\ \citenamefont
  {Bernevig}}]{elcoro2020magnetic}%
  \BibitemOpen
  \bibfield  {author} {\bibinfo {author} {\bibfnamefont {L.}~\bibnamefont
  {Elcoro}}, \bibinfo {author} {\bibfnamefont {B.~J.}\ \bibnamefont {Wieder}},
  \bibinfo {author} {\bibfnamefont {Z.~D.}\ \bibnamefont {Song}}, \bibinfo
  {author} {\bibfnamefont {Y.~F.}\ \bibnamefont {Xu}}, \bibinfo {author}
  {\bibfnamefont {B.}~\bibnamefont {Bradlyn}}, \ and\ \bibinfo {author}
  {\bibfnamefont {B.~A.}\ \bibnamefont {Bernevig}},\ }\bibfield  {title}
  {\enquote {\bibinfo {title} {Magnetic topological quantum chemistry},}\
  }\href@noop {} {\bibfield  {journal} {\bibinfo  {journal} {arXiv:2010.00598}\
  } (\bibinfo {year} {2020})}\BibitemShut {NoStop}%
\bibitem [{\citenamefont {Bouhon}\ \emph {et~al.}(2021)\citenamefont {Bouhon},
  \citenamefont {Lange},\ and\ \citenamefont {Slager}}]{Bouhon2021}%
  \BibitemOpen
  \bibfield  {author} {\bibinfo {author} {\bibfnamefont {A.}~\bibnamefont
  {Bouhon}}, \bibinfo {author} {\bibfnamefont {G.~F.}\ \bibnamefont {Lange}}, \
  and\ \bibinfo {author} {\bibfnamefont {R.-J.}\ \bibnamefont {Slager}},\
  }\bibfield  {title} {\enquote {\bibinfo {title} {Topological correspondence
  between magnetic space group representations and subdimensions},}\ }\href
  {\doibase 10.1103/PhysRevB.103.245127} {\bibfield  {journal} {\bibinfo
  {journal} {Phys. Rev. B}\ }\textbf {\bibinfo {volume} {103}},\ \bibinfo
  {pages} {245127} (\bibinfo {year} {2021})}\BibitemShut {NoStop}%
\bibitem [{\citenamefont {Lange}\ \emph {et~al.}(2021)\citenamefont {Lange},
  \citenamefont {Bouhon},\ and\ \citenamefont {Slager}}]{Lange2021}%
  \BibitemOpen
  \bibfield  {author} {\bibinfo {author} {\bibfnamefont {Gunnar~F.}\
  \bibnamefont {Lange}}, \bibinfo {author} {\bibfnamefont {Adrien}\
  \bibnamefont {Bouhon}}, \ and\ \bibinfo {author} {\bibfnamefont {Robert-Jan}\
  \bibnamefont {Slager}},\ }\bibfield  {title} {\enquote {\bibinfo {title}
  {Subdimensional topologies, indicators, and higher order boundary effects},}\
  }\href {\doibase 10.1103/PhysRevB.103.195145} {\bibfield  {journal} {\bibinfo
   {journal} {Phys. Rev. B}\ }\textbf {\bibinfo {volume} {103}},\ \bibinfo
  {pages} {195145} (\bibinfo {year} {2021})}\BibitemShut {NoStop}%
\bibitem [{\citenamefont {Freed}\ and\ \citenamefont
  {Moore}(2013)}]{freed2013twisted}%
  \BibitemOpen
  \bibfield  {author} {\bibinfo {author} {\bibfnamefont {D.~S.}\ \bibnamefont
  {Freed}}\ and\ \bibinfo {author} {\bibfnamefont {G.~W.}\ \bibnamefont
  {Moore}},\ }\bibfield  {title} {\enquote {\bibinfo {title} {Twisted
  equivariant matter},}\ }in\ \href {\doibase 10.1007/s00023-013-0236-x} {\emph
  {\bibinfo {booktitle} {Ann. Henri Poincaré}}},\ Vol.~\bibinfo {volume} {14}\
  (\bibinfo {organization} {Springer},\ \bibinfo {year} {2013})\ pp.\ \bibinfo
  {pages} {1927--2023}\BibitemShut {NoStop}%
\bibitem [{\citenamefont {Okuma}\ \emph {et~al.}(2019)\citenamefont {Okuma},
  \citenamefont {Sato},\ and\ \citenamefont {Shiozaki}}]{Okuma2019}%
  \BibitemOpen
  \bibfield  {author} {\bibinfo {author} {\bibfnamefont {N.}~\bibnamefont
  {Okuma}}, \bibinfo {author} {\bibfnamefont {M.}~\bibnamefont {Sato}}, \ and\
  \bibinfo {author} {\bibfnamefont {K.}~\bibnamefont {Shiozaki}},\ }\bibfield
  {title} {\enquote {\bibinfo {title} {Topological classification under
  nonmagnetic and magnetic point group symmetry: application of real-space
  {A}tiyah-{H}irzebruch spectral sequence to higher-order topology},}\ }\href
  {\doibase 10.1103/PhysRevB.99.085127} {\bibfield  {journal} {\bibinfo
  {journal} {Phys. Rev. B}\ }\textbf {\bibinfo {volume} {99}},\ \bibinfo
  {pages} {085127} (\bibinfo {year} {2019})}\BibitemShut {NoStop}%
\bibitem [{\citenamefont {Yu}\ \emph {et~al.}(2011)\citenamefont {Yu},
  \citenamefont {Qi}, \citenamefont {Bernevig}, \citenamefont {Fang},\ and\
  \citenamefont {Dai}}]{yu2011equivalent}%
  \BibitemOpen
  \bibfield  {author} {\bibinfo {author} {\bibfnamefont {R.}~\bibnamefont
  {Yu}}, \bibinfo {author} {\bibfnamefont {X.-L.}\ \bibnamefont {Qi}}, \bibinfo
  {author} {\bibfnamefont {B.~A.}\ \bibnamefont {Bernevig}}, \bibinfo {author}
  {\bibfnamefont {Z.}~\bibnamefont {Fang}}, \ and\ \bibinfo {author}
  {\bibfnamefont {X.}~\bibnamefont {Dai}},\ }\bibfield  {title} {\enquote
  {\bibinfo {title} {Equivalent expression of $\mathbb{Z}_2$ topological
  invariant for band insulators using the non-abelian {B}erry connection},}\
  }\href {\doibase 10.1103/PhysRevB.84.075119} {\bibfield  {journal} {\bibinfo
  {journal} {Phys. Rev. B}\ }\textbf {\bibinfo {volume} {84}},\ \bibinfo
  {pages} {075119} (\bibinfo {year} {2011})}\BibitemShut {NoStop}%
\bibitem [{\citenamefont {Alexandradinata}\ \emph {et~al.}(2014)\citenamefont
  {Alexandradinata}, \citenamefont {Dai},\ and\ \citenamefont
  {Bernevig}}]{alexandradinata2014wilson}%
  \BibitemOpen
  \bibfield  {author} {\bibinfo {author} {\bibfnamefont {A}~\bibnamefont
  {Alexandradinata}}, \bibinfo {author} {\bibfnamefont {X.}~\bibnamefont
  {Dai}}, \ and\ \bibinfo {author} {\bibfnamefont {B.~A.}\ \bibnamefont
  {Bernevig}},\ }\bibfield  {title} {\enquote {\bibinfo {title} {Wilson-loop
  characterization of inversion-symmetric topological insulators},}\ }\href
  {\doibase 10.1103/PhysRevB.89.155114} {\bibfield  {journal} {\bibinfo
  {journal} {Phys. Rev. B}\ }\textbf {\bibinfo {volume} {89}},\ \bibinfo
  {pages} {155114} (\bibinfo {year} {2014})}\BibitemShut {NoStop}%
\bibitem [{\citenamefont {Taherinejad}\ \emph {et~al.}(2014)\citenamefont
  {Taherinejad}, \citenamefont {Garrity},\ and\ \citenamefont
  {Vanderbilt}}]{Taherinejad2014}%
  \BibitemOpen
  \bibfield  {author} {\bibinfo {author} {\bibfnamefont {M.}~\bibnamefont
  {Taherinejad}}, \bibinfo {author} {\bibfnamefont {K.~F.}\ \bibnamefont
  {Garrity}}, \ and\ \bibinfo {author} {\bibfnamefont {D.}~\bibnamefont
  {Vanderbilt}},\ }\bibfield  {title} {\enquote {\bibinfo {title} {{Wannier
  center sheets in topological insulators}},}\ }\href {\doibase
  10.1103/PhysRevB.89.115102} {\bibfield  {journal} {\bibinfo  {journal} {Phys.
  Rev. B}\ }\textbf {\bibinfo {volume} {89}},\ \bibinfo {pages} {1--14}
  (\bibinfo {year} {2014})},\ \Eprint {http://arxiv.org/abs/1312.6940}
  {1312.6940} \BibitemShut {NoStop}%
\bibitem [{\citenamefont {Gresch}\ \emph {et~al.}(2017)\citenamefont {Gresch},
  \citenamefont {Aut\`es}, \citenamefont {Yazyev}, \citenamefont {Troyer},
  \citenamefont {Vanderbilt}, \citenamefont {Bernevig},\ and\ \citenamefont
  {Soluyanov}}]{Z2pack}%
  \BibitemOpen
  \bibfield  {author} {\bibinfo {author} {\bibfnamefont {D.}~\bibnamefont
  {Gresch}}, \bibinfo {author} {\bibfnamefont {G.}~\bibnamefont {Aut\`es}},
  \bibinfo {author} {\bibfnamefont {O.~V.}\ \bibnamefont {Yazyev}}, \bibinfo
  {author} {\bibfnamefont {M.}~\bibnamefont {Troyer}}, \bibinfo {author}
  {\bibfnamefont {D.}~\bibnamefont {Vanderbilt}}, \bibinfo {author}
  {\bibfnamefont {B.~A.}\ \bibnamefont {Bernevig}}, \ and\ \bibinfo {author}
  {\bibfnamefont {A.~A.}\ \bibnamefont {Soluyanov}},\ }\bibfield  {title}
  {\enquote {\bibinfo {title} {Z2pack: Numerical implementation of hybrid
  wannier centers for identifying topological materials},}\ }\href {\doibase
  10.1103/PhysRevB.95.075146} {\bibfield  {journal} {\bibinfo  {journal} {Phys.
  Rev. B}\ }\textbf {\bibinfo {volume} {95}},\ \bibinfo {pages} {075146}
  (\bibinfo {year} {2017})}\BibitemShut {NoStop}%
\bibitem [{\citenamefont {Soluyanov}\ and\ \citenamefont
  {Vanderbilt}(2011)}]{Soluyanov2011}%
  \BibitemOpen
  \bibfield  {author} {\bibinfo {author} {\bibfnamefont {A.~A.}\ \bibnamefont
  {Soluyanov}}\ and\ \bibinfo {author} {\bibfnamefont {D.}~\bibnamefont
  {Vanderbilt}},\ }\bibfield  {title} {\enquote {\bibinfo {title} {Computing
  topological invariants without inversion symmetry},}\ }\href {\doibase
  10.1103/PhysRevB.83.235401} {\bibfield  {journal} {\bibinfo  {journal} {Phys.
  Rev. B}\ }\textbf {\bibinfo {volume} {83}},\ \bibinfo {pages} {235401}
  (\bibinfo {year} {2011})}\BibitemShut {NoStop}%
\bibitem [{\citenamefont {Bouhon}\ \emph {et~al.}(2019)\citenamefont {Bouhon},
  \citenamefont {Black-Schaffer},\ and\ \citenamefont
  {Slager}}]{bouhon2019wilson}%
  \BibitemOpen
  \bibfield  {author} {\bibinfo {author} {\bibfnamefont {A.}~\bibnamefont
  {Bouhon}}, \bibinfo {author} {\bibfnamefont {A.~M.}\ \bibnamefont
  {Black-Schaffer}}, \ and\ \bibinfo {author} {\bibfnamefont {R.-J.}\
  \bibnamefont {Slager}},\ }\bibfield  {title} {\enquote {\bibinfo {title}
  {Wilson loop approach to fragile topology of split elementary band
  representations and topological crystalline insulators with time-reversal
  symmetry},}\ }\href {\doibase 10.1103/PhysRevB.100.195135} {\bibfield
  {journal} {\bibinfo  {journal} {Phys. Rev. B}\ }\textbf {\bibinfo {volume}
  {100}},\ \bibinfo {pages} {195135} (\bibinfo {year} {2019})}\BibitemShut
  {NoStop}%
\bibitem [{\citenamefont {Bradlyn}\ \emph {et~al.}(2019)\citenamefont
  {Bradlyn}, \citenamefont {Wang}, \citenamefont {Cano},\ and\ \citenamefont
  {Bernevig}}]{bradlyn2019disconnected}%
  \BibitemOpen
  \bibfield  {author} {\bibinfo {author} {\bibfnamefont {B.}~\bibnamefont
  {Bradlyn}}, \bibinfo {author} {\bibfnamefont {Z.}~\bibnamefont {Wang}},
  \bibinfo {author} {\bibfnamefont {J.}~\bibnamefont {Cano}}, \ and\ \bibinfo
  {author} {\bibfnamefont {B.~A.}\ \bibnamefont {Bernevig}},\ }\bibfield
  {title} {\enquote {\bibinfo {title} {Disconnected elementary band
  representations, fragile topology, and wilson loops as topological indices:
  An example on the triangular lattice},}\ }\href {\doibase
  10.1103/PhysRevB.99.045140} {\bibfield  {journal} {\bibinfo  {journal} {Phys.
  Rev. B}\ }\textbf {\bibinfo {volume} {99}},\ \bibinfo {pages} {045140}
  (\bibinfo {year} {2019})}\BibitemShut {NoStop}%
\bibitem [{\citenamefont {Mao}\ \emph {et~al.}(2011)\citenamefont {Mao},
  \citenamefont {Yamakage},\ and\ \citenamefont {Kuramoto}}]{Mao2011}%
  \BibitemOpen
  \bibfield  {author} {\bibinfo {author} {\bibfnamefont {S.}~\bibnamefont
  {Mao}}, \bibinfo {author} {\bibfnamefont {A.}~\bibnamefont {Yamakage}}, \
  and\ \bibinfo {author} {\bibfnamefont {Y.}~\bibnamefont {Kuramoto}},\
  }\bibfield  {title} {\enquote {\bibinfo {title} {Tight-binding model for
  topological insulators: Analysis of helical surface modes over the whole
  brillouin zone},}\ }\href {\doibase 10.1103/PhysRevB.84.115413} {\bibfield
  {journal} {\bibinfo  {journal} {Phys. Rev. B}\ }\textbf {\bibinfo {volume}
  {84}},\ \bibinfo {pages} {115413} (\bibinfo {year} {2011})}\BibitemShut
  {NoStop}%
\bibitem [{\citenamefont {Tyner}\ \emph {et~al.}(2020)\citenamefont {Tyner},
  \citenamefont {Sur}, \citenamefont {Puggioni}, \citenamefont {Rondinelli},\
  and\ \citenamefont {Goswami}}]{tyner2020topology}%
  \BibitemOpen
  \bibfield  {author} {\bibinfo {author} {\bibfnamefont {A.~C.}\ \bibnamefont
  {Tyner}}, \bibinfo {author} {\bibfnamefont {S.}~\bibnamefont {Sur}}, \bibinfo
  {author} {\bibfnamefont {D.}~\bibnamefont {Puggioni}}, \bibinfo {author}
  {\bibfnamefont {J.~M.}\ \bibnamefont {Rondinelli}}, \ and\ \bibinfo {author}
  {\bibfnamefont {P.}~\bibnamefont {Goswami}},\ }\bibfield  {title} {\enquote
  {\bibinfo {title} {Topology of three-dimensional {D}irac semimetals and
  generalized quantum spin {H}all systems without gapless edge modes},}\ }\href
  {https://doi.org/10.48550/arXiv.2012.12906} {\bibfield  {journal} {\bibinfo
  {journal} {arXiv:2012.12906v2}\ } (\bibinfo {year} {2020})}\BibitemShut
  {NoStop}%
\bibitem [{\citenamefont {Tyner}\ \emph {et~al.}(2021)\citenamefont {Tyner}
  \emph {et~al.}}]{tyner2021quantized}%
  \BibitemOpen
  \bibfield  {author} {\bibinfo {author} {\bibfnamefont {A.~C.}\ \bibnamefont
  {Tyner}} \emph {et~al.},\ }\bibfield  {title} {\enquote {\bibinfo {title}
  {Quantized non-abelian, {B}erry's flux and higher-order topology of
  {N}a$_3${B}i},}\ }\href {https://doi.org/10.48550/arXiv.2102.06207}
  {\bibfield  {journal} {\bibinfo  {journal} {arXiv:2102.06207}\ } (\bibinfo
  {year} {2021})}\BibitemShut {NoStop}%
\bibitem [{\citenamefont {Schindler}\ \emph {et~al.}(2018)\citenamefont
  {Schindler} \emph {et~al.}}]{schindler2018higher}%
  \BibitemOpen
  \bibfield  {author} {\bibinfo {author} {\bibfnamefont {F.}~\bibnamefont
  {Schindler}} \emph {et~al.},\ }\bibfield  {title} {\enquote {\bibinfo {title}
  {Higher-order topology in bismuth},}\ }\href {\doibase
  10.1038/s41567-020-0902-0} {\bibfield  {journal} {\bibinfo  {journal} {Nat.
  Phys.}\ }\textbf {\bibinfo {volume} {14}},\ \bibinfo {pages} {918--924}
  (\bibinfo {year} {2018})}\BibitemShut {NoStop}%
\bibitem [{\citenamefont {Rudenko}\ \emph {et~al.}(2017)\citenamefont
  {Rudenko}, \citenamefont {Katsnelson},\ and\ \citenamefont
  {Rold\'an}}]{Rudenko2017}%
  \BibitemOpen
  \bibfield  {author} {\bibinfo {author} {\bibfnamefont {A.~N.}\ \bibnamefont
  {Rudenko}}, \bibinfo {author} {\bibfnamefont {M.~I.}\ \bibnamefont
  {Katsnelson}}, \ and\ \bibinfo {author} {\bibfnamefont {R.}~\bibnamefont
  {Rold\'an}},\ }\bibfield  {title} {\enquote {\bibinfo {title} {Electronic
  properties of single-layer antimony: Tight-binding model, spin-orbit
  coupling, and the strength of effective coulomb interactions},}\ }\href
  {\doibase 10.1103/PhysRevB.95.081407} {\bibfield  {journal} {\bibinfo
  {journal} {Phys. Rev. B}\ }\textbf {\bibinfo {volume} {95}},\ \bibinfo
  {pages} {081407} (\bibinfo {year} {2017})}\BibitemShut {NoStop}%
\bibitem [{\citenamefont {Kim}\ \emph {et~al.}(2016)\citenamefont {Kim} \emph
  {et~al.}}]{kim2016topological}%
  \BibitemOpen
  \bibfield  {author} {\bibinfo {author} {\bibfnamefont {S.~H.}\ \bibnamefont
  {Kim}} \emph {et~al.},\ }\bibfield  {title} {\enquote {\bibinfo {title}
  {Topological phase transition and quantum spin hall edge states of antimony
  few layers},}\ }\href {\doibase 10.1038/srep33193} {\bibfield  {journal}
  {\bibinfo  {journal} {Sci. Rep.}\ }\textbf {\bibinfo {volume} {6}},\ \bibinfo
  {pages} {1--7} (\bibinfo {year} {2016})}\BibitemShut {NoStop}%
\bibitem [{\citenamefont {Zhu}\ \emph {et~al.}(2019)\citenamefont {Zhu} \emph
  {et~al.}}]{zhu2019evidence}%
  \BibitemOpen
  \bibfield  {author} {\bibinfo {author} {\bibfnamefont {S.-Y.}\ \bibnamefont
  {Zhu}} \emph {et~al.},\ }\bibfield  {title} {\enquote {\bibinfo {title}
  {Evidence of topological edge states in buckled antimonene monolayers},}\
  }\href {\doibase 10.1021/acs.nanolett.9b02444} {\bibfield  {journal}
  {\bibinfo  {journal} {Nano Lett.}\ }\textbf {\bibinfo {volume} {19}},\
  \bibinfo {pages} {6323--6329} (\bibinfo {year} {2019})}\BibitemShut {NoStop}%
\bibitem [{\citenamefont {Bieniek}\ \emph {et~al.}(2017)\citenamefont
  {Bieniek}, \citenamefont {Wo{\'z}niak},\ and\ \citenamefont
  {Potasz}}]{bieniek2017stability}%
  \BibitemOpen
  \bibfield  {author} {\bibinfo {author} {\bibfnamefont {M.}~\bibnamefont
  {Bieniek}}, \bibinfo {author} {\bibfnamefont {T.}~\bibnamefont
  {Wo{\'z}niak}}, \ and\ \bibinfo {author} {\bibfnamefont {P.}~\bibnamefont
  {Potasz}},\ }\bibfield  {title} {\enquote {\bibinfo {title} {Stability of
  topological properties of bismuth (1 1 1) bilayer},}\ }\href {\doibase
  10.1088/1361-648X/aa5e79} {\bibfield  {journal} {\bibinfo  {journal} {J.
  Condens. Matter Phys.}\ }\textbf {\bibinfo {volume} {29}},\ \bibinfo {pages}
  {155501} (\bibinfo {year} {2017})}\BibitemShut {NoStop}%
\bibitem [{\citenamefont {Hsu}\ \emph {et~al.}(2019)\citenamefont {Hsu} \emph
  {et~al.}}]{hsu2019topology}%
  \BibitemOpen
  \bibfield  {author} {\bibinfo {author} {\bibfnamefont {C.-H.}\ \bibnamefont
  {Hsu}} \emph {et~al.},\ }\bibfield  {title} {\enquote {\bibinfo {title}
  {Topology on a new facet of bismuth},}\ }\href {\doibase
  10.1073/pnas.1900527116} {\bibfield  {journal} {\bibinfo  {journal} {Proc.
  Natl. Acad. Sci.}\ }\textbf {\bibinfo {volume} {116}},\ \bibinfo {pages}
  {13255--13259} (\bibinfo {year} {2019})}\BibitemShut {NoStop}%
\bibitem [{\citenamefont {Hofmann}(2006)}]{hofmann2006surfaces}%
  \BibitemOpen
  \bibfield  {author} {\bibinfo {author} {\bibfnamefont {P.}~\bibnamefont
  {Hofmann}},\ }\bibfield  {title} {\enquote {\bibinfo {title} {The surfaces of
  bismuth: Structural and electronic properties},}\ }\href {\doibase
  10.1016/j.progsurf.2006.03.001} {\bibfield  {journal} {\bibinfo  {journal}
  {Prog. Surf. Sci.}\ }\textbf {\bibinfo {volume} {81}},\ \bibinfo {pages}
  {191--245} (\bibinfo {year} {2006})}\BibitemShut {NoStop}%
\bibitem [{\citenamefont {Wada}\ \emph {et~al.}(2011)\citenamefont {Wada},
  \citenamefont {Murakami}, \citenamefont {Freimuth},\ and\ \citenamefont
  {Bihlmayer}}]{MurakamiBiFilm}%
  \BibitemOpen
  \bibfield  {author} {\bibinfo {author} {\bibfnamefont {M.}~\bibnamefont
  {Wada}}, \bibinfo {author} {\bibfnamefont {S.}~\bibnamefont {Murakami}},
  \bibinfo {author} {\bibfnamefont {F.}~\bibnamefont {Freimuth}}, \ and\
  \bibinfo {author} {\bibfnamefont {G.}~\bibnamefont {Bihlmayer}},\ }\bibfield
  {title} {\enquote {\bibinfo {title} {Localized edge states in two-dimensional
  topological insulators: Ultrathin bi films},}\ }\href {\doibase
  10.1103/PhysRevB.83.121310} {\bibfield  {journal} {\bibinfo  {journal} {Phys.
  Rev. B}\ }\textbf {\bibinfo {volume} {83}},\ \bibinfo {pages} {121310}
  (\bibinfo {year} {2011})}\BibitemShut {NoStop}%
\bibitem [{\citenamefont {Rasche}\ \emph {et~al.}(2013)\citenamefont {Rasche}
  \emph {et~al.}}]{rasche2013stacked}%
  \BibitemOpen
  \bibfield  {author} {\bibinfo {author} {\bibfnamefont {B.}~\bibnamefont
  {Rasche}} \emph {et~al.},\ }\bibfield  {title} {\enquote {\bibinfo {title}
  {Stacked topological insulator built from bismuth-based graphene sheet
  analogues},}\ }\href {\doibase 10.1038/nmat3570} {\bibfield  {journal}
  {\bibinfo  {journal} {Nat. Mater.}\ }\textbf {\bibinfo {volume} {12}},\
  \bibinfo {pages} {422--425} (\bibinfo {year} {2013})}\BibitemShut {NoStop}%
\bibitem [{\citenamefont {Drozdov}\ \emph {et~al.}(2014)\citenamefont {Drozdov}
  \emph {et~al.}}]{drozdov2014one}%
  \BibitemOpen
  \bibfield  {author} {\bibinfo {author} {\bibfnamefont {I.~K.}\ \bibnamefont
  {Drozdov}} \emph {et~al.},\ }\bibfield  {title} {\enquote {\bibinfo {title}
  {One-dimensional topological edge states of bismuth bilayers},}\ }\href
  {\doibase 10.1038/nphys3048} {\bibfield  {journal} {\bibinfo  {journal} {Nat.
  Phys.}\ }\textbf {\bibinfo {volume} {10}},\ \bibinfo {pages} {664--669}
  (\bibinfo {year} {2014})}\BibitemShut {NoStop}%
\bibitem [{\citenamefont {Chen}\ \emph {et~al.}(2013)\citenamefont {Chen},
  \citenamefont {Wang},\ and\ \citenamefont {Liu}}]{Chen2013bilayer}%
  \BibitemOpen
  \bibfield  {author} {\bibinfo {author} {\bibfnamefont {L.}~\bibnamefont
  {Chen}}, \bibinfo {author} {\bibfnamefont {Z.~F.}\ \bibnamefont {Wang}}, \
  and\ \bibinfo {author} {\bibfnamefont {F.}~\bibnamefont {Liu}},\ }\bibfield
  {title} {\enquote {\bibinfo {title} {Robustness of two-dimensional
  topological insulator states in bilayer bismuth against strain and electrical
  field},}\ }\href {\doibase 10.1103/PhysRevB.87.235420} {\bibfield  {journal}
  {\bibinfo  {journal} {Phys. Rev. B}\ }\textbf {\bibinfo {volume} {87}},\
  \bibinfo {pages} {235420} (\bibinfo {year} {2013})}\BibitemShut {NoStop}%
\bibitem [{\citenamefont {Nayak}\ \emph {et~al.}(2019)\citenamefont {Nayak}
  \emph {et~al.}}]{nayak2019resolving}%
  \BibitemOpen
  \bibfield  {author} {\bibinfo {author} {\bibfnamefont {A.~K.}\ \bibnamefont
  {Nayak}} \emph {et~al.},\ }\bibfield  {title} {\enquote {\bibinfo {title}
  {Resolving the topological classification of bismuth with topological
  defects},}\ }\href {\doibase 10.1126/sciadv.aax6996} {\bibfield  {journal}
  {\bibinfo  {journal} {Sci. Adv.}\ }\textbf {\bibinfo {volume} {5}},\ \bibinfo
  {pages} {eaax6996} (\bibinfo {year} {2019})}\BibitemShut {NoStop}%
\bibitem [{\citenamefont {Takayama}\ \emph {et~al.}(2015)\citenamefont
  {Takayama}, \citenamefont {Sato}, \citenamefont {Souma}, \citenamefont
  {Oguchi},\ and\ \citenamefont {Takahashi}}]{TakayamBi}%
  \BibitemOpen
  \bibfield  {author} {\bibinfo {author} {\bibfnamefont {A.}~\bibnamefont
  {Takayama}}, \bibinfo {author} {\bibfnamefont {T.}~\bibnamefont {Sato}},
  \bibinfo {author} {\bibfnamefont {S.}~\bibnamefont {Souma}}, \bibinfo
  {author} {\bibfnamefont {T.}~\bibnamefont {Oguchi}}, \ and\ \bibinfo {author}
  {\bibfnamefont {T.}~\bibnamefont {Takahashi}},\ }\bibfield  {title} {\enquote
  {\bibinfo {title} {One-dimensional edge states with giant spin splitting in a
  bismuth thin film},}\ }\href {\doibase 10.1103/PhysRevLett.114.066402}
  {\bibfield  {journal} {\bibinfo  {journal} {Phys. Rev. Lett.}\ }\textbf
  {\bibinfo {volume} {114}},\ \bibinfo {pages} {066402} (\bibinfo {year}
  {2015})}\BibitemShut {NoStop}%
\bibitem [{\citenamefont {Lei}\ \emph {et~al.}(2016)\citenamefont {Lei} \emph
  {et~al.}}]{lei2016electronic}%
  \BibitemOpen
  \bibfield  {author} {\bibinfo {author} {\bibfnamefont {T.}~\bibnamefont
  {Lei}} \emph {et~al.},\ }\bibfield  {title} {\enquote {\bibinfo {title}
  {Electronic structure evolution of single bilayer bi (1 1 1) film on 3d
  topological insulator bi2se x te3- x surfaces},}\ }\href {\doibase
  10.1088/0953-8984/28/25/255501} {\bibfield  {journal} {\bibinfo  {journal}
  {J. Condens. Matter Phys.}\ }\textbf {\bibinfo {volume} {28}},\ \bibinfo
  {pages} {255501} (\bibinfo {year} {2016})}\BibitemShut {NoStop}%
\bibitem [{\citenamefont {Chang}\ \emph {et~al.}(2019)\citenamefont {Chang}
  \emph {et~al.}}]{chang2019band}%
  \BibitemOpen
  \bibfield  {author} {\bibinfo {author} {\bibfnamefont {T.-R.}\ \bibnamefont
  {Chang}} \emph {et~al.},\ }\bibfield  {title} {\enquote {\bibinfo {title}
  {Band topology of bismuth quantum films},}\ }\href {\doibase
  10.3390/cryst9100510} {\bibfield  {journal} {\bibinfo  {journal} {Crystals}\
  }\textbf {\bibinfo {volume} {9}},\ \bibinfo {pages} {510} (\bibinfo {year}
  {2019})}\BibitemShut {NoStop}%
\bibitem [{\citenamefont {Ito}\ \emph {et~al.}(2016)\citenamefont {Ito} \emph
  {et~al.}}]{BiFilmIto}%
  \BibitemOpen
  \bibfield  {author} {\bibinfo {author} {\bibfnamefont {S.}~\bibnamefont
  {Ito}} \emph {et~al.},\ }\bibfield  {title} {\enquote {\bibinfo {title}
  {Proving nontrivial topology of pure bismuth by quantum confinement},}\
  }\href {\doibase 10.1103/PhysRevLett.117.236402} {\bibfield  {journal}
  {\bibinfo  {journal} {Phys. Rev. Lett.}\ }\textbf {\bibinfo {volume} {117}},\
  \bibinfo {pages} {236402} (\bibinfo {year} {2016})}\BibitemShut {NoStop}%
\bibitem [{\citenamefont {Saito}\ \emph {et~al.}(2016)\citenamefont {Saito},
  \citenamefont {Sawahata}, \citenamefont {Komine},\ and\ \citenamefont
  {Aono}}]{BiFilmTB}%
  \BibitemOpen
  \bibfield  {author} {\bibinfo {author} {\bibfnamefont {K.}~\bibnamefont
  {Saito}}, \bibinfo {author} {\bibfnamefont {H.}~\bibnamefont {Sawahata}},
  \bibinfo {author} {\bibfnamefont {T.}~\bibnamefont {Komine}}, \ and\ \bibinfo
  {author} {\bibfnamefont {T.}~\bibnamefont {Aono}},\ }\bibfield  {title}
  {\enquote {\bibinfo {title} {Tight-binding theory of surface spin states on
  bismuth thin films},}\ }\href {\doibase 10.1103/PhysRevB.93.041301}
  {\bibfield  {journal} {\bibinfo  {journal} {Phys. Rev. B}\ }\textbf {\bibinfo
  {volume} {93}},\ \bibinfo {pages} {041301} (\bibinfo {year}
  {2016})}\BibitemShut {NoStop}%
\bibitem [{\citenamefont {Liu}\ and\ \citenamefont
  {Allen}(1995)}]{liu1995electronic}%
  \BibitemOpen
  \bibfield  {author} {\bibinfo {author} {\bibfnamefont {Y.}~\bibnamefont
  {Liu}}\ and\ \bibinfo {author} {\bibfnamefont {R.~E.}\ \bibnamefont
  {Allen}},\ }\bibfield  {title} {\enquote {\bibinfo {title} {Electronic
  structure of the semimetals {B}i and {S}b},}\ }\href {\doibase
  10.1103/PhysRevB.52.1566} {\bibfield  {journal} {\bibinfo  {journal} {Phys.
  Rev. B}\ }\textbf {\bibinfo {volume} {52}},\ \bibinfo {pages} {1566--1577}
  (\bibinfo {year} {1995})}\BibitemShut {NoStop}%
\bibitem [{\citenamefont {Teo}\ \emph {et~al.}(2008)\citenamefont {Teo},
  \citenamefont {Fu},\ and\ \citenamefont {Kane}}]{teo2008surface}%
  \BibitemOpen
  \bibfield  {author} {\bibinfo {author} {\bibfnamefont {J.~C.~Y.}\
  \bibnamefont {Teo}}, \bibinfo {author} {\bibfnamefont {L.}~\bibnamefont
  {Fu}}, \ and\ \bibinfo {author} {\bibfnamefont {C.~L.}\ \bibnamefont
  {Kane}},\ }\bibfield  {title} {\enquote {\bibinfo {title} {Surface states and
  topological invariants in three-dimensional topological insulators:
  Application to {B}i$_{1- x}${S}b$_x$},}\ }\href {\doibase
  10.1103/PhysRevB.78.045426} {\bibfield  {journal} {\bibinfo  {journal} {Phys.
  Rev. B}\ }\textbf {\bibinfo {volume} {78}},\ \bibinfo {pages} {045426}
  (\bibinfo {year} {2008})}\BibitemShut {NoStop}%
\bibitem [{\citenamefont {Golin}(1968)}]{golin}%
  \BibitemOpen
  \bibfield  {author} {\bibinfo {author} {\bibfnamefont {S.}~\bibnamefont
  {Golin}},\ }\bibfield  {title} {\enquote {\bibinfo {title} {Band structure of
  bismuth: Pseudopotential approach},}\ }\href {\doibase
  10.1103/PhysRev.166.643} {\bibfield  {journal} {\bibinfo  {journal} {Phys.
  Rev.}\ }\textbf {\bibinfo {volume} {166}},\ \bibinfo {pages} {643--651}
  (\bibinfo {year} {1968})}\BibitemShut {NoStop}%
\bibitem [{\citenamefont {Jain}\ \emph {et~al.}(2013)\citenamefont {Jain} \emph
  {et~al.}}]{Jain2013}%
  \BibitemOpen
  \bibfield  {author} {\bibinfo {author} {\bibfnamefont {A.}~\bibnamefont
  {Jain}} \emph {et~al.},\ }\bibfield  {title} {\enquote {\bibinfo {title}
  {Commentary: The materials project: A materials genome approach to
  accelerating materials innovation},}\ }\href {\doibase 10.1063/1.4812323}
  {\bibfield  {journal} {\bibinfo  {journal} {APL Materials}\ }\textbf
  {\bibinfo {volume} {1}},\ \bibinfo {pages} {011002} (\bibinfo {year}
  {2013})}\BibitemShut {NoStop}%
\bibitem [{\citenamefont {Giannozzi}\ \emph {et~al.}(2009)\citenamefont
  {Giannozzi} \emph {et~al.}}]{QE-2009}%
  \BibitemOpen
  \bibfield  {author} {\bibinfo {author} {\bibfnamefont {P.}~\bibnamefont
  {Giannozzi}} \emph {et~al.},\ }\bibfield  {title} {\enquote {\bibinfo {title}
  {Quantum espresso: a modular and open-source software project for quantum
  simulations of materials},}\ }\href {http://www.quantum-espresso.org}
  {\bibfield  {journal} {\bibinfo  {journal} {J. Phys. Condens. Matter}\
  }\textbf {\bibinfo {volume} {21}},\ \bibinfo {pages} {395502 (19pp)}
  (\bibinfo {year} {2009})}\BibitemShut {NoStop}%
\bibitem [{\citenamefont {Giannozzi}\ \emph {et~al.}(2017)\citenamefont
  {Giannozzi} \emph {et~al.}}]{QE-2017}%
  \BibitemOpen
  \bibfield  {author} {\bibinfo {author} {\bibfnamefont {P.}~\bibnamefont
  {Giannozzi}} \emph {et~al.},\ }\bibfield  {title} {\enquote {\bibinfo {title}
  {Advanced capabilities for materials modelling with quantum espresso},}\
  }\href {http://stacks.iop.org/0953-8984/29/i=46/a=465901} {\bibfield
  {journal} {\bibinfo  {journal} {J. Phys. Condens. Matter}\ }\textbf {\bibinfo
  {volume} {29}},\ \bibinfo {pages} {465901} (\bibinfo {year}
  {2017})}\BibitemShut {NoStop}%
\bibitem [{\citenamefont {Giannozzi}\ \emph {et~al.}(2020)\citenamefont
  {Giannozzi} \emph {et~al.}}]{QE-2020}%
  \BibitemOpen
  \bibfield  {author} {\bibinfo {author} {\bibfnamefont {P.}~\bibnamefont
  {Giannozzi}} \emph {et~al.},\ }\bibfield  {title} {\enquote {\bibinfo {title}
  {Quantum espresso toward the exascale},}\ }\href {\doibase 10.1063/5.0005082}
  {\bibfield  {journal} {\bibinfo  {journal} {J. Chem. Phys.}\ }\textbf
  {\bibinfo {volume} {152}},\ \bibinfo {pages} {154105} (\bibinfo {year}
  {2020})}\BibitemShut {NoStop}%
\bibitem [{\citenamefont {Perdew}\ \emph {et~al.}(1996)\citenamefont {Perdew},
  \citenamefont {Burke},\ and\ \citenamefont {Ernzerhof}}]{Perdew1996}%
  \BibitemOpen
  \bibfield  {author} {\bibinfo {author} {\bibfnamefont {J.~P.}\ \bibnamefont
  {Perdew}}, \bibinfo {author} {\bibfnamefont {K.}~\bibnamefont {Burke}}, \
  and\ \bibinfo {author} {\bibfnamefont {M.}~\bibnamefont {Ernzerhof}},\
  }\bibfield  {title} {\enquote {\bibinfo {title} {Generalized gradient
  approximation made simple},}\ }\href@noop {} {\bibfield  {journal} {\bibinfo
  {journal} {Phys. Rev. Lett.}\ }\textbf {\bibinfo {volume} {77}},\ \bibinfo
  {pages} {3865} (\bibinfo {year} {1996})}\BibitemShut {NoStop}%
\bibitem [{\citenamefont {Pizzi}\ \emph {et~al.}(2020)\citenamefont {Pizzi}
  \emph {et~al.}}]{Pizzi2020}%
  \BibitemOpen
  \bibfield  {author} {\bibinfo {author} {\bibfnamefont {P.}~\bibnamefont
  {Pizzi}} \emph {et~al.},\ }\bibfield  {title} {\enquote {\bibinfo {title}
  {Wannier90 as a community code: new features and applications},}\ }\href
  {\doibase 10.1088/1361-648x/ab51ff} {\bibfield  {journal} {\bibinfo
  {journal} {J. Phys. Condens. Matter}\ }\textbf {\bibinfo {volume} {32}},\
  \bibinfo {pages} {165902} (\bibinfo {year} {2020})}\BibitemShut {NoStop}%
\bibitem [{\citenamefont {Thouless}\ \emph {et~al.}(1982)\citenamefont
  {Thouless}, \citenamefont {Kohmoto}, \citenamefont {Nightingale},\ and\
  \citenamefont {den Nijs}}]{tknn1982}%
  \BibitemOpen
  \bibfield  {author} {\bibinfo {author} {\bibfnamefont {D.~J.}\ \bibnamefont
  {Thouless}}, \bibinfo {author} {\bibfnamefont {M.}~\bibnamefont {Kohmoto}},
  \bibinfo {author} {\bibfnamefont {M.~P.}\ \bibnamefont {Nightingale}}, \ and\
  \bibinfo {author} {\bibfnamefont {M.}~\bibnamefont {den Nijs}},\ }\bibfield
  {title} {\enquote {\bibinfo {title} {Quantized {H}all conductance in a
  two-dimensional periodic potential},}\ }\href {\doibase
  10.1103/PhysRevLett.49.405} {\bibfield  {journal} {\bibinfo  {journal} {Phys.
  Rev. Lett.}\ }\textbf {\bibinfo {volume} {49}},\ \bibinfo {pages} {405--408}
  (\bibinfo {year} {1982})}\BibitemShut {NoStop}%
\end{thebibliography}%

\end{document}